\documentclass[11pt]{article}
\usepackage{latexsym,epsfig}

\renewcommand{\theequation}{\arabic{section}.\arabic{equation}}
\begin{document}
%
%                                             Title/Authors/Abstract page
%                                             ---------------------------
\pagestyle{empty}                   
\begin{flushright}
 MS--TPI--98--19 \\
 August 1998 \\
 hep-th/9808155  \\
\mbox{}\\    \mbox{}\\    \mbox{}\\
\end{flushright}
\begin{center}
 {\Large \bf{Extended Iterative Scheme for QCD:} } \\
 {\Large \bf{the Four-Gluon Vertex} }
\end{center}
\mbox{}\\    \mbox{}\\    \mbox{}\\
\begin{center}
 L.\ Driesen and M.\ Stingl \\
 Institute for Theoretical Physics I, University of M\"unster \\
 D-48149 M\"unster (Westf.), Germany
\end{center}
\mbox{}\\    \mbox{}\\    \mbox{}\\    \mbox{}\\    \mbox{}\\
{\large \bf{Abstract:}} \
We study the self-consistency problem of the generalized Feynman rule
( nonperturbatively modified vertex of zeroth perturbative order ) for the 
4-gluon vertex function in the framework of an extended perturbation scheme 
accounting for non-analytic coupling dependence through the $\Lambda$ scale.
Tensorial structure is restricted to a minimal dynamically closed basis set. 
The self-consistency conditions are obtained at one loop, in Landau gauge, and
at the lowest approximation level $(r=1)$ of interest for QCD. At this level,
they are found to be linear in the nonperturbative 4-gluon coefficients, but 
strongly overdetermined due to the lack of manifest Bose symmetry in the 
relevant Dyson-Schwinger equation. The observed near decoupling from the 
2-and-3-point conditions permits least-squares quasisolutions for given 
2-and-3-point input within an effective one-parameter freedom. We present such
solutions for $N_{F}=2$ massless quarks and for the pure gluon theory, adapted
to the 2-and-3-point coefficients determined previously.
\newpage     \mbox{  }     \newpage
\pagestyle{plain}                                      % Sections 1 - 4
\setcounter{page}{1}
%

% Latex Source File for Sect. 1 of 4-gluon-vertex Paper
% -----------------------------------------------------
%
\section{The generalized Feynman rule $\Gamma_{\ 4V}^{[r,0]}$}
\setcounter{equation}{0}
The present paper continues, and brings to a provisional stage of
completion, the determination of generalized Feynman rules in an extended
perturbation scheme for QCD \cite{STI}, designed to account for the
strongly nonanalytic coupling dependence of correlation functions through
the renormalization-group invariant mass scale $\Lambda$. The generalized
rules, denoted $\Gamma^{[r,0]}$, are proper vertex functions of zeroth 
perturbative order ( no power corrections in the coupling $g^2$, as
indicated by the index $0$ ), but with a nonperturbative $\Lambda$ 
dependence which in turn is approximated systematically at a level 
characterized by the index $r$. In contrast to the ordinary Feynman rules, 
it is a nontrivial self-consistency problem for these extended rules to
reproduce themselves in the integral equations for vertex functions.
However, the divergence-related mechanism operative in the self-reproduction 
\cite{STI} does ensure that formation of $\Gamma^{[r,0]}$'s
remains rigorously restricted to the small number of vertices corresponding
to the ordinary Feynman rules -- the superficially divergent vertices.
In the companion paper \cite{VE3}, this self-consistency problem was set
up and solved for the vertices with up to three external legs, on the
lowest nontrivial approximation level ($r=1$, and one loop) of interest
for QCD.

Below, we complement this study by examining the highest superficially
divergent amplitude, the 4-vector (4-gluon)  vertex $\Gamma_{4V}$ \cite{DISS}.
There are several features 
that set this vertex function apart and motivate its
separate treatment. One is its sheer kinematical complication, due partly
to the large number of 6 Lorentz-scalar momentum variables, but mostly
to the exorbitantly lengthy tensor decompositions implied by its 4 color
and 4 Lorentz indices. In constructing the $r=1$ approximant $\Gamma
_{\ 4V}^{[r,0]}$ (section 2), one is thereby forced to adopt a 
theoretically motivated simplification of the full tensor structure. Another
special feature is that $\Gamma_{4V}$ represents the lowest QCD amplitude
which in the nonperturbative context exhibits the phenomenon of
{\em compensating poles} ( or, in the terminology of \cite{VE3}, of
negative {\em shadow} poles ), first noted in Abelian models in \cite{JCN}.
While these poles can be inferred uniquely already from the ``lower''
DS equations for $\Gamma_{3V}$ and $\Gamma_{2V}$ by suitable residue-
taking operations, it is now necessary to demonstrate that on the level 
of the 4-point equation they are in turn self-consistent ( section 3 ).
A natural by-product of this demonstration is then a ``reduced'' integral
equation for the remainder vertex $V_{4V}$ defined by
\begin{equation}                                             \label{V4def}
 \Gamma_{4V}^{[r,0]} \ = \ -(C_{1}^{[r]})_{2V,2V}
                         \ -(C_{2}^{[r]})_{2V,2V}
                         \ -(C_{3}^{[r]})_{2V,2V}
                         \ +V_{4V}^{[r,0]},
\end{equation}
where $C_{1,2,3}^{[r]}$ are the shadow-pole terms in the 3 channels of the
4-gluon amplitude. This amounts to rearranging the full connected-and-
amputated 4-gluon correlation function as
\begin{equation}                                             \label{T4dec}
 T_{4V} \ = \ A'_{1} \ + \ A'_{2} \ + \ A'_{3} \ + \ V_{4V},
\end{equation}
where $A'_{1,2,3}$ are ``softened'', i.e. one-shadow-irreducible, one-
gluon exchange diagrams formed by subtracting the shadow-pole term from
the ordinary dressed-one-gluon-exchange diagram $A_{1,2,3}$ in each 
of the 3 channels. The object $V_{4V}^{[r,0]}$ depends on only four Lorentz-
scalar variables, and establishing the self-consistency conditions for its
nonperturbative coefficients (section 4) is simpler than for the full
$\Gamma_{4V}$.
 
The latter conditions will exhibit a problem that presumably besets all
approximate treatments of vertex equations with three or more external
lines : if such vertices possess a symmetry among their external legs
such as Bose symmetry, their Dyson-Schwinger (DS) equations will in
general not display this symmetry manifestly, and any approximation to them
will produce unsymmetric terms. Forcing self-reproduction of a symmetric 
input by requiring these terms to vanish leads to {\em overdetermination} 
of the dynamical conditions. If this does not happen to be counteracted 
by underdetermination tendencies, as was the case for the $r=1$
three-point conditions derived in \cite{VE3}, one must settle for a
``compromise'' solution in the sense of least-squares error minimization
(section 5).-- Technically, the self-consistency conditions 
will be derived at the $r=1$ and one-loop $(l=1)$ levels, and 
in {\em Landau gauge}. The latter again provides considerable simplification
because the two ghost vertices at $l=1$ remain perturbative, and because
in this gauge a closed DS system can be written for the amplitudes with
only transverse (if any) gluon legs, whose tensorial complexity is
significantly lower. A disadvantage, of course, is that the self-consistency
( or lack of it ) for statements concerning longitudinal-gluon amplitudes, 
such as Slavnov-Taylor identities, cannot be checked directly 
from such a system. 

The 4-gluon vertex also occupies a special position in that it belongs
both to the class of superficially divergent ``basic'' vertices having
their own ( ordinary or extended ) Feynman rules, and to the class of
amplitudes capable of developing bound-state ( here, glueball ) poles --
a phenomenon otherwise restricted to the superficially {\em convergent}
higher vertices. In the present paper we are exclusively concerned with
the generalized Feynman rule, a {\em zeroth-order} quantity,
which like its ordinary counterpart contains no information as yet
about bound states. The latter, if present, arise from an infinite partial 
resummation of the {\em higher-order}, quasi-perturbative corrections
\begin{equation}                                              \label{hicor}
 [g(\nu)^{2}]^{p} \ \cdot \ \Gamma_{4V}^{[r,p)} (\ \{k\},\ \Lambda,\ \nu \ )
 \qquad (\ p \ = \ 1,2,3,\ldots \ ).
\end{equation}
As the notation recalls, these have their own nonperturbative
$\Lambda$ dependences, which may be complemented by an inversely
logarithmic ``perturbative'' dependence when reparametrizing the
quasi-perturbative series in terms of $\alpha_s (k^2)$ rather than 
$g^2 (\nu)$. Upon ladder or similar resummation, these may build up 
a pole in some Mandelstam variable ( the squared sum of some subset
of the momenta $\{k\}$ ) at some multiple of $\Lambda^{2}$ representing
the square of a bound-state mass. Such a pole is accompanied by a pair
of residue functions, each representing an infinite power series in 
$g^{2}$ starting at least with $p=1$. It is therefore an example of 
nonperturbative $\Lambda$ dependence embedded in an otherwise 
quasi-perturbative term ( starting at least at order $p=2$ ) 
of the amplitude. These characteristics should prevent confusion
between bound-state poles and the poles which in the present scheme
arise {\em as rational approximations to branch cuts} in the squares of
single-external-line momenta, and {\em in zeroth quasi-perturbative order}.
The 4-gluon function is unique in exhibiting both phenomena. It is
yet another matter ( and not touched upon in this article ) that in
a confining theory the calculation of {\em S-matrix elements} will
require a still larger set of generalized Feynman rules, comprising
hadron-to-quarks and hadron-to-gluons bound-state vertices for the outer
corners of S-matrix diagrams -- vertices of an intrinsically hybrid
nature, since their dependence on the squares of single-quark momenta
would have to be determined and approximated consistently with the 
generalized Feynman rules for the internal lines and vertices, whereas their
dependence on the total virtuality of the bound state is governed by
``resummed-perturbative'' bound-state dynamics.

%
%
% Latex Source File for Sect. 2 of 4-Gluon-Vertex Paper
%------------------------------------------------------
%
%
\section{Structure of the $r=1$ approximant}
\setcounter{equation}{0}
Due to the large number of possible tensor structures of a four-gluon amplitude
( e.g., 43 independent Lorentz tensors for a fully transverse amplitude,
each combined with 8 independent color tensors for\mbox{ $SU(3)_{C}$)}, 
one is forced, at the present stage, to adopt some theoretically
motivated restriction to a smaller subset of tensors. We briefly describe
such a restricted form, partly repeating material from appendix A of
\cite{STI} to keep the discussion self-contained. The amplitude
\begin{equation}                                           \label{V4V}
 \left(\, V^{[r,0]}_{\ 4V} (p_1,p_2,p_3,p_4) \,\right)^{\kappa\lambda
 \mu\nu}_{abcd}  \qquad (\, p_1+p_2+p_3+p_4 \,=\,0 \,)
\end{equation}
is expanded over an 18-member set of color $\otimes$ Lorentz tensors, 
of which 15 are linearly independent. They are formed from building blocks
\begin{equation}                                           \label{CTL}
 C^{(i)}_{abcd} \, L_{(j)}^{\kappa\lambda\mu\nu} \qquad
 (\, i=1 \ldots 6, \, j=1 \ldots 3 \,),
\end{equation}
where the color tensors $C^{(i)}$, fourth-rank objects over the adjoint
representation of $SU(3)_{C}$, are given by
\begin{eqnarray}
 C^{(1)}_{abcd} \,=\, \delta_{ab\,} \delta_{cd\,}, \quad & C^{(2)}_{abcd}
 \,=\,  \delta_{ac\,} \delta_{db\,}, \quad & C^{(3)}_{abcd} \,=\,
 \delta_{ad\,} \delta_{bc\,},                               \label{COLD} \\
 C^{(4)}_{abcd} \,=\,      f_{abn }      f_{cdn }, \quad & C^{(5)}_{abcd}
 \,=\,       f_{acn }      f_{dbn }, \quad & C^{(6)}_{abcd} \,=\,
      f_{adn }      f_{bcn },                               \label{COLF}
\end{eqnarray}
in terms of the $SU(3)$ structure constants $f_{abc}$. Only five of these
are linearly independent, since $C^{(4)}+C^{(5)}+C^{(6)} = 0$ by the
Jacobi identity. As compared to the most general structure, this set
omits three tensors containing a symmetric $SU(3)$ structure constant
$d_{abc}$. The Lorentz-tensor building blocks $L_{(j)}$ are the three
dimensionless ones,
\begin{equation}                                            \label{LOR}
 L_{(1)}^{\kappa\lambda\mu\nu}=\,\delta^{\kappa\lambda}\delta^{\mu\nu},\quad
 L_{(2)}^{\kappa\lambda\mu\nu}=\,\delta^{\kappa\mu}\delta^{\nu\lambda},\quad
 L_{(3)}^{\kappa\lambda\mu\nu}=\,\delta^{\kappa\nu}\delta^{\lambda\mu},
\end{equation}
which are linearly independent. In obvious ways, all these fall into
crossing triplets, where second and third members in each triplet arise
from the first member by the crossing operations,
\begin{eqnarray}
 `` \,s\, \rightarrow \,u\, `` \quad &:& \quad (2,3,4) \ \rightarrow
 \ (3,4,2),                                                 \label{S-U} \\
 `` \,s\, \rightarrow \,t\, `` \quad &:& \quad (2,3,4) \ \rightarrow
 \ (4,2,3),                                                 \label{S-T}
\end{eqnarray} 
the $s$ channel being by convention the $(1+2)\leftrightarrow(3+4)$
channel. Here $2$ stands for the set $(p_2,\lambda,b)$, etc.

The set (\ref{CTL}--\ref{LOR}) is distinguished in that it constitutes 
the {\em smallest dynamically closed subset} of the full tensor structure, 
i.e.:

  (i) It closes under Bethe-Salpeter (BS) iteration in each of the three
channels of the amplitude, as exemplified by the $s$-channel, color-space
multiplication,
\begin{equation}                                            \label{CMUL}
 \left(\, C^{(i)} \cdot C^{(j)} \,\right)_{abcd} \ =\ 
 \sum_{e,f=1}^{8} \,C^{(i)}_{abef} \,C^{(j)}_{efcd} \ ,
\end{equation}
and closes also under iteration of DS interaction diagrams, provided
the 3-gluon vertices involved have $f_{abc}$ color structure only. We
emphasized earlier \cite{VE3} that with 4-gluon color dependence
restricted to the five-dimensional basis (\ref{COLD}/\ref{COLF}), it
would be inconsistent to retain $d_{abc}$ color structure in the 3-gluon
vertex, since the tensors omitted on the 4-gluon level would also 
contribute to that structure via the hierarchical coupling in the 3-gluon 
DS equation.

 (ii) It closes under the crossing operations connecting the three 
channels, as given by (\ref{S-U}) and (\ref{S-T}).

(iii) It contains the perturbative zeroth-order vertex, namely,
\begin{equation}                                            \label{GANULL}
 \Gamma_{\ 4V}^{(0)pert} \ =\ \frac{3}{2} \,C^{(4)}\, \left(\,
 L_{(2)} - L_{(3)} \,\right) \ +\ \frac{1}{2} \, \left(\, C^{(5)} - 
 C^{(6)} \,\right) \left( -2L_{(1)} + L_{(2)} + L_{(3)} \,\right).
\end{equation}
( Without this property, there would exist a still smaller dynamically 
closed set comprising only the tensors (\ref{COLD}), but maintaining the
perturbative limit calls for inclusion of (\ref{COLF}) ). All these
statements are easy to verify by direct calculation.
 
For the choice (\ref{LOR}) of Lorentz tensors, additional motivation 
comes from the observation that these are the only ones accompanied by
invariant functions of zero mass dimension. The tensors omitted in
(\ref{LOR}) contain two or four powers of momentum, and are thus
associated with invariant functions of mass dimensions $-2$ or $-4$.
Together with the fact that their zeroth-order forms must have at least
one power of $\Lambda^2$ in each term, this severely restricts
possibilities for these functions to receive the divergent loop 
contributions required for self-reproduction, and it appears reasonable
then to start with a basis omitting such tensors.
 
Invariant functions for the $V_{4V}$ amplitude, we recall, have 
{\em zeroth-order} rational structure only with respect to the variables
$p_1{}^2 \ldots p_4{}^2$, since the entire zeroth-order rational structure
with respect to $s, \,u, \,t$ variables is borne by the compensating-pole 
terms of eq. (\ref{V4def}). Since our earlier demonstration of this applies, 
strictly speaking, only to the color-octet channels, we check in appendix
A of this paper that in fact it holds generally. Thus at order $r=1$,
all invariant functions involve the same, fully symmetric denominator
\begin{equation}                                              \label{DEN}
 \Delta^{[1,0]} \ =\ \left(\, p_1{}^2 \,+\, u_{2}'' \Lambda^2 \,\right)
                     \left(\, p_2{}^2 \,+\, u_{2}'' \Lambda^2 \,\right)
                     \left(\, p_3{}^2 \,+\, u_{2}'' \Lambda^2 \,\right)
                     \left(\, p_4{}^2 \,+\, u_{2}'' \Lambda^2 \,\right).
\end{equation}
Numerator polynomials in $\Lambda^2$ and the squared momenta, of mass
dimension 8, must conform to the restrictions of naive asymptotic
freedom. As seen in \cite{VE3} in the example of the 3-gluon vertex,
additional restrictions arise from the requirement that ``softened'' 
( one-shadow-irreducible ) one-gluon exchange diagrams, which represent
the {\em physical} one-gluon exchange mechanism, again should not generate
higher-than-perturbative degrees of divergence when inserted in loops 
( for an example, see the right-hand portion of diagram $(A3)_4$ in Fig.
10 ): this excludes terms in the rational approximant in which any one
squared momentum has a net positive power. Thus numerator polynomials
should not contain $s, \,t, \,u$ either, nor should they have terms of type
$\Lambda^{2} p_i^6$ or $\Lambda^{4} p_i^4$ still allowed by the 
primary restrictions. At level $r=1$, all zeroth-order rational structure
can then be expressed in terms of the single-pole quantities
\begin{equation}                                              \label{PIFA}
 \Pi_{i} \ = \ \frac{ \Lambda^2 }{ p_i{}^2 \,+\, u_2''\Lambda^2 }
 \qquad  (\ i \,=\,1\,,2\,,3\,,4 \ ) \, ,
\end{equation}
and the general $r=1$ approximant, with full Bose symmetry among the
four legs, turns out to involve seventeen dimensionless, real coefficients,
each preceding a product of a tensor of type (\ref{CTL}) and a combination
of one or more quantities (\ref{PIFA}).

In the following we use two different representations for $V^{[1,0]}_{4V}$.
The first has building blocks with manifest Bose symmetry but involving
linearly dependent color tensors internally; it uses and for our purpose 
defines the 17 independent coefficients. It is thus suitable for the 
{\em presentation} of the vertex as a generalized Feynman rule. 
This representation uses a color tensor
\begin{equation}                                               \label{TR4}
 C^{(S)}_{abcd} \ =\ {\rm tr} \left(\, t_a t_d t_b t_c \,\right)
                \ +\ {\rm tr} \left(\, t_a t_c t_b t_d \,\right)
\end{equation}
and its crossing partners $C^{(U)},C^{(T)}$ obtained via (\ref{S-U})
and (\ref{S-T}), where $t_n$ are the generator matrices of $SU(3)_{C}$
normalized by ${\rm tr}(t_mt_n)=\frac{1}{2}\delta_{mn}$. The latter trace is 
then used to supply another tensor triplet, proportional to (\ref{COLD}),
\begin{equation}                                               \label{TR2}
 {\rm tr} (t_a t_b)\, {\rm tr} (t_c t_d) \ =\ \frac{1}{4}
 C^{(1)}_{abcd},
\end{equation}
plus crossing partners. $SU(3)$ representation algebra \cite{SU3} gives
the relation of (\ref{TR4}) to (\ref{COLD}/\ref{COLF}) as
\begin{equation}                                               \label{CST}
 C^{(S)} \ =\ \frac{1}{12} \left[\, C^{(1)}+C^{(2)}+C^{(3)} \,\right]
         \ +\ \frac{1}{6 } \left[\, C^{(5)}-C^{(6)} \,\right],
\end{equation}
with analogous relations for $C^{(U)},C^{(T)}$. In this basis, which
has been used e.g. in work on the operator-product expansion \cite{OPE},
the vertex with four transverse gluon legs is written
\begin{eqnarray}
 \left[\, V^{[1,0]}_{\ 4T} (p_1 \ldots p_4) \,\right]^{\rho\sigma
 \tau\omega}_{abcd} \ = \ t^{\rho\kappa}(p_1) \,t^{\sigma\lambda}(p_2)
 \,t^{\tau\mu}(p_3) \,t^{\omega\nu}(p_4)                      \nonumber  \\ 
 \times\ \bigg\{ \,\left( \Gamma^{(0)pert}_{\ 4V} \right)^{\kappa\lambda
 \mu\nu}_{abcd} \ + \ \sum_{n=1}^{17} \,\zeta_{n} \, \Big(
 \,W^{[1,0]}_{n}(p_1^{\,2} \ldots p_4^{\,2}; \Lambda^2) \,\Big)^{\kappa
 \lambda\mu\nu}_{abcd} \bigg\},                              \label{VZETA}
\end{eqnarray}
the $t$'s now denoting transverse projectors in the external momenta.
Each of the building blocks $W_n$ comprises a color tensor (\ref{TR4})
or (\ref{TR2}) plus crossing partners, a combination of Lorentz tensors
(\ref{LOR}), and a combination of rational-function elements (\ref{PIFA}).
Each therefore has at least one overall power of $\Lambda^2$ and vanishes
in the perturbative limit, $\Lambda \rightarrow 0$.
A full listing of these building blocks is relegated to appendix B.1;
here we write out just one of them for illustration:
\begin{eqnarray}
 W_{\ 3}^{[1,0]} (\, p_1^{\,2} \ldots p_4^{\,2}; \Lambda^2 \,) \ &=& \ 
                                                           \nonumber    \\
    C^{(S)} \,\left(\, L_{(3)}-2L_{(1)}+L_{(2)} \,\right)\,
 \Big[\, \Pi_1\Pi_4+\Pi_2\Pi_3 &-& 2\left(\Pi_1\Pi_2+\Pi_3\Pi_4\right)
 \ +\  \Pi_1\Pi_3+\Pi_4\Pi_2 \,\Big]                       \nonumber    \\
 +\ C^{(U)} \,\left(\, L_{(1)}-2L_{(2)}+L_{(3)} \,\right)\,
 \Big[\, \Pi_1\Pi_2+\Pi_3\Pi_4 &-& 2\left(\Pi_1\Pi_3+\Pi_4\Pi_2\right)
 \ +\  \Pi_1\Pi_4+\Pi_2\Pi_3 \,\Big]                       \nonumber    \\
 +\ C^{(T)} \,\left(\, L_{(2)}-2L_{(3)}+L_{(1)} \,\right)\,
 \Big[\, \Pi_1\Pi_3+\Pi_4\Pi_2 &-& 2\left(\Pi_1\Pi_4+\Pi_2\Pi_3\right)
 \ +\  \Pi_1\Pi_2+\Pi_3\Pi_4 \,\Big] \,                       \label{WEE3}
\end{eqnarray}
Note that these 17 building blocks are linearly independent of each other; 
the linear dependence of the basis tensors $C^{(S,U,T)}$ affects only their 
{\em internal} structure, which upon rewriting in terms of independent
tensors would lose its manifest crossing ( and therefore Bose ) symmetry.
However, as a representation of the {\em total} $V_{4V}$, (\ref{VZETA}) is
based on dependent color tensors.

The second representation is more suitable for the self-consistency
{\em calculation} of the zeroth-order vertex. To permit tensor-by-tensor
comparison of the two sides of a DS equation, such a representation
must employ linearly independent color tensors, and consequently forgo
manifest Bose symmetry. However, since the way a $4V$ vertex appears
in a DS interaction diagram singles out one of the three channels 
( compare, e.g., the $4V$ vertex in diagrams $(A3)_4$ or $(C2)_4$ 
of Fig.10 ), it is useful to maintain partial symmetry properties 
within one channel.
A basis of this kind, adapted to the $s$ channel, has color tensors
\begin{eqnarray}
 C^{(A)} \,\equiv\, C^{(1)}, \qquad C^{(B)} \,=\, C^{(2)} & + & C^{(3)},
 \qquad C^{(E)} \,=\, C^{(5)} - C^{(6)},                   \label{COLABE} \\
 C^{(C)} \,=\, C^{(2)} - C^{(3)}, \qquad C^{(D)} \,& \equiv &\, C^{(4)}
 \ \ \left( \,= -C^{(5)}-C^{(6)} \,\right),                 \label{COLCD}  
\end{eqnarray}
and Lorentz tensors ( in $D=4-2\epsilon$ Euclidean dimensions ),
\begin{eqnarray}
 L_{(0)} \,=\, \frac{1}{D} L_{(1)}, \qquad L_{(+)} \, &=& \,
 \frac{1}{2}\left(L_{(2)}+L{(3)}\right)-\frac{1}{D}L_{(1)},  \label{LOP}\\
 L_{(-)} \, &=& \, \frac{1}{2}\left(L_{(2)}- L_{(3)}\right). \label{LON}
\end{eqnarray}
The latter have been chosen to display, under $s$-channel tensor
multiplication, the projector-like properties,
\begin{equation}                                             \label{LPROJ}
 L_{(m)}^{\kappa\lambda\rho\sigma} \, L_{(n)}^{\rho\sigma\mu\nu}
 \ = \ \delta_{mn} \, L_{(m)}^{\kappa\lambda\mu\nu} 
 \qquad (\, m,n \,=\, +,0,- \,).
\end{equation}
If we denote by the symbol
\begin{equation}                                             \label{PICL}
 \{ \ \pi_{12}, \ \pi_{34}; \ \pi_{s}\ \} \qquad
 (\  \mathrm{each} \  \pi \,=\, +1 \, \mathrm{or} \, -1 \ )
\end{equation}
the class of four-point amplitudes having parities $\pi_{12}, \pi_{34},
\pi_{s}$ under the $1 \leftrightarrow 2$, $3 \leftrightarrow 4$, and
$(1,2) \leftrightarrow (3,4)$ permutations respectively, then of the
fifteen product tensors $C^{(i)}L_{(j)}$ formed from (\ref{COLABE}/
\ref{COLCD}) and (\ref{LOP}/\ref{LON}) there are eight in class
$\{+,+;+\}$ and seven in class $\{-,-;+\}$. In the decomposition
\begin{eqnarray}
 \left(\ V_{\ 4T}^{[1,0]}(p_1 \ldots p_4)\ \right)_{abcd}^{\rho\sigma
 \tau\omega} \  = \  t^{\rho\kappa}(p_1) \,t^{\sigma\lambda}(p_2) 
 \,t^{\tau\mu}(p_3) \,t^{\omega\nu}(p_4)                    \nonumber  \\
 \times \ \sum_{i=A \ldots E} \ \sum_{j=+,0,-}\ C^{(i)}_{abcd} \,
 L_{(j)}^{\kappa\lambda\mu\nu} \ F_{i,j}^{[1,0]}(\,p_1{}^2 \ldots
 p_4{}^2; \,\Lambda^2 \,) \, ,                               \label{VFIJ}
\end{eqnarray}
the invariant functions $F_{i,j}$ therefore come in two types,
\begin{eqnarray}
 F^{[1,0]}_{(i,j)\in\{+,+;+\}} \ &=& \ \eta_{i,j,0} \,+\, \eta_{i,j,1}
 \left( \Pi_1+\Pi_2+\Pi_3+\Pi_4 \right) \,+\, \eta_{i,j,2} \left( \Pi_1
 +\Pi_2 \right) \left( \Pi_3+\Pi_4 \right)              \nonumber       \\ 
 \quad &+& \eta_{i,j,4} \left( \Pi_1\Pi_2 + \Pi_3\Pi_4 \right)
 \,+\, \eta_{i,j,5} \left[\, \Pi_1\Pi_2(\Pi_3+\Pi_4) \,+\, (\Pi_1
 +\Pi_2)\Pi_3\Pi_4 \,\right]                            \nonumber       \\
 \quad &+& \eta_{i,j,6} \left( \Pi_1\Pi_2\Pi_3\Pi_4 \right) \, ;
                                                        \label{FIJP}    \\   
 F^{[1,0]}_{(i,j)\in\{-,-;+\}} \ &=& \ \eta_{i,j,3} \left( \Pi_1-\Pi_2
 \right) \left( \Pi_3-\Pi_4 \right) \, .                \label{FIJN}    
\end{eqnarray}
The perturbative limit (\ref{GANULL}) fixes the coefficients
$\eta_{i,j,0}$ of (\ref{FIJP}) as
\begin{equation}                                            \label{NETA}
 \eta_{i,j,0} \ =\ 3\delta_{i,D}\delta_{j,-} \,+\, \delta_{i,E} \left[\,
 \delta_{j,+} - (D-1)\delta_{j,0} \,\right] \, .
\end{equation}
This representation then starts with a total of $8\times5\,+\,7\times1
\,=\,47$ real, dimensionless coefficients. With no further restrictions
on the latter, it would be adequate for quantities having partial Bose
symmetry with respect to one distinguished channel, such as a two-particle
irreducible kernel. For the full $V_{4V}$ amplitude, imposing complete
Bose symmetry produces 30 relations between the $\eta$'s so they can
ultimately be expressed linearly in terms of the 17 independent $\zeta$'s
of (\ref{VZETA}). The listing of these linear conversion equations is
relegated to appendix B.2.

The r.h.s. of a DS equation such as Fig.2 or Fig.10 below, which singles
out the leftmost external leg in an unsymmetric way, does not even
exhibit the reduced symmetry of (\ref{FIJP}/\ref{FIJN}) ( although it
does have a permutation symmetry between the three rightmost legs ).
Upon using (\ref{FIJP}/\ref{FIJN}) as input, the output therefore has
still lower symmetry, and enforcing the self-reproduction of a
symmetric vertex turns out to imply, in addition to 47 self-consistency
conditions proper, 7 more conditions for the vanishing of terms excluded
by the partial Bose symmetry, leading, after conversion of $\eta$'s to
$\zeta$'s, to a {\em total of 54 linear conditions on the 17
independent coefficients}. Moreover, one finds that unwanted contributions
to the eight constants (\ref{NETA}) arise which represent low-$r$
approximation errors in the perturbative renormalization constant, 
a phenomenon already encountered in \cite{VE3} for the three-point
vertices.
 
Finally note that with respect to any one of the four $p_i^{\,2}$, the
reduced vertex permits a partial-fraction (p.f.) decomposition, such as
\begin{equation}                                             \label{EEPF}
 V_{\ 4T}^{[r,0]} \ =\ E_{0,T}^{[r]} (p_2^{\,2},p_3^{\,2},p_4^{\,2}) \ +\ 
 \sum_{n=1}^{r} \,\left( \frac{\Lambda^2}{p_1^{\,2} + u_{2n}\Lambda^2}
 \right)\, E_{n,T}^{[r]} (p_2^{\,2},p_3^{\,2},p_4^{\,2}) \, ,
\end{equation}
where by omitting $p_1^{\,2}$ dependence in the $E_0$ term we have encoded
the restriction of no net positive powers of $p_1^{\,2}$. Such a
decomposition is useful technically in connection with the treatment
of the compensating poles.

%
%
% Latex Source File for Sect. 3 of 4-Gluon-Vertex Paper
%------------------------------------------------------
%
%
\section{Self-consistency of compensating poles}
\setcounter{equation}{0}
It was demonstrated in \cite{VE3} that from the lower ( i.e., $\Gamma
_{3V}$ and $\Gamma_{2V}$ ) equations alone one may already infer the
existence, and determine the residues, of certain pole terms in $\Gamma
_{4V}$ with respect to the three Mandelstam variables, and that these 
turn out ( as did similar poles in longitudinal channels in the work of
refs. \cite{JCN} ) to be compensating terms cancelling unphysical
singularities (``shadow'' poles ) in the one-gluon reducible terms of
$T_{4V}$. The compensating poles are structures of tree topology which, 
since their internal shadow lines are not propagators of any of the 
elementary QCD fields, nevertheless can be present in the 1PI
amplitude. This result is recalled in eq. (\ref{V4def}) and in Fig.1(a), 
where each double-wiggly line, at level $[r,0]$, stands for a set of $r$ 
shadow poles in the corresponding Mandelstam variable.

To understand the nature of these structures more fully, we now need
to discuss how they reproduce self-consistently in the DS equation for
$\Gamma_{4V}$ itself. As with all zeroth-order nonperturbative terms,
this will essentially occur through the hierarchical DS coupling, which
transfers the building materials for the compensating terms down from
the higher vertices in the equation. The argument is lengthy, but conveys
an idea of how, by analogy, such terms establish themselves in still
higher vertex functions. We therefore present it also {\em in lieu of}
a general N-legs-amplitudes proof, whose length would be out of
proportion to the insight gained. ( The reader not interested in the 
details of this self-consistency result, and willing to accept it on faith,
may proceed directly to the reduced DS equation for $V_{4V}$ given
diagrammatically in Fig.10, which is the starting point for sect.4.) 

{\bf (\,1\,)} The equation for $\Gamma_{4V}$ is written diagrammatically in
Fig.2, in a ``hybrid'' form in which interaction terms on the r.h.s.
have not been resolved down to the level of proper vertices: the $T'$
amplitudes appearing there are amputated functions 1PI in the channels
defined by their right-hand external legs, but otherwise still contain
1PR terms.

To keep technical complications to a minimum, we invoke all available 
simplifications: we disregard fermions, i.e. the term $(E)_4$ of
Fig.2, and consider zeroth-order self-consistency only at one loop, where
term $(D)_4$ does not yet contribute, and in Landau gauge, so that
the ghost term $(B)_4$ may also be omitted. We are then dealing,
first, with terms $(C)_4, \ (C')_4, \ (C'')_4,$ involving
the four-gluon function itself in the form of three $T'_{4V}$ amplitudes,
each one-gluon irreducible in the channel defined by its two right-hand
external legs. The structure of such a $T'_{4V}$ as following from
Fig.1(a) is recalled in Fig.1(b): it has one compensating ( negative
shadow ) pole in the distinguished channel, and ``softened'' 
\mbox{( one-shadow irreducible )} one-gluon exchanges in the two 
other channels.

{\bf (\,2\,)} Second, we have term $(A)_4$ involving a five-gluon
amplitude $T'_{5V}$, which by definition is one-gluon irreducible in
the four $2V \leftrightarrow 3V$ channels accessible through its three
right-hand external legs, namely the ``horizontal'' channel
\begin{equation}                                            \label{HOR5}
 (\,5,\,6\,) \ \leftrightarrow \ (\,2,\,3,\,4\,) \qquad (\,i=1\,)\, ,
\end{equation}
and the three ``tilted'' channels
\begin{equation}                                            \label{TILT5}
 (2,3) \, \leftrightarrow \, (4,5,6) \ (i=2), \qquad
 (3,4) \, \leftrightarrow \, (5,6,2) \ (i=3), \qquad
 (4,2) \, \leftrightarrow \, (3,5,6) \ (i=4).
\end{equation}
The integers $i$ assigned refer, by convention, to a numbering $i=1 \ldots
10$ of the ten $2V \leftrightarrow 3V$ channels of a five-point amplitude.
Thus $T'_{5V}$ has those one-gluon reducible terms removed that would
produce a one-gluon reducible graph when introduced into term $(A)_4$.
A representation of $T'_{5V}$ suitable for our purpose is given
diagrammatically in Fig.3; it is again hybrid in that it involves
$T'_{4V}$ amplitudes in addition to fully 1-gluon irreducible functions
$\Gamma_{4V}$ and $\Gamma_{5V}$. The derivation of this representation
is a technical matter that we relegate to appendix C. ( Its apparent
asymmetry with respect to the $(5,6)$ pair of legs is resolved by realizing
that there exists an equivalent representation with the roles of $\Gamma
_{4V}$ and $T'_{4V}$ interchanged ).

Insertion of Fig.3 into the $(A)_4$ term of Fig.2 then decomposes
that term into a piece $(A)_{4,\Gamma}$ involving the 1-gluon
irreducible $\Gamma_{5V}$, plus six pieces with triangle-graph topology.
We do not draw this decomposition separately.

{\bf (\,3\,)} Now take residues in the $\Gamma_{4V}$ equation of Fig.2 with
respect to the squared momentum, $p_1{}^2$, of the leftmost external leg.
This is entirely analogous to the residue comparison performed in
\cite{VE3} for the $\Gamma_{3V}$ equation, and leads to the analogous
conclusion: the zeroth-order pole terms in $p_1{}^2$ of the r.h.s. can
arise only from the term $(A)_{4,\Gamma}$, and the $\Gamma_{5V}$
in that term must therefore contain zeroth-order poles $(s_1 + u_{2n}
\Lambda^2)^{-1}$ in the Mandelstam variable of channel (\ref{HOR5}),
\begin{equation}                                              \label{SHOR}
 s_1 \ =\ (-p_5 -p_6\,)^{2} \ =\ (\,p_2 + p_3 + p_4\,)^{2}\, ,
\end{equation}
which equals $p_1{}^2$ by momentum conservation. Note that terms with
an inverse polynomial in $p_1{}^2$ produced by the {\em loop integration}
of term $(A)_{4,\Gamma}$, if any, would by the standard formulas of
dimensional integration have to arise from {\em convergent} parts of the
integral, and therefore would not trigger the self-consistency mechanism
for zeroth-order terms; they would be part of the {\em first-order},
quasi-perturbative correction - a remark indeed applying to all instances
of generation of zeroth-order terms.

The residues of $\Gamma_{5V}$ at those poles must be proportional to the
residue functions ${\cal E}_{\,n}^{[r]} (p_2,p_3,p_4)\ (1 \leq n \leq r)$
in the p.f. decomposition of $\Gamma_{4V}^{[r,0]}$ analogous to eq.
(\ref{EEPF}). On the other side they must be proportional to the residue
functions $B_{\,n}^{[r]} (p_5,p_6)\ (1 \leq n \leq r)$ of the three-
point vertex $\Gamma_{3V}$; this follows e.g. from the way the same
$\Gamma_{5V}$ amplitude enters in the two-DS-loops term of the three-
point equation ( term $(\rm D)_3$ in Fig.1 of \cite{VE3} ). For these
pole terms of $\Gamma_{5V}$, the loop of $(A)_{4,\Gamma}$ then turns
into loops of self-energy type already encountered in the self-consistency
problem of $\Gamma_{2V}^{[r,0]}$, which fixes the proportionality
factors: the internal lines in the zeroth-order pole terms of
$\Gamma_{5V}$ with respect to $s_1$ are associated with factors
\begin{equation}                                              \label{MSHAD}
 -\,S_{\,n}^{[r]}(p_1) \ =\ - \frac{1}{u_{r,2n+1}} \, \frac{ t(p_1) }
{p_1{}^2 + u_{r,2n}\Lambda^2} \qquad (\, n=1 \ldots r \,) \, ,
\end{equation}
and therefore constitute minus a shadow line as defined in Fig.3(b) 
of \cite{VE3}.

{\bf (\,4\,)} To demonstrate self-consistency of the three negative-shadow 
terms in eq.(\ref{V4def}), we now feed the decomposition of Fig.1(a) into
the various contributing terms on the r.h.s. of Fig.2 and verify that,       
upon appeal to information from the ``lower'' DS equations, those terms
cooperate so that their sum splits off explicitly the same triplet
of negative-shadow terms; after cancelling them from both sides of the
equation, the remainder then constitutes a DS equation for the reduced
amplitude $V_{4V}$.

The first place to feed in Fig.1(a) is the $\Gamma_{5V}$ poles just 
identified, with their ${\cal E}_{\,n}^{[r]}$ residue functions. This 
turns each of them into a pole with the corresponding reduced
$E_{\,n}^{[r]}$ instead, minus a triplet of terms with {\em two} shadow
poles each, involving in addition to the $B_{\,n}^{[r]}\ (n \geq 1)$
a set of functions $B_{m,n}^{[r]}$ defined by further decomposition 
of these,
\begin{equation}                                              \label{BMN}
 B_{\,m}^{[r]}(p,q) \ =\ B_{m,0}^{[r]}(q) \,+\, \sum_{n=1}^{r} \,
 \left(\frac{\Lambda^2}{p^2+u_{r,2n}\Lambda^2}\right)\, B_{m,n}^{[r]}(q) 
\end{equation}
Invoking now the full Bose symmetry of $\Gamma_{5V}$, we see that this
vertex has a shadow-poles structure given, in a condensed notation, by
\begin{eqnarray}
 \Gamma_{5V} &=& \ \ V_{5V}                                    \nonumber  \\
             & & - \ \sum_{i=1}^{10}\, \Big\{\, \sum_{n  } \,B_{\,n}^{[r]}
 \,S_{\,n}^{[r]} \,E_{\,n}^{[r]} \,\Big\}_{i}            \label{G5SHA}    \\
             & & + \ \sum_{k=1}^{15}\, \Big\{\, \sum_{m,n} \,B_{\,m}^{[r]}
 \,S_{\,m}^{[r]} \,B_{m,n}^{[r]} \,S_{\,n}^{[r]} \,B_{\,n}^{[r]} \,\Big\}
 _{k} \ ,                                                      \nonumber
\end{eqnarray}
where $i$ enumerates the 10 channels, while $k$ enumerates the 15 
different 2+1+2 partitions of the 5 legs. This structure is shown
diagrammatically in Fig.4. Note that the symmetrization is achieved with
only 15, not 30, of the latter terms: since each of them has shadow lines 
in two different channels, these are enough to supply the triplet of 
contributions mentioned before to pole structure in all of the 10
Mandelstam variables. The amplitude $V_{5V}$ defined by eq.(\ref{G5SHA}) 
will be referred to as the reduced 5-gluon vertex. As a superficially 
convergent vertex with all tree-like structures removed by construction,
it consists only of superficially convergent loops. Since these,
as emphasized repeatedly, do not support the self-reproduction of
zeroth-order terms, it is an exact statement that
\begin{equation}                                         \label{V5NULL}
 V_{\,5V}^{[r,0]} \  = \  0  \qquad  (\,\mathrm{all}\ r\,)\, .
\end{equation}

{\bf (\,5\,)} The next place to feed in the assertion of Fig.1(a), or its 
consequence of Fig.1(b), is the six terms of triangle topology,
arising from $(A)_4$ of Fig.2 upon insertion of the 1-gluon-reducible
terms in the 2nd and 3rd lines of Fig.3. Together with eq.(\ref{G5SHA})
and after some regrouping of terms, this is equivalent to using a new
representation of the $T'_{5V}$ amplitude given diagrammatically
in Fig.5. We again relegate details of its derivation to appendix C
and note only that this representation, apart from the fully 
one-shadow-irreducible gluon-exchange graphs in its 2nd and 3rd lines,
has a shadow-line content isolated in four terms: the negative-shadow
term in the first line, which is identical to the $i=1$ term in the 
2nd line of eq.(\ref{G5SHA}) and will play a special role by supplying
in Fig.2 the singularities of $V_{4V}$ in its $p_1^{\,2}$ variable,
and the 3 negative terms in the last two lines, which involve a
3-gluon-1-shadow auxiliary amplitude $\Xi_n$ as defined in Fig.7.

{\bf (\,6\,)} Lastly, we should introduce Fig.1(b) into the $(C)_4$,
$(C')_4$, and $(C'')_4$ terms of Fig.2. This will turn each of
them into a graph having a $V_{4V}$ instead of $T_{4V}$, plus a pair of
one-shadow-irreducible gluon-exchange terms giving equal contributions
and for which the factor of $\frac{1}{2}$ therefore gets cancelled,
plus a negative term with one shadow line, as displayed in Fig.8.
( Note that an ordinary gluon line connecting to at least one
{\em bare} vertex, such as the $\Gamma_{\,4V}^{(0)pert}$ in Fig.8,
has no shadow content ).

{\bf (\,7\,)} In the crucial step of the argument, we now refer back to the
{\em three-point} vertex equation, maintaining for consistency the
simplifications analogous to those of step (\,1\,) above, and again
feed the assertion of Fig.1 into the $T'_{4V}$ term of that equation
( term $(A)_3$ in Fig.1 of \cite{VE3}, where the $T'_{4V}$ is denoted
$T'_s$ ). We then take residues with respect to the squared momentum
of one of the two external legs {\em other than} the ``distinguished''
leg entering from the left ( one of the variables $p_1^{\,2}$ or
$p_3^{\,2}$ in Fig.1 of \cite{VE3} ). Note that this kind of residue
information stood unused so far, since in \cite{VE3} we only exploited
residue comparison with respect to the distinguished leg. The result
is the diagrammatic relation of Fig.9, a DS equation of sorts for the
partial amplitudes $B_n$ ($n\geq1$) of $\Gamma_{3V}$. ( If we do not
invoke the simplifications, this relation of course gets additional
ghost-loop, fermion-loop, and two-gluon-loops terms with their own
$\Xi$-type amplitudes, as alluded to in the last line of Fig.9 ).

But this relation is precisely what is needed to turn the 3 negative
shadow-line terms which arose in step (\,6\,) from the $(C)_4$ through
$(C'')_4$ terms of Fig.2, plus the 3 negative $\Xi$ terms that arose
from term $(A)_4$ upon insertion of the $\Xi$ graphs of Fig.5, into
the 3 negative one-shadow terms on the r.h.s. of Fig.1(a). We have
therefore attained our goal: we have shown that upon introducing the
decomposition of Fig.1(a) with its 3 compensating-pole terms, either
directly or through its immediate consequence of Fig.1(b), into wherever
a four-point building block occurs in the interaction terms of Fig.2,
these interaction terms cooperate so that their sum splits off explicitly
those 3 compensating terms again. Note that in so doing we appealed
to residue information not only from the lower ( 3-point and 2-point )
DS equations, but now also from the 4-point
equation itself, which led us to infer the shadow-reducibility structure
(\ref{G5SHA}) of the next higher, 5-gluon vertex.

We may now cancel the 3 compensating terms on both sides of the
$\Gamma_{4V}$ equation and thus establish the DS equation of Fig.10
for the reduced vertex $V_{4V}$, which is the main result of this section.

{\bf (\,8\,)} As a by-product, we may rewrite eq.(\ref{G5SHA}) for the
shadow content of $\Gamma_{5V}$. To each of the 10 terms with negative
signs in the second line, we may add a triplet of terms of structure
$\sum_m \Gamma_{3V} D B_m S_m B_m$, with one ordinary-gluon and one shadow 
line, such that the resulting quartet of terms form precisely the set
of compensating poles needed to cancel the four shadow lines present
in the one-gluon-exchange diagram $\{ \Gamma_{3V} \,D\, \Gamma_{4V} \}_i$
of the full connected-and-amputated 5-gluon amplitude $T_{5V}$ ( one
along the D propagator line, the other three hidden, by Fig.1(a), in the
$\Gamma_{4V}$ vertex ). The same 30 terms, now regrouped into 15 suitable
pairs, may then be subtracted from the 15 terms in the third line of
(\ref{G5SHA}), and it is straightforward to check with the aid of
Fig.6(b) that each resulting triplet then forms the set of compensating 
poles needed to cancel the one two-shadows term and two single-shadow
terms present in a double-gluon-exchange diagram $\{ \Gamma_{3V} D
\Gamma_{3V} D \Gamma_{3V} \}_k$ of $T_{5V}$. Since the total operation
has not changed the equation, we conclude, with a view to eq. (\ref{T5DEC})
of the appendix, that
\begin{eqnarray}                                         
 \Gamma_{5V} &=& V_{5V} \ + \ \big[ \ \textrm{complete set}   \nonumber  \\
& & \textrm{of compensating poles for all 25 one-gluon-reducible terms of} 
 \  T_{5V} \ \big]\, .                               \label{G5COMP}
\end{eqnarray}
Thus the final result for the next higher amplitude is again simple:
the full $T_{5V}$ is completely one-shadow irreducible, consisting as it
does of a $V_{5V}$ with full extended irreducibility, plus 25 ``softened''
( i.e. one-shadow irreducible ) but still one-gluon reducible diagrams.

{\bf (\,9\,)} Finally note that term $(A1)_4$ of the reduced 
$V_{4V}$ equation
of Fig. 10, upon taking residues at its horizontal-channel shadow poles
and invoking once again the $\Gamma_{2V}^{[r,0]}$ self-consistency
conditions, instantly reproduces the $E_{n}^{[r]}$ terms with $n\,\geq\,1$
of the p.f. decomposition (\ref{EEPF}). The remaining, further-reduced
equation for the partial amplitude $E_{0}^{[r]}$ of (\ref{EEPF}) at $r=1$
is what we shall work with in practice; it should be kept in mind that
this equation still has a term comprising the regular-at-poles 
remainders of $(A1)_4$ of Fig.10.

%
%
% Latex Source File for Section 4 of 4-Gluon-Vertex Paper
%--------------------------------------------------------
%
%
\section{Self-consistency conditions}
\setcounter{equation}{0}
\subsection{The overdetermined system}
A glance at Fig.10 shows that the reduced  DS equation for $V_{4T}$, or
for its $E_{0,T}$ component, is {\em linear} in the four-point unknowns,
though still nonlinear in the ``lower'', 2-and-3-point vertices. This
simplification results from (i) the restriction to a one-loop calculation
and (ii) our use of the ordinary DS system without Bethe-Salpeter
resummation. The self-consistency equations for the $\zeta$ coefficients
of eq. (\ref{VZETA}) will thus form a linear system but with a matrix 
( determined by the divergent parts of diagrams $(A1)_4$,
$(A3)_4$, and $(C2)_4$ of Fig.10 ) 
and with inhomogeneous terms ( determined by the divergent 
parts of the other diagrams of Fig.10 ) depending nonlinearly on the
2-point and 3-point coefficients.

As discussed in \cite{VE3}, the zeroth-order self-consistency system 
as a whole exhibits a {\em scaling property}, arising from the scheme-
insensitive character of the self-reproduction mechanism. It implies
that the system only determines ratios of the nonperturbative coefficients
to certain fixed powers of one of them. The latter, by convention, was
chosen to be $x_1$, a coefficient of the 3-transverse-gluons vertex.
We need to maintain this convention for the 4-gluon quantities; thus
for the parametrization (\ref{VZETA}) we introduce rescaled parameters
\begin{equation}                                              \label{SCAL}
 \tilde{\Lambda}^{2} \ =\ |x_1| \,\Lambda^2 \, , \qquad \qquad
 \tilde{\zeta}_i  \ =\ \zeta_i / |x_1|^{n_i} \quad (\,i=1 \ldots 17\,) \, ,
\end{equation}
where the integers $n_i$ are,
\begin{eqnarray}                                              \label{ENNS}
 n_1   &=&  n_7 \,=\, n_{13} \,=\, 1 \, ,           \nonumber         \\  
 n_2   &=&  n_3 \,=\, n_6    \,=\, n_8 \,=\, n_9 \,=\, n_{12} \,=\,
 n_{14}  \,=\, n_{17} \,=\, 2 \, ,                  \nonumber         \\
 n_4   &=&  n_{10} \,=\, n_{15} \,=\, 3 \, ,        \nonumber         \\
 n_5   &=&  n_{11} \,=\, n_{16} \,=\, 4 \, .
\end{eqnarray}               
All RG-invariant masses are obtained in terms of $\tilde{\Lambda}^2$ 
rather than $\Lambda^2$.

For the actual evaluation of the zeroth-order terms of diagrams 
$(A1)_4, \ (A3)_4, \ (C2)_4$ of Fig.10,
one uses parametrization (\ref{VFIJ})
with its independent tensors and $\eta$ coefficients. In contrast to what
happens ( at level $r=1$ ) in the 3-point systems, the immediate self-
reproduction of the entire $p_1^{\,2}$-singular portion $E_{1,T}$ of eq.
(\ref{EEPF}), achieved by extracting the pole term at $p_1^{\,2} = - u_2
\Lambda^2 = - \tilde{u}_2 \tilde{\Lambda}^2$ from diagrams $(A1)_4$ 
and appealing to the 2-gluon self-consistency conditions, is found not to
provide substantial relief for the overdetermination caused by the lack
of manifest Bose symmetry. After reimposing the symmetry by expressing 
all $\eta$'s in terms of the $\zeta$ coefficients of the fully Bose-
symmetric form (\ref{VZETA}), one is still faced with 54 linear equations
for these 17 basic coefficients, with matrix elements and inhomogeneities
depending nonlinearly ( though polynomially ) on the lower-vertex
coefficients. The complete system is listed in Appendix D. To convey an 
impression of its structure, we write 
here just one typical equation: self-reproduction of the coefficient
$\eta_{A,0,4}$ in the invariant function (\ref{FIJP}) associated with 
the $C^{(A)} \otimes L_{(0)}$ tensor structure gives, after conversion
to $\zeta$'s by means of appendix B.2 and multiplication by $\beta_0$ and
$u_3^{\,2}$, the condition
\begin{eqnarray}
\ & & \ \frac{ 9}{2} \left(u_3{}^2 x_3\right)         \,\zeta_1 \ +\ 
   \frac{ 9}{2} \left(\frac{3}{8}u_3{}^2-u_3x_4\right)\,\zeta_2 \ +\ 
   \frac{ 3}{2} \left((\ \frac{9}{8}-\beta_0)u_3{}^2 - 
                           \frac{3}{2}u_3x_4 \right)  \,\zeta_3   \nonumber\\
\ & & + \frac{27}{4}\left( u_3 x_1 \right)            \,\zeta_4 \ +\ 
   \frac{ 9}{4} \left(\frac{3}{2}u_3{}^2+u_3x_4\right)\,\zeta_6 \ +\ 
\frac{ 3}{2} \left(( -\frac{9}{4}+\beta_0)u_3{}^2 -
                             \frac{3}{2}u_3x_4 \right)            \nonumber\\
\ & &\ \times\left(\zeta_8-2\zeta_9+2\zeta_{12}-\zeta_{14}+\zeta_{17}\right)\ 
     +\  u_3\,\left(\frac{99}{4}x_1-\frac{15}{8}x_3-N_{F}z_3\right)\,
                                         (\zeta_{10}-\zeta_{15})  \nonumber\\
\ &=& 9\,     \left( -\frac{3}{2}u_3{}^2x_3{}^2 + 4u_3x_1x_3x_4
  + (\frac{1}{2}u_3-x_2)x_4{}^2 - 2x_1x_4x_5 \right)              \nonumber\\
\ & & +\, N_{F}\, \left(\frac{u_3}{w_3}\right)^2\, \frac{2}{3}z_4\,
             \left(w_3z_1z_3-w_3z_4+z_1z_5-z_2z_4\right)\, .    \label{ZETEQ}
\end{eqnarray}
Here $N_C=3$ has been used. The $N_F$ factors identify contributions 
from the quark-loop diagrams $(E1)_4, \ (E1')_4$
of Fig.10. Note that for the quark-vertex coefficients $z_i$, the
simplified numbering of eqs. (4.6) of \cite{VE3} has been employed.

As far as we are aware, no presently existing mathematical software tool
for nonlinear algebraic systems is capable of dealing directly with a 
system of this size and complicated, partially overdetermined structure. 
The only strategy currently practical for dealing with the total system 
is to take advantage of the ``near decoupling'' 
of the 4-point system: of the 17 $\zeta$
coefficients, only two combinations, $Z_1$ and $Z_2$, enter the 2-and-3-
point system (through some of the 3-gluon vertex conditions). One then
treats these as additional unknowns in the lower-vertices system and,
after invoking the scaling property, determines them to within an
one-parameter freedom. Subsequently, their definitions are appended to
the linear system for the 4-point $\zeta$'s as two extra conditions,
bringing the total to 56 equations: this represents only a minor increase
in the anyway massive overdetermination of that system. The extra
equations are, in rescaled form,
\begin{eqnarray}
 \frac{15}{32}\, \big( 3\tilde{\zeta}_1 - \tilde{\zeta}_7 \big)
 &=& \tilde{Z}_1 \ =\ \frac{Z_1}{x_1}                      \label{Z1T}  \\
 \frac{15}{32}\, \big( 3\tilde{\zeta}_2 +3\tilde{\zeta}_3 - \tilde{\zeta}_8
 - \tilde{\zeta}_9 \big) &=& \tilde{Z}_2 \ =\ \frac{Z_2}{x_1^{\,2}} \, ,
                                                           \label{Z2T}
\end{eqnarray}
where the $\tilde{Z}_1,\tilde{Z}_2$ on the r.h.s. stand for the values 
produced by the lower-vertices system. The structure of the four-point 
system then is,
\begin{eqnarray}
 \sum_{k=1}^{17} M_{ik} \big( \{\tilde{u}\},\{\tilde{w}\};\{\tilde{x}\},
 \{\tilde{z}\} \big) \, \tilde{\zeta}_k &=& b_i \big( \{\tilde{u}\},
 \{\tilde{w}\};\{\tilde{x}\},\{\tilde{z}\} \big)           \label{4SYS} \\
 (\ i \,=\, 1 \ldots 56 \ )\, , & &                           \nonumber
\end{eqnarray}
where the $56\times17$ matrix $M$, as well as the $56$-component 
inhomogeneity $b$, depend nonlinearly on the rescaled coefficient sets
$\{\tilde{u}\},\{\tilde{w}\}$ of the gluon and quark two-point functions
and on the sets $\{\tilde{x}\},\{\tilde{z}\}$ of the gluon and quark
three-point functions. Exceptions are the 55th and 56th rows of $M$,
given by the l.h. sides of eqs. (\ref{Z1T}/\ref{Z2T}), which contain only
pure numbers.

For a quasi-solution in the least-squares sense, one minimizes the
quadratic deviation between both sides of (\ref{4SYS}). The conditions
\begin{equation}                                           \label{LSQ}
 \frac{\partial}{\partial\tilde{\zeta}_n}\, \Big\{ \ \sum_{i=1}^{56}
 \,\big( \sum_k M_{ik}\tilde{\zeta}_k - b_i \big)^2 \ \Big\}
 \ =\ 0 \qquad (\ n \,=\, 1 \ldots 17 \ )
\end{equation}
lead, in standard fashion, to a linear system with the $17\times17$ 
matrix $M^T M$,
\begin{equation}                                           \label{LINSYS}
 \sum_{k=1}^{17} \big( M^T M \big)_{nk} \tilde{\zeta}_k \ =\ 
 \big( M^T b \big)_n \qquad (\ n \,=\, 1 \ldots 17 \ ) \, .
\end{equation}
Although the structure of $M$ is too complicated to be handled 
analytically, we have checked numerically that in the physically 
acceptable parameter range where all 2-and-3-point coefficients are real,
the matrix $M^TM$ is always invertible. ( The calculations have been
performed using the MAPLE V computer-algebra system ). As a measure 
of the deviation one may consider
\begin{equation}                                          \label{DELSQ} 
 \chi \ =\ \sqrt{ \frac{1}{56} \ \sum_{i=1}^{56}\,
 \Big( \sum_k M_{ik}\tilde{\zeta}_k - b_i \Big)^2 }
\end{equation}
as compared to a typical r.h.s. or l.h.s. of eqs. (\ref{4SYS}). Also,
the quantities $\tilde{Z}_1,\tilde{Z}_2$ when recalculated from the
quasi-solution according to eqs. (\ref{Z1T}/\ref{Z2T}) will be different from
their input values from the lower-vertices system, and the differences may
serve as rough indicators of the overall degree of difficulty in satisfying
zeroth-order self-consistency requirements from ( ordinary ) DS equations
with the rather simple structure of the $r=1$ set of
approximants. If they are large, one may alternatively try to enforce
conditions (\ref{Z1T}/\ref{Z2T}) exactly by assigning a large weight $w$
(e.g. $w=1000$) to the $i=55$ and $i=56$ terms of eq. (\ref{LSQ}). The
$Z_{1,2}$ from the two systems will then match, but the quality of the
$54$-term remainder solution will in general deteriorate.
\subsection{Typical solution for $N_F=2$}
The entire system of course shares the effective one-parameter freedom
from the ``decoupled'' lower-vertices system. In the presence
of fermions, this was parametrized \cite{VE3} by the coefficient 
$\tilde{w}_1$. We only consider here the physically most interesting 
parameter range where not only all vertex coefficients are
real, but also the values of the self-energy coefficients $u$, $w$ imply
the presence of a complex-conjugate pole pair in both the gluon and the
fermion propagator. Over this range, as noted earlier, most of the other
coefficients of the system do not vary substantially.

Table 4.1 gives least-squares values of the rescaled coefficients 
$\tilde{\zeta}$ corresponding to the ``typical'' solution of table 5.1
of \cite{VE3}, which at $\mid w_1 \mid = \mid w_2 \mid = 0.6749$ ( the
signs of $w_1,w_2$ do not affect the 4-point system ) is in about the 
middle of this parameter range. The two solutions shown are
obtained either without extra weighting ($w=1$) or with exact enforcement
($w=1000$) of the 55th and 56th conditions.
\begin{table}[!hbp] 
\vspace{3mm}
\begin{center}
\begin{tabular}{|c||c|c|c|c|c|c|c|c|c|}
\hline
   n  &  1  &  2  &  3  &  4  &  5  &  6  &  7  &  8  &  9              \\
\hline 
$\tilde{\zeta}_n\,(w=10^3)$ & $56.22$ & $646.67$ & $-513.11$ & $611.96$ 
 & $962.88$ & $271.12$ & $-168.19$ & $-803.84$ & $871.27$               \\
\hline
$\tilde{\zeta}_n\,(w=1   )$ & $-181.48$ & $-98.04$ & $-113.83$ & $297.85$
 & $-2747.9$ & $-37.55$ & $172.27$ & $-319.20$ & $35.75$                \\
\hline \hline
   n  & 10  & 11  & 12  & 13  & 14  &  15 & 16  & 17  & \               \\
\hline
$\tilde{\zeta}_n\,(w=10^3)$ & $64.77$ & $438.55$ & $-202.04$ & $-510.04$ 
 & $84.67$ & $-561.86$ & $-7481.9$ & $209.80$ & \                       \\
\hline
$\tilde{\zeta}_n\,(w=1   )$ & $1318.4$ & $-1757.5$ & $-26.14$ & $76.28$ 
 & $190.68$ & $367.79$ & $5446.0$ & $-317.7$ & \                        \\
\hline
\end{tabular}

\vspace{2mm}

        {\bf Table 4.1.} Coefficients $\tilde{\zeta}_n$ for 
              $N_F=2$ at typical value $\tilde{w}_1 = 0.67$
\end{center}
\end{table}
The resulting coefficients are generally rather large in absolute value, 
given the fact that from their very context one would expect them to be
numbers of order unity ( give or take an order of magnitude ). This may
partly be attributed to the choice of basis -- the factors of $\frac{1}{4},
\ \frac{1}{6}, \ \frac{1}{12}$ in the tensors (\ref{TR2}/\ref{CST}) 
employed in the $W_n$ building blocks -- and to the rescaling
(\ref{SCAL}) : it is conspicuous that the large coefficients mostly have
large exponents $n_i$ in (\ref{ENNS}). Although these effects may account
for one to two orders of magnitude, several coefficients still remain
implausibly large. We interpret this as the direct result of the 
considerable ``pressure'' generated by the overdetermination of the 
system.

The typical 2-and-3-point solution of \cite{VE3} implies values of
\begin{equation}                                             \label{ZVALF}
 \tilde{Z}_1 \ =\ 157.9, \qquad \tilde{Z}_2 \ =\ 156.2 \qquad (N_F=2)
\end{equation}
for the parameter combinations of eqs. (\ref{Z1T}/\ref{Z2T}) respectively.
When not enforced ( i.e. at $w=1$), their values recalculated from the
$\tilde{\zeta}'s$ of table 1, $-335.95$ and $-165.07$, are strongly
mismatched and even of the opposite sign. It is then hardly surprising
to find that this solution has $\chi \approx 317$ in the sense of 
(\ref{DELSQ}) -- of the same magnitude as, or even slightly larger than,
a typical $|b_i|$ in (\ref{4SYS}), which is exemplified by (\ref{ZVALF})
and of the order of a hundred.
Enforcing the values (\ref{ZVALF}) would then be expected 
to strongly alter the solution, which is indeed what happens -- 
several of the coefficients even change signs, and $\chi$ deteriorates 
further to $\approx 349$. This is in marked contrast 
to the purely gluonic case discussed in the next subsection, 
and implies that in the presence of ( massless ) fermions the 4-point
solution at the $r=1$ level is probably a poor one. The presence of the
massless-quark DS loops evidently makes the total system at low $r$ much
harder to satisfy; in particular these loops seem to work in the
direction of forcing a mismatch at the (2+3)-point-to-4-point interface
that can be alleviated only gradually at higher $r$. 
\subsection{Typical solution for pure-gluon theory}
For pure Yang-Mills theory ($N_F=0$) the typical solution, now 
parametrized by the gluonic coefficient $\tilde{x}_3$ as discussed 
in \cite{VE3} and taken at $\tilde{x}_3 = 1$, leads to the 4-point 
$\tilde{\zeta}$\,'s given in table 4.2.
\begin{table}[!hbp]
\vspace{3mm}
\begin{center}
\begin{tabular}{|c||c|c|c|c|c|c|c|c|c|}
\hline
   n  &  1  &  2  &  3  &  4  &  5  &  6  &  7  &  8  &  9              \\
\hline
$\tilde{\zeta}_n\,(w=10^3)$ & $0.814$ & $6.439$ & $-4.733$ & $-57.35$ 
 & $350.05$ & $-3.057$ & $8.672$ & $-30.03$ & $0.931$                   \\
\hline
$\tilde{\zeta}_n\,(w=1   )$ & $-0.432$ & $9.285$ & $-4.789$ & $-54.834$ 
 & $255.08$ & $-2.512$ & $6.398$ & $-21.50$ & $1.275$                   \\
\hline \hline
   n  & 10  & 11  & 12  & 13  & 14  &  15 & 16  & 17  & \               \\
\hline
$\tilde{\zeta}_n\,(w=10^3)$ & $96.02$ & $-328.14$ & $-0.060$ & $3.354$ 
 & $-4.212$ & $13.74$ & $-121.67$ & $-9.060$ & \                        \\
\hline
$\tilde{\zeta}_n\,(w=1   )$ & $56.75$ & $-110.07$ & $-0.518$ & $2.595$ 
 & $0.461$ & $-19.15$ & $124.0$ & $-8.744$ & \                          \\
\hline 
\end{tabular}

\vspace{2mm}

           {\bf Table 4.2.} Coefficients $\tilde{\zeta}_n$ for 
           pure-gluon theory at typical value $\tilde{x}_3 = 1.0$
\end{center}
\end{table}

Two features deserve comment. First, the coefficients overall are of a
distinctly more plausible order of magnitude, entirely understandable
from the above-discussed simple mechanisms; in particular, the three
coefficients of sizes more than a hundred are precisely those with
$n_i=4$ in (\ref{ENNS}). Second, and more remarkably, the input 
$\tilde{Z}_{1,2}$ values from the lower-point system,
\begin{equation}                                             \label{ZVALG}
 \tilde{Z}_1 \ =\ -2.92, \qquad \tilde{Z}_2 \ =\ 16.04 \qquad (N_F=0)
\end{equation}
are matched reasonably well by the four-point quasisolution even for $w=1$, 
i.e. without being enforced exactly: when recalculated they come out as
 -3.60 and +15.80 respectively. Moreover, the $\chi$ value of $\approx 0.11$
is now only about ten percent of the typical $|b_i|$ of (\ref{4SYS}),
which in this case is of order unity. Exact enforcement of the values
(\ref{ZVALG}) then leads to no appreciable deterioration in $\chi$.
This gives one confidence that {\em for the pure Yang-Mills system} 
the least-squares, four-point quasisolution 
does work reasonably well even on the $r=1$ level of approximation -- 
better, in fact, than one would expect in view of the still very simple 
and rigid structure of approximants at this level.

%
%
% LATEX source file for sect.6 of four-gluon-vertex paper
%--------------------------------------------------------
%
%
\section{Conclusion}
We have verified that self-consistent determination of a generalized 
Feynman rule for the highest superficially divergent and kinematically
most complex QCD vertex is possible in principle, provided the 
overdetermination resulting from lack of manifest Bose symmetry in
the relevant DS equation is dealt with by a least-squares procedure.
Within the well-defined and clearly visible limitations of our calculation, 
available indicators suggest that the least-squares solution at level 
$r=1$ is good for the pure gluon theory but quantitatively poor 
in the presence of massless quarks.

Among the limitations, the one most obviously in need of improvement
may be the use of an ``ordinary'' DS system: starting from BS-resummed
dynamical equations for $N \geq 3$ would presumably shift more important
physical effects into the lower loop orders, and would provide partial
( though not complete ) relief from the pressures of overdetermination.
Similar effects can probably be expected from use of a tensor basis 
larger than our ``dynamically minimal'' one; this would moreover allow
$d_{abc}$ color dependence in the three-gluon vertex to be treated
consistently. Going beyond the $r=1$ and $l=1$ levels of approximation
would have the algebraic complication rising steeply, as usual in
iterative schemes for QFT, but is presumably the only consistent way
of better satisfying perturbative limits and other desirable secondary
conditions. The above results do, however, give one confidence that
the method as a whole is viable, and produces enough nontrivial features
to render these improvements worth pursuing.

\newpage                                               % Appendices A - D
\begin{appendix}
\renewcommand{\theequation}{\Alph{section}.\arabic{equation}}
\begin{center}
{\Large \bf{Appendix}}
\end{center}
%
%
% Latex Source File for Appendix A of 4-Gluon-Vertex Paper
%---------------------------------------------------------
%
%
\section{Absence of non-compensating zeroth-order poles}
\setcounter{equation}{0}
Here we ask whether the four-gluon generalized Feynman rule $\Gamma_{4V}
^{[r,0]}$ may contain, in addition to the "compensating" poles in its three 
crossed color-octet channels as identified in sect.\ 2.2 and made explicit 
in eq.\ (\ref{V4def}), still other poles {\it of zeroth perturbative 
order} in its Mandelstam variables. Such poles would be absent 
in the perturbative limit $\Lambda \to 0$, so their residue factors would be 
proportional to at least one power of $\Lambda$. In e.g.\ the $s$ channel, 
such a pole would be of the form
\begin{equation}                                        \label{mandelst-pole}
\frac{\Lambda\Phi^T(p_1,p_2)\Lambda\Phi(p_3,p_4)}
     {P^2+b\Lambda^2}
\end{equation}
with a dimensionless residue function (or vector of functions) $\Phi$ 
carrying the color and Lorentz quantum numbers of the channel considered. 
This contrasts with the case of the compensating pole, where residue 
comparison in the $\Gamma_{3V}$ equation forces the residue 
factors to be proportional to $B_n$'s -- quantities of mass dimension +1 
which by themselves need not vanish as $\Lambda \to 0$ since 
in their defining equation ( eq. (2.14) of ref. \cite{VE3} ) they are 
already accompanied by a $\Lambda^2$ factor.

The Bethe-Salpeter amplitudes $\Lambda\Phi$ must satisfy a well-known 
normalization condition \cite{NORM}, which in a condensed notation, and 
again in the most simplified form omitting ghost and fermion terms, reads
\begin{eqnarray}                                            \label{bs-norm}
 \bigg\{\ g_0^2\frac{1}{2}\int\Lambda\Phi
 \Big(\frac{\partial}{\partial P^\mu}\big(DD\big)\Big)
 \Lambda\Phi & & \bigg\}_{P^2=-b\Lambda^2}             \nonumber \\
 +\bigg\{g_0^4\frac{1}{4}\int\int\Lambda\Phi\big(DD\big)
 \Big(\frac{\partial}{\partial P^\mu}K_s\Big)\big(DD\big)
 \Lambda\Phi & & \bigg\}_{P^2=-b\Lambda^2} = \,-2P^\mu.
\end{eqnarray}
This is depicted in Fig.11.    %\ \ref{bs_norm_fig}. 
Fulfillment of this condition 
by purely zeroth-order quantities is possible only if the loop integrals 
on the l.h.s.\ diverge and thereby trigger the divergence-related $1/g^2$
mechanism. With the $\Phi$ functions behaving at worst like constants 
at large loop momenta q, and with the easily checked behavior of 
\begin{equation}
\frac{\partial}{\partial P^\mu}\big(DD\big)={\cal O}(q^{-5})\qquad , \qquad
\frac{\partial}{\partial P^\mu}K_s={\cal O}(q^{-1})
\end{equation}
(note that terms in $K_s$ behaving like constants at large $q$'s, such as 
the $\Gamma^{(0)pert}$ term, drop out after differentiation), it is however 
clear that the integrals are convergent, and do not provide the 
$\frac{1}{\epsilon}$ factors necessary for the mechanism to work. 
Thus poles of type (\ref{mandelst-pole}) in $T_{4V}^{[r,0]}$ are ruled out.

Note that the "compensating" poles , by contrast, {\it can} satisfy 
(\ref{bs-norm}) as zeroth-order quantities, since there $\Lambda\Phi$ 
is replaced by a $B_n$ ( compare eq. (2.21) of \cite{VE3} ) which at 
large $q$ behaves like $q^1$, and therefore provides divergences on the 
l.h.s.\ of (\ref{bs-norm}).

Note also that if one does not insist on the $\Phi$ being quantities of 
zeroth perturbative order, (\ref{bs-norm}) may be satisfied, at suitable 
eigenvalues $-b\Lambda^2$ of $P^2$, by quantities which have no overall
$\Lambda$ factor but are power series 
starting at least at first order in $g^2$. This, as emphasized 
in the introduction, is the case for ordinary bound-state poles, which
arise from a Bethe-Salpeter partial resummation of the perturbation series 
for $\Gamma_{4V}$.

%
%
%Latex Source File for Appendix B of 4-Gluon-Vertex Paper
%--------------------------------------------------------
%
\section{Parametrizations of the reduced four-gluon vertex}
\setcounter{equation}{0}
\subsection{Building blocks for the symmetric parametrization}
To list the building blocks $W^{[1,0]}_{\,n}$ of the fully Bose-symmetric 
$V_{4V}$ representation (\ref{VZETA}), we use the following combinations
of single-pole quantities (\ref{PIFA}):
\begin{eqnarray}
 G_{(1) }(p_1^{\,2} \ldots p_4^{\,2})  &=& \Pi_1+\Pi_2+\Pi_3+\Pi_4 \ ;  \\
 G_{(2s)}(p_1^{\,2} \ldots p_4^{\,2}) \ = \ \Pi_1\Pi_2 + \Pi_3\Pi_4 \ &,& \ 
 G_{(2u)}(p_1^{\,2} \ldots p_4^{\,2}) \ = \ \Pi_1\Pi_3 + \Pi_4\Pi_2 \ ,  
                                                             \nonumber  \\   
 G_{(2t)}(p_1^{\,2} \ldots p_4^{\,2})  &=& \ \Pi_1\Pi_4 + \Pi_2\Pi_3 \ ;\\  
 G_{(3) }(p_1^{\,2} \ldots p_4^{\,2}) \ =\  \Pi_1\Pi_2\Pi_3 &+&
            \Pi_1\Pi_2\Pi_4\,+\,\Pi_1\Pi_3\Pi_4\,+\,\Pi_2\Pi_3\Pi_4 \ ;  \\
 G_{(4) }(p_1^{\,2} \ldots p_4^{\,2})  &=& \Pi_1\Pi_2\Pi_3\Pi_4      \ . 
\end{eqnarray}
Here $G_{(1)}$, $G_{(3)}$, and $G_{(4)}$ are Bose-symmetric in themselves, 
while $G_{2s}$, $G_{2u}$, and $G_{2t}$ form a crossing triplet.
In terms of these,
\begin{eqnarray}
 W^{[1,0]}_{\,1 } \ : &=& \Big\{\ C^{(S)}\ \big[L_{(3)}-2L_{(1)}+L_{(2)}\big]
       \nonumber \\   & & \ +\big(\ perms.\,s\rightarrow u, \,s\rightarrow t
                           \ \big)\ \Big\} \ G_{(1)} \ ;                  \\
 W^{[1,0]}_{\,2 } \ : &=& \Big\{\ C^{(S)}\ \big[L_{(3)}-2L_{(1)}+L_{(2)}\big]
       \nonumber \\   & & \ +\big(\ perms.\,s\rightarrow u, \,s\rightarrow t
           \ \big)\ \Big\} \ \big( G_{(2s)}+G_{(2u)}+G_{(2t)} \big)\ ;    \\
 W^{[1,0]}_{\,3 } \ : &=& \Big\{\ C^{(S)}\ \big[L_{(3)}-2L_{(1)}+L_{(2)}\big]
       \nonumber \\   & & \times\big[G_{(2t)}-2G_{(2s)}+G_{(2u)}\big]\ \Big\}
       \ + \ \Big(\ perms.\,s\rightarrow u, \,s\rightarrow t\ \Big)\ ;    \\
 W^{[1,0]}_{\,4 } \ : &=& \Big\{\ C^{(S)}\ \big[L_{(3)}-2L_{(1)}+L_{(2)}\big]
       \nonumber \\   & & \ +\big(\ perms.\,s\rightarrow u, \,s\rightarrow t
                           \ \big)\ \Big\} \ G_{(3)} \ ;                  \\
 W^{[1,0]}_{\,5 } \ : &=& \Big\{\ C^{(S)}\ \big[L_{(3)}-2L_{(1)}+L_{(2)}\big]
       \nonumber \\   & & \ +\big(\ perms.\,s\rightarrow u, \,s\rightarrow t
                           \ \big)\ \Big\} \ G_{(4)} \ ;                  \\
 W^{[1,0]}_{\,6 } \ : &=& \Big\{\frac{1}{4}C^{(1)}\ \big[L_{(1)}+L_{(2)}+L_{(
3)}\big]\nonumber\\   & & \times \big[G_{(2t)}-2G_{(2s)}+G_{(2u)}\big]\ \Big\}
       \ + \ \Big(\ perms.\,s\rightarrow u, \,s\rightarrow t\ \Big)\ ;    \\
 W^{[1,0]}_{\,7 } \ : &=& \Big\{\frac{1}{4}C^{(1)}\ \big[L_{(3)}-2L_{(1)}+L_{(
2)}\big]\nonumber\\   & & \ +\big(\ perms.\,s\rightarrow u, \,s\rightarrow t
                           \ \big)\ \Big\} \ G_{(1)} \ ;                  \\ 
 W^{[1,0]}_{\,8 } \ : &=& \Big\{\frac{1}{4}C^{(1)}\ \big[L_{(3)}-2L_{(1)}+L_{(
2)}\big]\nonumber\\   & & \ +\big(\ perms.\,s\rightarrow u, \,s\rightarrow t
           \ \big)\ \Big\} \ \big( G_{(2s)}+G_{(2u)}+G_{(2t)} \big)\ ;    \\
 W^{[1,0]}_{\,9 } \ : &=& \Big\{\frac{1}{4}C^{(1)}\ \big[L_{(3)}-2L_{(1)}+L_{(
2)}\big]\nonumber\\   & & \times \big[G_{(2t)}-2G_{(2s)}+G_{(2u)}\big] \ \Big\}
       \ + \ \Big(\ perms.\,s\rightarrow u, \,s\rightarrow t\ \Big)\ ;    \\
 W^{[1,0]}_{\,10} \ : &=& \Big\{\frac{1}{4}C^{(1)}\ \big[L_{(3)}-2L_{(1)}+L_{(
2)}\big]\nonumber\\   & & \ +\big(\ perms.\,s\rightarrow u, \,s\rightarrow t
                           \ \big)\ \Big\} \ G_{(3)} \ ;                  \\
 W^{[1,0]}_{\,11} \ : &=& \Big\{\frac{1}{4}C^{(1)}\ \big[L_{(3)}-2L_{(1)}+L_{(
2)}\big]\nonumber\\   & & \ +\big(\ perms.\,s\rightarrow u, \,s\rightarrow t
                           \ \big)\ \Big\} \ G_{(4)} \ ;                  \\
 W^{[1,0]}_{\,12} \ : &=& \Big\{\frac{1}{4}C^{(1)}\ \big[L_{(1)}+L_{(2)}+L_{(
3)}\big]\nonumber\\   & & \times \big[G_{(2t)}-2G_{(2s)}+G_{(2u)}\big]\ \Big\}
       \ + \ \Big(\ perms.\,s\rightarrow u, \,s\rightarrow t\ \Big)\ ;    \\
 W^{[1,0]}_{\,13} \ : &=& \frac{1}{4}\big[C^{(1)}+C^{(2)}+C^{(3)}\big]\,
                          \big[L_{(1)}+L_{(2)}+L_{(3)}\big] \ G_{(1)} \ ; \\
 W^{[1,0]}_{\,14} \ : &=& \frac{1}{4}\big[C^{(1)}+C^{(2)}+C^{(3)}\big]\,
                          \big[L_{(1)}+L_{(2)}+L_{(3)}\big] 
       \nonumber \\   & & \times\big[G_{(2t)}+G_{(2s)}+G_{(2u)}\big] \ ;    \\ 
 W^{[1,0]}_{\,15} \ : &=& \frac{1}{4}\big[C^{(1)}+C^{(2)}+C^{(3)}\big]\,
                          \big[L_{(1)}+L_{(2)}+L_{(3)}\big] \ G_{(3)} \ ; \\
 W^{[1,0]}_{\,16} \ : &=& \frac{1}{4}\big[C^{(1)}+C^{(2)}+C^{(3)}\big]\,
                          \big[L_{(1)}+L_{(2)}+L_{(3)}\big] \ G_{(4)} \ ; \\
 W^{[1,0]}_{\,17} \ : &=& \frac{1}{4}\big[C^{(1)}+C^{(2)}+C^{(3)}\big]\ 
                          \Big\{\,\big[L_{(3)}-2L_{(1)}+L_{(2)}\big]\ G_{(2s)}
       \nonumber \\   & & \ +\big(\ perms.\,s\rightarrow u, \,s\rightarrow t
                          \ \big)\ \Big\} \ .
\end{eqnarray}
\subsection{Relation between parametrizations}
%\clearpage\noindent
%
To obtain the expressions for the $\eta$ parameters of (\ref{FIJP}) and
(\ref{FIJN}) in terms of the 17 $\zeta$ coefficients, one rewrites
the $W^{[1,0]}_{\,n}$ building blocks ( as listed above ) in eq. 
(\ref{VZETA}) in terms of the linearly independent color tensors
(\ref{COLABE}/\ref{COLCD}) and the $s$-channel-adapted Lorentz tensors
(\ref{LOP}/\ref{LON}) and compares coefficients with eq. (\ref{VFIJ}).
The ensuing relations are,
\begin{eqnarray}
      \eta_{A,0,1} & = & -\; \frac{3}{2}\,\zeta_{\,7} \;+\;
                             \frac{3}{2}\,\zeta_{\,13} \quad  \\%*[1mm] 
      \eta_{A,0,2} & = & -\; \frac{3}{4}\,\zeta_{\,3} \;-\;
                             \frac{3}{2}\,\zeta_{\,8} \;-\;
                             \frac{3}{2}\,\zeta_{\,9} \;+\;
                             \frac{3}{2}\,\zeta_{\,12} \;+\;
                             \frac{3}{2}\,\zeta_{\,14} \;+\;
                             \frac{3}{4}\,\zeta_{\,17}  \quad  \\%*[1mm]
      \eta_{A,0,4} & = & \frac{3}{2}\,\zeta_{\,3} \;-\;
                             \frac{3}{2}\,\zeta_{\,8} \;+\;
                             3\,\zeta_{\,9} \;-\;
                             3\,\zeta_{\,12} \;+\;
                             \frac{3}{2}\,\zeta_{\,14} \;-\;
                             \frac{3}{2}\,\zeta_{\,17}         \\%*[1mm]
      \eta_{A,0,5} & = & -\; \frac{3}{2}\,\zeta_{\,10} \;+\;
                             \frac{3}{2}\,\zeta_{\,15}         \\%*[1mm] 
      \eta_{A,0,6} & = & -\; \frac{3}{2}\,\zeta_{\,11} \;+\;
                             \frac{3}{2}\,\zeta_{\,16}         \\%*[1mm]
      \eta_{A,+,1} & = & -\; \frac{1}{2}\,\zeta_{\,7} \;+\;
                             \frac{1}{2}\,\zeta_{\,13}         \\%*[1mm]
      \eta_{A,+,2} & = & \frac{1}{4}\,\zeta_{\,3} \;+\;
                             \frac{1}{2}\,\zeta_{\,8} \;+\;
                             \frac{1}{2}\,\zeta_{\,9} \;+\;
                             \frac{1}{2}\,\zeta_{\,12} \;+\;
                             \frac{1}{2}\,\zeta_{\,14} \;-\;
                             \frac{1}{4}\,\zeta_{\,17}         \\%*[1mm]
      \eta_{A,+,4} & = & -\; \frac{1}{2}\,\zeta_{\,3} \;+\;
                             \frac{1}{2}\,\zeta_{\,8} \;-\;
                             \zeta_{\,9} \;-\;
                             \zeta_{\,12} \;+\;
                             \frac{1}{2}\,\zeta_{\,14} \;+\;
                             \frac{1}{2}\,\zeta_{\,17}         \\%*[1mm]
      \eta_{A,+,5} & = & \frac{1}{2}\,\zeta_{\,10} \;+\;
                             \frac{1}{2}\,\zeta_{\,15}         \\%*[1mm] 
      \eta_{A,+,6} & = & \frac{1}{2}\,\zeta_{\,11} \;+\;
                             \frac{1}{2}\,\zeta_{\,16}         \\%*[1mm]
      \eta_{A,-,3} & = & \frac{3}{4}\,\zeta_{\,3} \;-\;
                             \frac{3}{4}\,\zeta_{\,17}         \\%*[1mm]
      \eta_{B,0,1} & = & \frac{3}{4}\,\zeta_{\,7} \;+\;
                             \frac{3}{2}\,\zeta_{\,13}         \\%*[1mm]
      \eta_{B,0,2} & = & -\; \frac{3}{4}\,\zeta_{\,3} \;+\;
                             \frac{3}{4}\,\zeta_{\,8} \;-\;
                             \frac{3}{8}\,\zeta_{\,9} \;-\;
                             \frac{3}{4}\,\zeta_{\,12} \;+\;
                             \frac{3}{2}\,\zeta_{\,14} \;+\;
                             \frac{3}{4}\,\zeta_{\,17}         \\%*[1mm]
      \eta_{B,0,4} & = & \frac{3}{2}\,\zeta_{\,3} \;+\;
                             \frac{3}{4}\,\zeta_{\,8} \;+\;
                             \frac{3}{4}\,\zeta_{\,9} \;+\;
                             \frac{3}{2}\,\zeta_{\,12} \;+\;
                             \frac{3}{2}\,\zeta_{\,14} \;-\;
                             \frac{3}{2}\,\zeta_{\,17}         \\%*[1mm]
      \eta_{B,0,5} & = & \frac{3}{4}\,\zeta_{\,10} \;+\;
                             \frac{3}{2}\,\zeta_{\,15}         \\%*[1mm] 
      \eta_{B,0,6} & = & \frac{3}{4}\,\zeta_{\,11} \;+\;
                             \frac{3}{2}\,\zeta_{\,16}         \\%*[1mm]
      \eta_{B,+,1} & = & -\; \frac{1}{4}\,\zeta_{\,7} \;+\;
                             \frac{1}{2}\,\zeta_{\,13}         \\%*[1mm]
      \eta_{B,+,2} & = & \frac{1}{4}\,\zeta_{\,3} \;-\;
                             \frac{1}{4}\,\zeta_{\,8} \;+\;
                             \frac{1}{8}\,\zeta_{\,9} \;-\;
                             \frac{1}{4}\,\zeta_{\,12} \;+\;
                             \frac{1}{2}\,\zeta_{\,14} \;-\;
                             \frac{1}{4}\,\zeta_{\,17}         \\%*[1mm]
      \eta_{B,+,4} & = & -\; \frac{1}{2}\,\zeta_{\,3} \;-\;
                             \frac{1}{4}\,\zeta_{\,8} \;-\;
                             \frac{1}{4}\,\zeta_{\,9} \;+\;
                             \frac{1}{2}\,\zeta_{\,12} \;+\;
                             \frac{1}{2}\,\zeta_{\,14} \;+\;
                             \frac{1}{2}\,\zeta_{\,17}         \\%*[1mm]
      \eta_{B,+,5} & = & -\; \frac{1}{4}\,\zeta_{\,10} \;+\;
                             \frac{1}{2}\,\zeta_{\,15}         \\%*[1mm] 
      \eta_{B,+,6} & = & -\; \frac{1}{4}\,\zeta_{\,11} \;+\;
                             \frac{1}{2}\,\zeta_{\,16}         \\%*[1mm] 
      \eta_{B,-,3} & = & \frac{3}{4}\,\zeta_{\,3} \;+\;
                             \frac{9}{8}\,\zeta_{\,9} \;-\;
                             \frac{3}{4}\,\zeta_{\,17}          \\%*[1mm]
      \eta_{C,0,3} & = & -\; \frac{9}{8}\,\zeta_{\,9} \;-\;    
                             \frac{9}{4}\,\zeta_{\,12}          \\%*[1mm]
%
%\clearpage\noindent
      \eta_{C,+,3} & = & \frac{3}{8}\,\zeta_{\,9} \;-\;
                             \frac{3}{4}\,\zeta_{\,12}          \\%*[1mm] 
      \eta_{C,-,1} & = & -\; \frac{3}{4}\,\zeta_{\,7}           \\%*[1mm]
      \eta_{C,-,2} & = & -\; \frac{3}{4}\,\zeta_{\,8} \;+\;
                             \frac{3}{8}\,\zeta_{\,9}           \\%*[1mm] 
      \eta_{C,-,4} & = & -\; \frac{3}{4}\,\zeta_{\,8} \;-\;
                             \frac{3}{4}\,\zeta_{\,9}           \\%*[1mm]
      \eta_{C,-,5} & = & -\; \frac{3}{4}\,\zeta_{\,10}          \\%*[1mm]
      \eta_{C,-,6} & = & -\; \frac{3}{4}\,\zeta_{\,11}          \\%*[1mm] 
      \eta_{D,0,3} & = & \frac{9}{4}\,\zeta_{\,3} \;+\;
                             \frac{9}{2}\,\zeta_{\,6}           \\%*[1mm]
      \eta_{D,+,3} & = & -\; \frac{3}{4}\,\zeta_{\,3} \;+\;
                             \frac{3}{2}\,\zeta_{\,6}           \\%*[1mm]
      \eta_{D,-,1} & = & \frac{3}{2}\,\zeta_{\,1}               \\%*[1mm] 
      \eta_{D,-,2} & = & \frac{3}{2}\,\zeta_{\,2} \;-\;
                             \frac{3}{4}\,\zeta_{\,3}           \\%*[1mm]
      \eta_{D,-,4} & = & \frac{3}{2}\,\zeta_{\,2} \;+\;
                             \frac{3}{2}\,\zeta_{\,3}           \\%*[1mm]
      \eta_{D,-,5} & = & \frac{3}{2}\,\zeta_{\,4}               \\%*[1mm]
      \eta_{D,-,6} & = & \frac{3}{2}\,\zeta_{\,5}               \\%*[1mm]
      \eta_{E,0,1} & = & -\; \frac{3}{2}\,\zeta_{\,1}           \\%*[1mm]
      \eta_{E,0,2} & = & -\; \frac{3}{2}\,\zeta_{\,2} \;-\;
                             \frac{3}{4}\,\zeta_{\,3} \;+\;
                             \frac{3}{2}\,\zeta_{\,6} 
                               \qquad\qquad                     \\%*[1mm]
      \eta_{E,0,4} & = & -\; \frac{3}{2}\,\zeta_{\,2} \;+\;
                             \frac{3}{2}\,\zeta_{\,3} \;-\;
                             3\,\zeta_{\,6}                     \\%*[1mm]
      \eta_{E,0,5} & = & -\; \frac{3}{2}\,\zeta_{\,4}           \\%*[1mm] 
      \eta_{E,0,6} & = & -\; \frac{3}{2}\,\zeta_{\,5}           \\%*[1mm]
      \eta_{E,+,1} & = & \frac{1}{2}\,\zeta_{\,1}               \\%*[1mm]
      \eta_{E,+,2} & = & \frac{1}{2}\,\zeta_{\,2} \;+\;
                             \frac{1}{4}\,\zeta_{\,3} \;+\;
                             \frac{1}{2}\,\zeta_{\,6}           \\%*[1mm] 
      \eta_{E,+,4} & = & \frac{1}{2}\,\zeta_{\,2} \;-\;
                             \frac{1}{2}\,\zeta_{\,3} \;-\;
                             \zeta_{\,6}                        \\%*[1mm]
      \eta_{E,+,5} & = & \frac{1}{2}\,\zeta_{\,4}               \\%*[1mm] 
      \eta_{E,+,6} & = & \frac{1}{2}\,\zeta_{\,5}               \\%*[1mm]
%
%\clearpage\noindent
      \eta_{E,-,3} & = & -\; \frac{3}{4}\,\zeta_{\,3} \ .  
                     \qquad \qquad  
\end{eqnarray}

%
%
%Latex Source File for Appendix C of Four-Gluon Vertex Paper
%-----------------------------------------------------------
%
\section{Representations of the amplitude $T'_{5V}$}
\setcounter{equation}{0}
Here we sketch derivations for two different representations of the
partially irreducible 5-gluon amplitude $T'_{5V}$ needed in the 
argument of sect. 3.

{\bf (\,1\,)} To establish the representation given in Fig.3, start from 
the full connected and amputated 5-gluon amplitude ( off-shell 5-gluon T
matrix ), $T_{5V}$. It has 25 one-gluon-reducible terms ( dressed tree
diagrams ): one of structure $\Gamma_{3V}\,D\,\Gamma_{4V}$, reducible
along a single internal gluon line, for each of the 10 different
2+3 partitions ( or $2V\leftrightarrow3V$ channels ) of the 5 legs,
and one of structure $\Gamma_{3V}\,D\,\Gamma_{3V}\,D\,\Gamma_{3V}$,
reducible along two different internal lines, for each of the 15
different 2+1+2 partitions of the 5 legs. In the condensed notation
adopted for eq. (\ref{G5SHA}), we have
\begin{eqnarray}                                          \label{T5DEC}
 T_{5V} &=& \ \ \Gamma_{5V}                               \nonumber   \\
        & & + \ \sum_{i=1}^{10} \big\{ \,\Gamma_{3V}\,D\,\Gamma_{4V}\,
            \big\}_{i}                                                \\
        & & + \ \sum_{k=1}^{15} \big\{ \,\Gamma_{3V}\,D\,\Gamma_{3V}\,
            D\,\Gamma_{3V}\, \big\}_{k} \ .                \nonumber
\end{eqnarray}
Of these one-gluon-reducible terms, 13 are reducible in at least one
of the four channels (\ref{HOR5}/\ref{TILT5}), and therefore by definition
are to be excluded from the $T'_{5V}$ amplitude, leaving the latter to
consist of the fully one-gluon-irreducible piece ( proper vertex ) 
$\Gamma_{5V}$, plus 12 dressed-tree diagrams still reducible in other
channels -- 6 from the second and 6 from the third line of (\ref{T5DEC}).   

The 6 one-gluon exchange terms from the second line of (\ref{T5DEC}) 
are those in which one of the two legs no. 5 and 6 entering from the left
in Fig.3 connects to the $\Gamma_{3V}$, while the other connects to the
$\Gamma_{4V}$. The 3 terms in which leg no. 6 connects to a $\Gamma_{3V}$
constitute the second line of Fig.3. The 3 terms in which leg no. 5
connects to the $\Gamma_{3V}$ may each be taken together with a suitable
pair from among the 6 terms retained of the third line of (\ref{T5DEC}),
in such a way as to replace their $\Gamma_{4V}$'s with $T'_{4V}$'s
whose primes refer to the channels defined by their external-line pairs.
( Remember that, by definition, $T'_{4V}\ =\ \Gamma_{4V}$ plus 2 
one-gluon exchange terms in the channels other than the distinguished 
channel to which the prime refers ). This establishes the third line of
Fig.3.

Of course, in this regrouping the roles of legs 5 and 6 may be
interchanged, leading to a representation equivalent to Fig.3 in which
the $T'_{4V}$'s and $\Gamma_{4V}$'s have their roles exchanged. This
explains the apparent asymmetry of Fig.3.
\bigskip
{\bf (\,2\,)} The representation of $T'_{5V}$ given in Fig.5 is established
by $(i)$ applying the decomposition of eq. (\ref{G5SHA}) to the
$\Gamma_{5V}$ in the first line of Fig.3, $(ii)$ inserting the
decompositions of Fig.1(a) and Fig.1(b) into the $\Gamma_{4V}$ and the
$T'_{4V}$ terms of Fig.3 respectively, and $(iii)$ performing
decompositions of the remaining ordinary gluon lines into ``softened''
one-gluon exchanges and shadow-pole terms as indicated in Fig.6. ( As for
Fig.6(b), it lumps together for simplicity two decompositions, 
corresponding to its first and second lines, that actually occur
separately ). The six last terms of Fig.3 then give rise to:

(a) 6 softened one-gluon-exchange graphs of the type of term (B) of
Fig.6, shown in the second line of Fig.5,

(b) 6 graphs with two softened one-gluon exchanges each, of the type of
term (E) of Fig.6, shown in the third line of Fig.5,

(c) 6 one-shadow-line graphs as in term (C) of Fig.6, which cancel 6 
of the 10 negative terms coming from the second line of the $\Gamma_{5V}$
decomposition (\ref{G5SHA}), leaving just those four of the 10 negative
terms that possess a shadow line in one of the four channels 
(\ref{HOR5}/\ref{TILT5}),

(d) 12 graphs with two shadow lines each of the type of term (H) of
Fig.6 -- 9 from the $\Gamma_{4V}$ and 3 from the $T'_{4V}$ terms -- 
with minus signs, which cancel 12 of the 15 terms from the third line
of eq. (\ref{G5SHA}), leaving just those three of the 15 terms that have
one shadow in the ``horizontal'' channel (\ref{HOR5}) and the other in
one of the three ``tilted'' channels (\ref{TILT5}),

(e) 18 graphs with one softened-gluon exchange and one shadow line
each: 12 with minus signs from both the $\Gamma_{4V}$ and $T'_{4V}$
terms, and 6 with plus signs from the $T'_{4V}$ terms alone, and which
actually cancel 6 of the minus-sign graphs, leaving a net count of
6 terms with one softened-gluon and one shadow line and with minus signs.

Now the 6 terms surviving from (e), the 3 two-shadows terms surviving 
from (d), and 3 of the 4 one-shadow terms from (c) may be lumped 
together in three groups by defining the auxiliary amplitudes $\Xi_n$
as in Fig.7; they then give the three graphs in the last two lines
of Fig.5. The one remaining one-shadow graph with minus sign from (c)
appears explicitly in the first line of Fig.5, which is therefore
fully established.

\clearpage
%
% Latex Source File for Appendix D of 4-Gluon-Vertex Paper
%---------------------------------------------------------
%       ( Only for hep-th submission, not for print publication )
%       ---------------------------------------------------------
%
%
\section{Listing of $V_{4V}$ self-consistency equations} 
\setcounter{equation}{0}
\thispagestyle{empty}
In the following, the 54 self-consistency conditions (\ref{4SYS}), {\em
multiplied by} $\beta_0 u_3^{\,2}$ to rid them of ubiquitous factors from
the self-consistency mechanism, are listed according
to the $C^{(i)} \otimes L_{(j)}$ tensor combination ($i=A \ldots E, \ j=0,
+,-$) in whose invariant function they arise. For the quark-vertex
coefficients $z_i$, the simplified numbering of eqs. (4.6) of \cite{VE3}
is employed. 
\vspace*{2mm}
\subsection{$C_A$--$L_{0}$ equations:}
\vspace{6mm}
\fbox{\parbox{14.8cm}{
\begin{small}
\begin{eqnarray*} & & \hspace*{-5mm} \frac{9}{4}\,u_3^2x_3 \,
                      \Big( \,5\,\zeta_2 \,+\, 
                \zeta_3 \,+\, 8\,\zeta_6 \,-\, 6\,\zeta_8 \,+\, 
                      12\,\zeta_9 \,-\, 12\,\zeta_{12} \,+\, 
                      6\,\zeta_{14} \,-\, 6\,\zeta_{17} \,\Big) 
                       \,-\, \frac{45}{4}\,u_3x_4\,\zeta_4         \\*[1mm]
      & & \hspace*{-5mm} +\; \Big( \frac{3}{2}\,\beta_0\,u_3^2 \,+\, 
                      \frac{27}{2}\,u_3x_4 \Big)
             \,\Big( \,\zeta_{10} \,-\, \zeta_{15} \,\Big) \,+\, 
              \Big( \frac{45}{4}\,u_3x_1 \,-\, 
             \frac{15}{8}\,u_3x_3 \,-\, N_F\,u_3z_3 \Big)
             \,\Big( \,\zeta_{11} \,-\, \zeta_{16} \,\Big) 
\end{eqnarray*}   
\[  \qquad\qquad = \quad -\; 27\,u_3^2x_3^3 \,+\, 
                       54\,u_3x_3x_4^2 \,-\, 27\,x_4^2x_5 \]       \\*
\end{small} }}
\begin{equation} \label{skgv1} 
\end{equation}\\
\fbox{\parbox{14.8cm}{
\begin{small}
\begin{eqnarray*} & & \hspace*{-6mm} \frac{9}{2}\,u_3^2x_3\,\zeta_1 \,+\, 
             \Big( \frac{27}{16}\,u_3^2 \,-\, \frac{9}{2}\,u_3x_4
                 \Big)\,\zeta_2 \,-\, \Big( \frac{3}{2}\,
                    \beta_0\,u_3^2 \,-\, \frac{27}{16}\,u_3^2
                  \,+\, \frac{9}{4}\,u_3x_4 \Big) \,\zeta_3 
                    \,+\, \frac{27}{4}\,u_3x_1\,\zeta_4           \\*[1mm]
      & & \hspace*{-6mm} +\; \Big( 
                    \frac{27}{8}\,u_3^2 \,+\, \frac{9}{4}\,u_3x_4 
                \Big) \,\zeta_6 \,+\, \Big( \frac{3}{2}\,
                    \beta_0\,u_3^2 \,-\, \frac{27}{8}\,u_3^2
                  \,-\, \frac{9}{4}\,u_3x_4 \Big) \,
                  \Big( \,\zeta_8 \,-\, 2\,\zeta_9 \,+\, 2\,\zeta_{12}
                \,-\, \zeta_{14} \,+\, \zeta_{17} \,\Big)         \\*[1mm]
      & & \hspace*{-6mm} +\;
            \Big( \frac{99}{4}\,u_3x_1 \,-\, \frac{15}{8}\,u_3x_3 
                    \,-\, N_F\,u_3z_3 \Big)
             \,\Big( \,\zeta_{10} \,-\, \zeta_{15} \,\Big)        
\end{eqnarray*}   
\begin{eqnarray*}  \qquad\qquad & = & -\; \frac{27}{2}\,u_3^2x_3^2 \,+\, 
                       36\,u_3x_1x_3x_4 \,+\, \frac{9}{2}\,u_3x_4^2
                       \,-\, 9\,x_2x_4^2  \,-\, 18\,x_1x_4x_5      \\*[1mm]
             & & +\; N_F\,\Big( \frac{u_3}{w_3} \Big)^2 \,
                  \Big( \frac{2}{3}\,w_3z_1z_3z_4
                     \,-\, \frac{2}{3}\,w_3z_4^2
                    \,+\, \frac{2}{3}\,z_1z_4z_5
                       \,-\, \frac{2}{3}\,z_2z_4^2 \Big)
\end{eqnarray*}
\end{small} }}
\begin{equation} \label{skgv2}
\end{equation}
\clearpage\noindent
\fbox{\parbox{14.8cm}{
\begin{small}
\begin{eqnarray*} & & \hspace*{-6mm} \frac{9}{2}\,u_3^2x_3\,\Big( \,2\,\zeta_1 
                 \,-\, 3\,\zeta_7 \,+\, 3\,\zeta_{13} \,\Big) \,-\, 
               9\,u_3x_4\,\zeta_2 \,+\, \Big( \,\frac{3}{4}\,
                    \beta_0\,u_3^2 \,+\, \frac{9}{4}\,u_3x_4                  
   \,\Big) \,\zeta_3 \,-\, \frac{9}{4}\,u_3x_1\,\zeta_4 \\*[1mm]
      & & \hspace*{-6mm} +\; \Big( \frac{27}{16}\,u_3^2 \,+\, 
           \frac{63}{8}\,u_3x_4 \Big) \,\zeta_6 \,+\, \Big( \frac{3}{4}\,
                \beta_0\,u_3^2 \,+\, \frac{27}{4}\,u_3x_4 \Big) \,
                  \Big( \,2\,\zeta_8 \,+\, 2\,\zeta_9 \,-\, 2\,\zeta_{12}
                \,-\, 2\,\zeta_{14} \,-\, \zeta_{17} \,\Big) \\*[1mm]
      & & \hspace*{-6mm} +\; \Big( \frac{45}{4}\,u_3x_1 
             \,-\, \frac{15}{8}\,u_3x_3 \,-\, N_F\,u_3z_3 \Big)
             \,\Big( \,\zeta_{10} \,-\, \zeta_{15} \,\Big)        
\end{eqnarray*}   
\begin{eqnarray*}  \quad\qquad & = & -\; \frac{45}{2}\,u_3^2x_3^2 \,+\, 
                       36\,u_3x_1x_3x_4 \,+\, 27\,u_3x_4^2
                       \,-\, 9\,x_2x_4^2  \,-\, 18\,x_1x_4x_5 \\*[1mm]
             & & +\; N_F\,\Big( \frac{u_3}{w_3} \Big)^2 \,
                  \Big( -\,\frac{1}{3}\,w_3z_1z_3z_4
                     \,+\, \frac{1}{3}\,w_3z_4^2
                    \,-\, \frac{1}{3}\,z_1z_4z_5
                       \,+\, \frac{1}{3}\,z_2z_4^2 \Big)
\end{eqnarray*}
\end{small} }}
\begin{equation} \label{skgv3}
\end{equation}\\
\fbox{\parbox{14.8cm}{
\begin{small}
\begin{eqnarray*} & & \hspace*{-10mm} -\;\frac{27}{4}\,u_3x_4\,\zeta_1 \,-\, 
             \frac{9}{2} \,u_3x_1\,\zeta_2 \,-\, \Big( \frac{27}{2}\,
                    u_3x_1 \,-\, \frac{15}{8}\,u_3x_3
                  \,-\, N_F\,u_3z_3 \Big) \,\zeta_3 
                       \,+\, \frac{9}{4}\,u_3x_1\,\zeta_6 \\*[1mm]
      & & \hspace*{-10mm} +\;\Big( \frac{3}{2}\,\beta_0\,u_3^2 
                        \,+\, \frac{27}{2}\,u_3x_4 
                \Big) \,\Big( \,\zeta_7 \,-\, \zeta_{13} \,\Big)
                    \,+\, \Big( \frac{45}{4}\,u_3x_1 \,-\, 
                 \frac{15}{8}\,u_3x_3 \,-\, N_F\,u_3z_3 \Big) \,
                  \Big( \,\zeta_8 \,-\, 2\,\zeta_9 \Big.  \\*[1mm]
      & & \hspace*{-10mm} \Big. \! +\;  2\,\zeta_{12}
                \,-\, \zeta_{14} \,+\, \zeta_{17} \,\Big)        
\end{eqnarray*}   
\begin{eqnarray*}  \qquad\qquad & = & -\; \frac{9}{2}\,u_3^2x_3 \,+\, 
                       18\,u_3x_1^2x_3 \,+\, 27\,u_3x_1x_4
                       \,-\, 9\,x_1^2x_5  \,-\, 18\,x_1x_2x_4 \\*[1mm]
             & & +\; N_F\,\Big( \frac{u_3}{w_3} \Big)^2 \,
                  \Big( \frac{2}{3}\,w_3z_1z_4
                     \,-\, \frac{2}{3}\,w_3z_1^2z_3
                    \,+\, \frac{2}{3}\,z_1z_2z_4
                       \,-\, \frac{2}{3}\,z_1^2z_5 \Big)
\end{eqnarray*}
\end{small} }}
\begin{equation} \label{skgv4}
\end{equation}\\
\fbox{\parbox{14.8cm}{
\begin{small}
\begin{eqnarray*} & & \hspace*{-18mm} \Big( \frac{27}{16}\,u_3^2 \,-\,
               \frac{9}{4}\,u_3x_4 \Big) \,\zeta_1 \,-\, 
               9\,u_3x_1 \,\zeta_2 \,+\, \Big( \,\frac{63}{8}\,
                    u_3x_1 \,-\, \frac{15}{16}\,u_3x_3 \,-\, 
               \frac{1}{2}\,N_F \,u_3z_3 \,\Big) \,\zeta_3 \\*[1mm]
      & & \hspace*{-18mm} +\;\frac{63}{8}\,u_3x_1\,\zeta_6 \,+\, \Big( 
                    \frac{3}{2}\,\beta_0\,u_3^2 \,-\, \frac{27}{8}\,u_3^2 
                \Big) \,\Big( \,\zeta_7 \,-\, \zeta_{13} \,\Big) \\*[1mm]
      & & \hspace*{-18mm} +\; \Big( \frac{99}{8}\,u_3x_1 \,-\, 
             \frac{15}{16}\,u_3x_3 \,-\, \frac{1}{2}\,N_F\,u_3z_3 
            \Big) \, \Big( \,2\,\zeta_8 \,+\, 2\,\zeta_9 
           \,-\, 2\,\zeta_{12} \,-\, 2\,\zeta_{14} \,-\, \zeta_{17} \,\Big) 
\end{eqnarray*}   
\begin{eqnarray*}  \qquad\qquad & = & -\; \frac{9}{2}\,u_3^2x_3 \,+\, 
                       18\,u_3x_1^2x_3 \,+\, 18\,u_3x_1x_4
                       \,-\, 9\,x_1^2x_5  \,-\, 18\,x_1x_2x_4 \\*[1mm]
             & & +\; N_F\,\Big( \frac{u_3}{w_3} \Big)^2 \,
                  \Big( -\, \frac{1}{3}\,w_3z_1z_4
                     \,+\, \frac{1}{3}\,w_3z_1^2z_3
                    \,-\, \frac{1}{3}\,z_1z_2z_4
                       \,+\, \frac{1}{3}\,z_1^2z_5 \Big)
\end{eqnarray*}
\end{small} }}
\begin{equation} \label{skgv5}
\end{equation}\\
\subsection{$C_A$--$L_{+}$ equations:}
\vspace{6mm}
\fbox{\parbox{14.8cm}{
\begin{small}
\begin{eqnarray*} & & \hspace*{-5mm} \frac{1}{4}\,u_3^2x_3 \,
                      \Big( \,15\,\zeta_2 \,-\, 
               13\, \zeta_3 \,+\, 24\,\zeta_6 \,-\, 2\,\zeta_8 \,+\, 
                 4\,\zeta_9 \,+\, 4\,\zeta_{12} \,-\, 2\,\zeta_{14} \,-\, 
                         2\,\zeta_{17} \,\Big) \,-\, 
                             \frac{15}{4} \,u_3x_4\,\zeta_4  \\*[1mm]
      & & \hspace*{-5mm} -\; \Big( \frac{1}{2}\,\beta_0\,u_3^2 \,-\, 
                      \frac{1}{2}\,u_3x_4 \Big)
             \,\Big( \,\zeta_{10} \,+\, \zeta_{15} \,\Big) \,-\, 
              \Big( \frac{15}{4}\,u_3x_1 \,-\, 
             \frac{5}{8}\,u_3x_3 \,-\, \frac{1}{3} \,N_F\,u_3z_3 \Big)
             \,\Big( \,\zeta_{11} \,+\, \zeta_{16} \,\Big) 
\end{eqnarray*}
\vspace*{1mm}
\[  \qquad\qquad = \quad -\; 9\,u_3^2x_3^3 \,+\, 
                       18\,u_3x_3x_4^2 \,-\, 9\,x_4^2x_5 \]        \\*
\end{small} }}
\begin{equation} \label{skgv6} 
\end{equation}\\
\fbox{\parbox{14.8cm}{
\begin{small}
\begin{eqnarray*} & & \hspace*{-14mm} \frac{7}{2}\,u_3^2x_3\,\zeta_1 \,-\, 
             \Big( \frac{3}{16}\,u_3^2 \,+\, \frac{7}{2}\,u_3x_4
                 \Big)\,\zeta_2 \,+\, \Big( \frac{1}{2}\,
                    \beta_0\,u_3^2 \,-\, \frac{3}{16}\,u_3^2
                  \,-\, \frac{7}{4}\,u_3x_4 \Big) \,\zeta_3 
                        \,-\, \frac{1}{4}\,u_3x_1\,\zeta_4 \\*[1mm]
      & & \hspace*{-14mm} +\; \Big( 
                    \frac{3}{8}\,u_3^2 \,+\, \frac{13}{4}\,u_3x_4 
                \Big) \,\zeta_6 \,-\, \Big( \frac{1}{2}\,
                    \beta_0\,u_3^2 \,-\, \frac{3}{8}\,u_3^2
                  \Big) \,\Big( \,\zeta_8 \,-\, 2\,\zeta_9 
                    \,-\, 2\,\zeta_{12}
                \,+\, \zeta_{14} \,+\, \zeta_{17} \,\Big) \\*[1mm]
      & & \hspace*{-14mm} -\;
            \Big( \frac{13}{4}\,u_3x_1 \,-\, \frac{5}{8}\,u_3x_3 
                    \,-\, \frac{1}{3} \,N_F\,u_3z_3 \Big)
             \,\Big( \,\zeta_{10} \,+\, \zeta_{15} \,\Big)        
\end{eqnarray*}   
\begin{eqnarray*}  \qquad\qquad & = & -\; \frac{17}{2}\,u_3^2x_3^2 \,+\, 
                       12\,u_3x_1x_3x_4 \,+\, \frac{11}{2}\,u_3x_4^2
                       \,-\, 3\,x_2x_4^2  \,-\, 6\,x_1x_4x_5 \\*[1mm]
             & & +\; N_F\,\Big( \frac{u_3}{w_3} \Big)^2 \,
                  \Big( -\,\frac{2}{9}\,w_3z_1z_3z_4
                     \,+\, \frac{2}{9}\,w_3z_4^2
                    \,-\, \frac{2}{9}\,z_1z_4z_5
                       \,+\, \frac{2}{9}\,z_2z_4^2 \Big)
\end{eqnarray*}
\end{small} }}
\begin{equation} \label{skgv7}
\end{equation}\\
\fbox{\parbox{14.8cm}{
\begin{small}
\begin{eqnarray*} & & \hspace*{-8mm} \frac{1}{2}\,u_3^2x_3\,\Big( \,4\,\zeta_1 
                 \,-\, \zeta_7 \,-\,\zeta_{13} \,\Big) \,+\, 
             \Big( \,\frac{3}{8}\,u_3^2 \,-\, 2\,u_3x_4
                 \,\Big) \,\zeta_2 \,-\, \Big( \,\frac{1}{4}\,
                    \beta_0\,u_3^2 \,+\, \frac{3}{8}\,u_3^2
                  \,+\, \frac{3}{4}\,u_3x_4 \Big) \,\zeta_3 \\*[1mm]
      & & \hspace*{-8mm} -\;\frac{7}{4}\,u_3x_1\,\zeta_4 \,+\, \Big( 
                    \frac{15}{16}\,u_3^2 \,+\, \frac{11}{8}\,u_3x_4 
                \Big) \,\zeta_6 \,-\, \Big( \frac{1}{4}\,
                    \beta_0\,u_3^2 \,-\, \frac{1}{4}\,u_3^2 \Big) \,
                  \Big( \,2\,\zeta_8 \,+\, 2\,\zeta_9 \Big. \\*[1mm]
      & & \hspace*{-8mm} \!\Big. +\, 2\,\zeta_{12}
                \,+\, 2\,\zeta_{14} \,-\, \zeta_{17} \,\Big) \,-\,
            \Big( \frac{15}{4}\,u_3x_1 \,-\, \frac{5}{8}\,u_3x_3 
                    \,-\, \frac{1}{3}\,N_F\,u_3z_3 \Big)
             \,\Big( \,\zeta_{10} \,+\, \zeta_{15} \,\Big)        
\end{eqnarray*}   
\begin{eqnarray*}  \qquad\qquad & = & -\; \frac{11}{2}\,u_3^2x_3^2 \,+\, 
                       12\,u_3x_1x_3x_4 \,+\, \frac{5}{2}\,u_3x_4^2
                       \,-\, 3\,x_2x_4^2  \,-\, 6\,x_1x_4x_5 \\*[1mm]
             & & +\; N_F\,\Big( \frac{u_3}{w_3} \Big)^2 \,
                  \Big( \frac{1}{9}\,w_3z_1z_3z_4
                     \,-\, \frac{1}{9}\,w_3z_4^2
                    \,+\, \frac{1}{9}\,z_1z_4z_5
                       \,-\, \frac{1}{9}\,z_2z_4^2 \Big)
\end{eqnarray*}
\end{small} }}
\begin{equation} \label{skgv8}
\end{equation}\\
\fbox{\parbox{14.8cm}{
\begin{small}
\begin{eqnarray*} & & \hspace*{-19mm} \Big( \frac{3}{4}\,u_3^2 \,-\, 
                \frac{1}{4}\,u_3x_4 \Big) \,\zeta_1 \,-\, 
             \frac{1}{4} \,u_3x_1\,\Big( \,14\,\zeta_2 \,-\,
                      13\,\zeta_6 \,\Big) \\*[1mm]
      & & \hspace*{-19mm} +\; \Big( 2\,
                    u_3x_1 \,-\, \frac{5}{8}\,u_3x_3 \,-\, 
               \frac{1}{3}\,N_F\,u_3z_3 \,\Big) \,\zeta_3 
                         \,-\, \Big( \frac{1}{2}
                       \,\beta_0\,u_3^2 \,-\, \frac{1}{2}\,u_3x_4 
                \Big) \,\Big( \,\zeta_7 \,+\, \zeta_{13} \,\Big) \\*[1mm]
      & & \hspace*{-19mm} -\; \Big( \frac{15}{4}\,u_3x_1 \,-\, 
                 \frac{5}{8}\,u_3x_3 \,-\, 
                      \frac{1}{3}\,N_F\,u_3z_3 \Big) \,
                  \Big( \,\zeta_8 \,-\, 2\,\zeta_9 \,-\, 2\,\zeta_{12}
                \,+\, \zeta_{14} \,+\, \zeta_{17} \,\Big)        
\end{eqnarray*}   
\begin{eqnarray*}  \quad\qquad & = & -\; \frac{3}{2}\,u_3^2x_3 \,+\, 
                       6\,u_3x_1^2x_3 \,+\, 5\,u_3x_1x_4
                       \,-\, 3\,x_1^2x_5  \,-\, 6\,x_1x_2x_4 \\*[1mm]
             & & +\; N_F\,\Big( \frac{u_3}{w_3} \Big)^2 \,
                  \Big( -\,\frac{2}{9}\,w_3z_1z_4
                     \,+\, \frac{2}{9}\,w_3z_1^2z_3
                    \,-\, \frac{2}{9}\,z_1z_2z_4
                       \,+\, \frac{2}{9}\,z_1^2z_5 \Big)
\end{eqnarray*}
\end{small} }}
\begin{equation} \label{skgv9}
\end{equation}\\
\fbox{\parbox{14.8cm}{
\begin{small}
\begin{eqnarray*} & & \hspace*{-16mm} \Big( \frac{3}{16}\,u_3^2 \,-\, 
                \frac{7}{4}\,u_3x_4 \Big) \,\zeta_1 \,-\, \frac{1}{8} 
                \,u_3x_1\,\Big( \,16\,\zeta_2 \,-\, 11\,\zeta_6 \,\Big)
                                                \\*[1mm]
      & & \hspace*{-16mm} -\; \Big( \frac{21}{8}\,
                    u_3x_1 \,-\, \frac{5}{16}\,u_3x_3 \,-\, 
                 \frac{1}{6}\,N_F\,u_3z_3 \,\Big) \,\zeta_3
                          \,-\, \Big( \frac{1}{2}\,
                    \beta_0\,u_3^2 \,-\, \frac{3}{8}\,u_3^2 
                \Big) \,\Big( \,\zeta_7 \,+\, \zeta_{13} \,\Big) \\*[1mm]
      & & \hspace*{-16mm} -\; \Big( \frac{13}{8}\,u_3x_1 \,-\, 
                 \frac{5}{16}\,u_3x_3 \,-\, 
                       \frac{1}{6}\,N_F\,u_3z_3 \Big) \,
                  \Big( \,2\,\zeta_8 \,+\, 2\,\zeta_9 \,+\, 2\,\zeta_{12}
                \,+\, 2\,\zeta_{14} \,-\, \zeta_{17} \,\Big)        
\end{eqnarray*}   
\begin{eqnarray*}  \qquad\qquad & = & -\; \frac{3}{2}\,u_3^2x_3 \,+\, 
                       6\,u_3x_1^2x_3 \,+\, 8\,u_3x_1x_4
                       \,-\, 3\,x_1^2x_5  \,-\, 6\,x_1x_2x_4 \\*[1mm]
             & & +\; N_F\,\Big( \frac{u_3}{w_3} \Big)^2 \,
                  \Big( \frac{1}{9}\,w_3z_1z_4
                     \,-\, \frac{1}{9}\,w_3z_1^2z_3
                    \,+\, \frac{1}{9}\,z_1z_2z_4
                       \,-\, \frac{1}{9}\,z_1^2z_5 \Big)
\end{eqnarray*}
\end{small} }}
\begin{equation} \label{skgv10}
\end{equation}\\
\subsection{$C_A$--$L_{-}$ equations:}
\vspace{6mm}
\fbox{\parbox{14.8cm}{
\begin{small}
\begin{eqnarray*} & & \hspace*{-16mm} \frac{3}{2}\,u_3^2x_3 \,\zeta_1 \,-\, 
             \Big( \frac{9}{16}\,u_3^2 \,+\, \frac{3}{2}\,u_3x_4 
                 \Big) \,\zeta_2 \,+\, \Big( 
             \frac{3}{4}\,\beta_0\,u_3^2 \,+\, \frac{9}{16}\,u_3^2 
                \,-\, \frac{3}{4}\,u_3x_4 \Big) \,\zeta_3 
                 \,+\, \frac{3}{2}\,u_3x_1 \zeta_4 \\*[1mm]
      & & \hspace*{-16mm} -\; \Big( \frac{9}{16} \,u_3^2 \,-\,
                      \frac{15}{8} \,u_3x_4 \Big) \,\zeta_6 
           \,-\, \frac{3}{4}\,\beta_0\,u_3^2 \,\zeta_{17} 
\end{eqnarray*}   
\begin{eqnarray*}  \qquad\qquad & = & -\; 3\,u_3^2x_3^2 \,+\, 3\,u_3x_4^2
                 \,+\, N_F\,\Big( \frac{u_3}{w_3} \Big)^2 \,
                  \Big( -\,\frac{1}{3}\,w_3z_1z_3z_4
                     \,+\, \frac{1}{3}\,w_3z_4^2 \Big. \\*[1mm]
           & &  \Big. \!-\; \frac{1}{3}\,z_1z_4z_5
                       \,+\, \frac{1}{3}\,z_2z_4^2 \Big)
\end{eqnarray*}
\end{small} }}
\begin{equation} \label{skgv11}
\end{equation}
\fbox{\parbox{14.8cm}{
\begin{small}
\begin{eqnarray*} & & \hspace*{-13mm} -\;\Big( \frac{9}{16} \,u_3^2 \,+\, 
           \frac{3}{2}\,u_3x_4 \Big) \,\zeta_1 \,+\, 
              \frac{3}{2}\,u_3x_1 \,\zeta_2 \,-\, \Big( 
               \frac{39}{8}\,u_3x_1 \,-\,\frac{15}{16}\,u_3x_3 
              \,-\, \frac{1}{2} \,N_F\,u_3z_3 \Big) \,\zeta_3 \\*[1mm]
      & & \hspace*{-13mm} -\; \frac{15}{8}\,u_3x_1\,\zeta_6 \,+\,
               \Big( \frac{45}{8}\,u_3x_1 \,-\, 
                      \frac{15}{16}\,u_3x_3 \,-\, \frac{1}{2}\,N_F 
                      \,u_3z_3 \Big) \,\zeta_{17}  
\end{eqnarray*}   
\begin{eqnarray*}  \qquad\qquad\qquad\qquad & = &  3\,u_3x_1x_4
                 \,+\, N_F\,\Big( \frac{u_3}{w_3} \Big)^2 \,
                  \Big( \frac{1}{3}\,w_3z_1z_4
                     \,-\, \frac{1}{3}\,w_3z_1^2z_3 \Big. \\*[1mm]
           & &  \Big. \!+\; \frac{1}{3}\,z_1z_2z_4
                       \,-\, \frac{1}{3}\,z_1^2z_5 \Big)
\end{eqnarray*}
\end{small} }}
\begin{equation} \label{skgv12} 
\end{equation}\\
\subsection{$C_B$--$L_{0}$ equations:}
\vspace{6mm}
\fbox{\parbox{14.8cm}{
\begin{small}
\begin{eqnarray*} & & \hspace*{-5mm} \frac{9}{4}\,u_3^2x_3 \,
                      \Big( \,5\,\zeta_2 \,-\, 
               3\,\zeta_3 \,+\, 8\,\zeta_6 \,-\, 2\,\zeta_8 \,+\, 
                      4\,\zeta_9 \,-\, 2\,\zeta_{12} \,+\, 
                      \zeta_{14} \,-\, 2\,\zeta_{17} \,\Big) 
                       \,-\, \frac{45}{4}\,u_3x_4\,\zeta_4 \\*[1mm]
      & & \hspace*{-5mm} -\; \Big( \frac{3}{4}\,\beta_0\,u_3^2 \,-\, 
                      \frac{9}{2}\,u_3x_4 \Big) \,\zeta_{10} \,-\, 
              \Big( \frac{45}{8}\,u_3x_1 \,-\, 
             \frac{15}{16}\,u_3x_3 \,-\, \frac{1}{2}\,N_F\,u_3z_3 \Big)
             \,\Big( \,\zeta_{11} \,+\, 2\,\zeta_{16} \,\Big) \\*[1mm]
      & & \hspace*{-5mm} -\; \Big( \frac{3}{2}\,\beta_0\,u_3^2 \,+\, 
                      \frac{9}{4}\,u_3x_4 \Big) \,\zeta_{15} 
\end{eqnarray*}
\[  \qquad\qquad = \quad -\; 27\,u_3^2x_3^3 \,+\, 
                       54\,u_3x_3x_4^2 \,-\, 27\,x_4^2x_5 \]       \\*
\end{small} }}
\begin{equation} \label{skgv13} 
\end{equation}\\
\fbox{\parbox{14.8cm}{
\begin{small}
\begin{eqnarray*} & & \hspace*{-13mm} \frac{9}{2}\,u_3^2x_3\,\zeta_1 \,+\, 
             \Big( \frac{27}{16}\,u_3^2 \,-\, \frac{9}{2}\,u_3x_4
                 \Big)\,\zeta_2 \,-\, \Big( \frac{3}{2}\,
                    \beta_0\,u_3^2 \,+\, \frac{27}{16}\,u_3^2
               \Big) \,\zeta_3 \,-\, \frac{27}{4}\,u_3x_1\,\zeta_4 \\*[1mm]
      & & \hspace*{-13mm} +\; \Big( 
                    \frac{27}{8}\,u_3^2 \,+\, \frac{9}{4}\,u_3x_4 
                \Big) \,\zeta_6 \,-\,\frac{9}{4}\,
                    u_3^2x_3 \,\Big( \,2\,\zeta_7 \,-\,\zeta_{13} \,\Big)
                  \,-\, \Big( \frac{3}{4}\,\beta_0\,u_3^2
                  \,-\, \frac{9}{2}\,u_3x_4 \Big) \,
                  \Big( \,\zeta_8 \,+\, \zeta_9 \,\Big) \\*[1mm]
      & & \hspace*{-13mm} \!-\; \Big( \frac{45}{8}\,u_3x_1 \,-\, 
               \frac{15}{16}\,u_3x_3 \,-\, 
                \frac{1}{2}\,N_F\,u_3z_3 \Big)
             \,\Big( \,\zeta_{10} \,+\, 2\,\zeta_{15} \,\Big) \\*[1mm]
      & & \hspace*{-13mm} \!-\;
                \Big( \frac{3}{2}\,\beta_0\,u_3^2
                  \,+\, \frac{9}{4}\,u_3x_4 \Big) \,
                  \Big( \,\zeta_{12} \,+\, \zeta_{14} \,\Big) \,+\,
                \Big( \frac{3}{2}\,\beta_0\,u_3^2
                  \,-\, \frac{9}{4}\,u_3x_4 \Big) \,\zeta_{17}        
\end{eqnarray*}   
\begin{eqnarray*}  \qquad\qquad & = & -\; \frac{27}{2}\,u_3^2x_3^2 \,+\, 
                       36\,u_3x_1x_3x_4 \,+\, \frac{9}{2}\,u_3x_4^2
                       \,-\, 9\,x_2x_4^2  \,-\, 18\,x_1x_4x_5 \\*[1mm]
             & & +\; N_F\,\Big( \frac{u_3}{w_3} \Big)^2 \,
                  \Big( \frac{2}{3}\,w_3z_1z_3z_4
                     \,-\, \frac{2}{3}\,w_3z_4^2
                    \,+\, \frac{2}{3}\,z_1z_4z_5
                       \,-\, \frac{2}{3}\,z_2z_4^2 \Big)
\end{eqnarray*}
\end{small} }}
\begin{equation} \label{skgv14}
\end{equation}\\
\fbox{\parbox{14.8cm}{
\begin{small}
\begin{eqnarray*} & & \hspace*{-14mm} \frac{9}{8}\,u_3^2x_3\,
                      \Big( \,8\,\zeta_1 
                 \,-\, 2\,\zeta_7 \,+\, \zeta_{13} \,\Big) \,-\, 
               9\,u_3x_4\,\zeta_2 \,+\, \Big( \,\frac{3}{4}\,
                    \beta_0\,u_3^2 \,-\, \frac{27}{8}\,u_3x_4 
                \,\Big) \,\zeta_3 \,-\, \frac{9}{4}\,u_3x_1
                                            \,\zeta_4 \\*[1mm]
      & & \hspace*{-14mm} +\;\Big( 
                    \frac{27}{16}\,u_3^2 \,+\, \frac{63}{8}\,u_3x_4 
                \Big) \,\zeta_6 \,-\, \Big( \frac{3}{4}\,
                    \beta_0\,u_3^2 \,-\, \frac{9}{4}\,u_3x_4 \Big) \,
                  \zeta_8 \,+\, \Big( \frac{3}{8}\,
                    \beta_0\,u_3^2 \,+\,\frac{9}{4}\,u_3x_4 \Big) \,
                                    \zeta_9  \\*[1mm]
      & & \hspace*{-14mm} -\;\Big( \frac{27}{8}\,u_3x_1 \,-\, 
              \frac{15}{16}\,u_3x_3 \,-\, \frac{1}{2}\,N_F\,u_3z_3 \Big)
             \,\zeta_{10} \,+\, \Big( \frac{3}{4}\,\beta_0\,u_3^2 \,-\,
                  \frac{27}{16} \,u_3^2 \,-\,\frac{9}{8}\,u_3x_4 \Big) 
                                    \,\zeta_{12} \\*[1mm]
      & & \hspace*{-14mm} -\; \Big( \frac{3}{2}\,\beta_0\,u_3^2 \,-\,
                  \frac{27}{32} \,u_3^2 \,+\,\frac{9}{8}\,u_3x_4 \Big) 
                    \,\zeta_{14} \,-\, \Big( \frac{99}{8}\,u_3x_1 \,-\, 
              \frac{15}{8}\,u_3x_3 \,-\,N_F\,u_3z_3 \Big)
                                    \,\zeta_{15} \\*[1mm]
      & & \hspace*{-14mm} -\; \Big( \frac{3}{4}\,\beta_0\,u_3^2 
                      \,+\,\frac{9}{8}\,u_3x_4 \Big) \,\zeta_{17}         
\end{eqnarray*}   
\begin{eqnarray*}  \quad\qquad & = & -\; \frac{45}{2}\,u_3^2x_3^2 \,+\, 
                       36\,u_3x_1x_3x_4 \,+\, \frac{27}{2}\,u_3x_4^2
                       \,-\, 9\,x_2x_4^2  \,-\, 18\,x_1x_4x_5 \\*[1mm]
             & & +\; N_F\,\Big( \frac{u_3}{w_3} \Big)^2 \,
                  \Big( -\,\frac{1}{3}\,w_3z_1z_3z_4
                     \,+\, \frac{1}{3}\,w_3z_4^2
                    \,-\, \frac{1}{3}\,z_1z_4z_5
                       \,+\, \frac{1}{3}\,z_2z_4^2 \Big)
\end{eqnarray*}
\end{small} }}
\begin{equation} \label{skgv15}
\end{equation}\\
\fbox{\parbox{14.8cm}{
\begin{small}
\begin{eqnarray*} & & \hspace*{-18mm} -\;\frac{27}{4}\,u_3x_4\,\zeta_1 \,-\, 
             \frac{9}{4} \,u_3x_1\,\Big( \,2\,\zeta_2 \,-\, \zeta_6
                  \,\Big) \,-\, \Big( \frac{45}{4}\,
                    u_3x_1 \,-\, \frac{15}{8}\,u_3x_3
                  \,-\, N_F\,u_3z_3 \,\Big) \,\zeta_3 \\*[1mm]
      & & \hspace*{-18mm} -\;\frac{3}{4}\,\beta_0\,u_3^2\,\zeta_7 \,-\, 
                \Big(  \frac{9}{8}\,u_3x_1 \,-\, \frac{15}{16}\,u_3x_3 
                \,-\, \frac{1}{2}\,N_F\,u_3z_3 \Big) 
               \,\Big( \,\zeta_8 \,+\, \zeta_9 \,\Big) \\*[1mm]
      & & \hspace*{-18mm} -\; \Big( \frac{27}{2}\,u_3x_1 \,-\, 
                 \frac{15}{8}\,u_3x_3 \,-\, N_F\,u_3z_3 \Big) \,
                  \Big( \,\zeta_{12} \,+\, \zeta_{14} \,\Big)
                    \,-\, \Big( \frac{3}{2} \,\beta_0\,u_3^2
                    \,-\,\frac{27}{16}\,u_3^2 \,\Big) \,\zeta_{13} \\*[1mm] 
      & & \hspace*{-18mm} +\; \Big( 9\,u_3x_1 \,-\, 
                 \frac{15}{8}\,u_3x_3 \,-\, N_F\,u_3z_3 \Big) 
                               \,\zeta_{17}
\end{eqnarray*}   
\begin{eqnarray*}  \qquad\qquad & = & -\; \frac{9}{2}\,u_3^2x_3 \,+\, 
                       18\,u_3x_1^2x_3 \,+\, 27\,u_3x_1x_4
                       \,-\, 9\,x_1^2x_5  \,-\, 18\,x_1x_2x_4 \\*[1mm]
             & & +\; N_F\,\Big( \frac{u_3}{w_3} \Big)^2 \,
                  \Big( \frac{2}{3}\,w_3z_1z_4
                     \,-\, \frac{2}{3}\,w_3z_1^2z_3
                    \,+\, \frac{2}{3}\,z_1z_2z_4
                       \,-\, \frac{2}{3}\,z_1^2z_5 \Big)
\end{eqnarray*}
\end{small} }}
\begin{equation} \label{skgv16}
\end{equation}\\
\fbox{\parbox{14.8cm}{
\begin{small}
\begin{eqnarray*} & & \hspace*{-8mm} \Big( \frac{27}{16}\,u_3^2 \,-\,
               \frac{9}{4}\,u_3x_4 \Big) \,\zeta_1 \,-\, 
               \frac{9}{8}\,u_3x_1 \,\Big( \,8\,\zeta_2 \,-\, 
                   7\,\zeta_6 \,\Big) \,+\, \Big( \,\frac{9}{4}\,
                    u_3x_1 \,-\, \frac{15}{16}\,u_3x_3 \,-\, 
               \frac{1}{2}\,N_F \,u_3z_3 \,\Big) \,\zeta_3 \\*[1mm]
      & & \hspace*{-8mm} -\; \Big( 
               \frac{3}{4}\,\beta_0\,u_3^2 \,-\, \frac{9}{4}\,u_3x_4 
                \Big) \,\zeta_7 \,-\, \Big( \frac{27}{8}\,
                    u_3x_1 \,-\, \frac{15}{16}\,u_3x_3 \,-\, 
               \frac{1}{2}\,N_F \,u_3z_3 \Big) \,\zeta_8  \\*[1mm]
      & & \hspace*{-8mm} +\; \Big( \frac{81}{16}\,
                    u_3x_1 \,-\, \frac{15}{32}\,u_3x_3 \,-\, 
               \frac{1}{4}\,N_F \,u_3z_3 
                \Big) \,\zeta_9 \,+\, \Big( \frac{9}{2}\,
                    u_3x_1 \,-\, \frac{15}{16}\,u_3x_3 \,-\, 
               \frac{1}{2}\,N_F \,u_3z_3 \,\Big) \,\zeta_{12} \\*[1mm]
      & & \hspace*{-8mm} -\; \Big( 
               \frac{3}{2}\,\beta_0\,u_3^2 \,-\, \frac{27}{32}\,u_3^2 
                \,+\, \frac{9}{8}\,u_3x_4 \Big) \,\zeta_{13} \,-\, 
              \Big( \frac{99}{8}\,u_3x_1 \,-\, \frac{15}{8}\,u_3x_3 
                 \,-\,N_F\,u_3z_3 \Big) \,\zeta_{14} \\*[1mm]
      & & \hspace*{-8mm} -\; \Big( \frac{27}{4}\,
                   u_3x_1 \,-\, \frac{15}{16}\,u_3x_3 \,-\, 
               \frac{1}{2}\,N_F \,u_3z_3 \Big) \,\zeta_{17}
\end{eqnarray*}   
\begin{eqnarray*}  \qquad\qquad & = & -\; \frac{9}{2}\,u_3^2x_3 \,+\, 
                       18\,u_3x_1^2x_3 \,+\, 18\,u_3x_1x_4
                       \,-\, 9\,x_1^2x_5  \,-\, 18\,x_1x_2x_4 \\*[1mm]
             & & +\; N_F\,\Big( \frac{u_3}{w_3} \Big)^2 \,
                  \Big( -\, \frac{1}{3}\,w_3z_1z_4
                     \,+\, \frac{1}{3}\,w_3z_1^2z_3
                    \,-\, \frac{1}{3}\,z_1z_2z_4
                       \,+\, \frac{1}{3}\,z_1^2z_5 \Big)
\end{eqnarray*}
\end{small} }}
\begin{equation} \label{skgv17}
\end{equation}\\
\subsection{$C_B$--$L_{+}$ equations:}
\vspace{6mm}
\fbox{\parbox{14.8cm}{
\begin{small}
\begin{eqnarray*} & & \hspace*{-6mm} \frac{1}{4}\,u_3^2x_3 \,
                      \Big( \,15\,\zeta_2 \,-\, 
               \zeta_3 \,+\, 24\,\zeta_6 \,-\, 14\,\zeta_8 \,+\, 
            28\,\zeta_9 \,-\, 26\,\zeta_{12} \,+\, 13\,\zeta_{14} \,-\, 
                    14\,\zeta_{17} \,\Big) \,-\, 
                        \frac{15}{4} \,u_3x_4\,\zeta_4  \\*[1mm]
      & & \hspace*{-6mm} +\; \Big( \frac{1}{4}\,\beta_0\,u_3^2 \,+\, 
                      \frac{7}{2}\,u_3x_4 \Big)
             \,\zeta_{10} \,+\, \Big( \frac{15}{8}\,u_3x_1 \,-\, 
             \frac{5}{16}\,u_3x_3 \,-\, \frac{1}{6} \,N_F\,u_3z_3 \Big)
             \,\Big( \,\zeta_{11} \,-\, 2\,\zeta_{16} \,\Big) \\*[1mm]
      & & \hspace*{-6mm} -\; \Big( \frac{1}{2}\,\beta_0\,u_3^2 \,+\, 
                      \frac{13}{4}\,u_3x_4 \Big) \,\zeta_{15} 
\end{eqnarray*}   
\[  \qquad\qquad = \quad -\; 9\,u_3^2x_3^3 \,+\, 
                       18\,u_3x_3x_4^2 \,-\, 9\,x_4^2x_5 \]          \\*
\end{small} }}
\begin{equation} \label{skgv18} 
\end{equation}\\
\fbox{\parbox{14.8cm}{
\begin{small}
\begin{eqnarray*} & & \hspace*{-14mm} \frac{1}{4}\,u_3^2x_3\,
                      \Big( \,14\,\zeta_1 
               \,-\,14\,\zeta_7 \,+\,13\,\zeta_{13}\,\Big) \,-\, 
             \Big( \frac{3}{16}\,u_3^2 \,+\, \frac{7}{2}\,u_3x_4
                 \Big)\,\zeta_2 \,+\, \Big( \frac{1}{2}\,
                    \beta_0\,u_3^2 \,+\, \frac{3}{16}\,u_3x_4
                            \Big) \,\zeta_3 \\*[1mm]
      & & \hspace*{-14mm} -\;\frac{1}{4}\,u_3x_1\,\zeta_4 \,+\, \Big( 
                    \frac{3}{8}\,u_3^2 \,+\, \frac{13}{4}\,u_3x_4 
                \Big) \,\zeta_6 \,+\, \Big( \frac{1}{4}\,
                    \beta_0\,u_3^2 \,+\, \frac{7}{2}\,u_3x_4 \Big) 
                        \Big( \,\zeta_8 \,+\,\zeta_9\,\Big) \\*[1mm]
      & & \hspace*{-14mm} +\; \Big( \frac{15}{8}\,u_3x_1 \,-\, 
                 \frac{5}{16}\,u_3x_3  \,-\, \frac{1}{6} \,N_F\,u_3z_3 
                       \Big) \,\Big( \,\zeta_{10} \,-\,2\,\zeta_{15}
                          \,\Big) \\*[1mm]
      & & \hspace*{-14mm} -\; \Big( \frac{1}{2}\,\beta_0\,u_3^2 
                 \,+\, \frac{13}{4}\,u_3x_4 \Big) 
                  \,\Big( \,\zeta_{12} \,+\,\zeta_{14} \,\Big) 
                     \,-\, \Big( \frac{1}{2}\,\beta_0\,u_3^2 \,+\, 
                  \frac{7}{4}\,u_3x_4 \Big) \, \zeta_{17}
\end{eqnarray*}   
\begin{eqnarray*}  \qquad\qquad & = & -\; \frac{17}{2}\,u_3^2x_3^2 \,+\, 
                       12\,u_3x_1x_3x_4 \,+\, \frac{11}{2}\,u_3x_4^2
                       \,-\, 3\,x_2x_4^2  \,-\, 6\,x_1x_4x_5 \\*[1mm]
             & & +\; N_F\,\Big( \frac{u_3}{w_3} \Big)^2 \,
                  \Big( -\,\frac{2}{9}\,w_3z_1z_3z_4
                     \,+\, \frac{2}{9}\,w_3z_4^2
                    \,-\, \frac{2}{9}\,z_1z_4z_5
                       \,+\, \frac{2}{9}\,z_2z_4^2 \Big)
\end{eqnarray*}
\end{small} }}
\begin{equation} \label{skgv19}
\end{equation}\\
\fbox{\parbox{14.8cm}{
\begin{small}
\begin{eqnarray*} & & \hspace*{-18mm} \frac{1}{8}\,u_3^2x_3\,
                      \Big( \,16\,\zeta_1 
               \,-\,14\,\zeta_7 \,+\,13\,\zeta_{13}\,\Big) \,+\, 
             \Big( \frac{3}{8}\,u_3^2 \,-\, 2\,u_3x_4
                 \Big)\,\zeta_2 \,-\, \Big( \frac{1}{4}\,
                    \beta_0\,u_3^2 \,+\, \frac{1}{8}\,u_3x_4
                            \Big) \,\zeta_3 \\*[1mm]
      & & \hspace*{-18mm} -\;\frac{7}{4}\,u_3x_1\,\zeta_4 \,+\, \Big( 
                    \frac{15}{16}\,u_3^2 \,+\, \frac{11}{8}\,u_3x_4 
                \Big) \,\zeta_6 \,+\, \Big( \frac{1}{4}\,
                    \beta_0\,u_3^2 \,-\, \frac{3}{8}\,u_3^2 
                     \,+\, \frac{7}{4}\,u_3x_4 \Big) \,\zeta_8  \\*[1mm]
      & & \hspace*{-18mm} -\; \Big( \frac{1}{8}\,
                    \beta_0\,u_3^2 \,-\, \frac{3}{4}\,u_3^2 
                 \,-\, \frac{7}{4}\,u_3x_4 \Big) \,\zeta_9 \,+\,    
            \Big( \frac{29}{8}\,u_3x_1 \,-\, \frac{5}{16}\,u_3x_3 
                    \,-\, \frac{1}{6} \,N_F\,u_3z_3 \Big)
                         \,\zeta_{10} \\*[1mm]
      & & \hspace*{-18mm} +\; \Big( \frac{1}{4}\,
                    \beta_0\,u_3^2 \,-\, \frac{15}{16}\,u_3^2 
                 \,-\, \frac{13}{8}\,u_3x_4 \Big) \,\zeta_{12} \,-\, 
                 \Big( \frac{1}{2}\,\beta_0\,u_3^2 \,-\, 
                \frac{15}{32}\,u_3^2 \,+\,\frac{13}{8}\,u_3x_4 \Big) \,
                                     \zeta_{14} \\*[1mm]
      & & \hspace*{-18mm} -\; \Big( \frac{43}{8}\,
                    u_3x_1 \,-\, \frac{5}{8}\,u_3x_3 \,-\, 
                \frac{1}{3}\,N_F\,u_3z_3 \Big) \,\zeta_{15} \,+\, 
                 \Big( \frac{1}{4}\,\beta_0\,u_3^2 \,-\, 
                \frac{3}{8}\,u_3^2 \,-\,\frac{7}{8}\,u_3x_4 \Big) \,
                                     \zeta_{17}
\end{eqnarray*}   
\begin{eqnarray*}  \qquad\qquad & = & -\; \frac{11}{2}\,u_3^2x_3^2 \,+\, 
                       12\,u_3x_1x_3x_4 \,+\, \frac{5}{2}\,u_3x_4^2
                       \,-\, 3\,x_2x_4^2  \,-\, 6\,x_1x_4x_5 \\*[1mm]
             & & +\; N_F\,\Big( \frac{u_3}{w_3} \Big)^2 \,
                  \Big( \frac{1}{9}\,w_3z_1z_3z_4
                     \,-\, \frac{1}{9}\,w_3z_4^2
                    \,+\, \frac{1}{9}\,z_1z_4z_5
                       \,-\, \frac{1}{9}\,z_2z_4^2 \Big)
\end{eqnarray*}
\end{small} }}
\begin{equation} \label{skgv20}
\end{equation}\\
\fbox{\parbox{14.8cm}{
\begin{small}
\begin{eqnarray*} & & \hspace*{-9mm} \Big( \frac{3}{4}\,u_3^2 \,-\, 
                \frac{1}{4}\,u_3x_4 \Big) \,\zeta_1 \,-\, 
             \frac{1}{4}\,u_3x_1\,\Big( \,14\,\zeta_2 \,-\,
                      13\,\zeta_6 \,\Big) \,+\, \Big( \frac{15}{4}\,
                    u_3x_1 \,-\, \frac{5}{8}\,u_3x_3 \,-\, 
               \frac{1}{3}\,N_F\,u_3z_3 \,\Big) \,\zeta_3 \\*[1mm]
      & & \hspace*{-9mm} +\; \Big( \frac{1}{4}\,\beta_0\,u_3^2 
                 \,-\,\frac{3}{4}\,u_3^2 \Big) \,\zeta_7 
                   \,+\, \Big( \frac{43}{8}\,u_3x_1 \,-\, 
                 \frac{5}{16}\,u_3x_3 \,-\, 
                      \frac{1}{6}\,N_F\,u_3z_3 \Big) \,\Big( 
             \,\zeta_8 \,+\,\zeta_9 \,\Big) \\*[1mm]
      & & \hspace*{-9mm} -\; \Big( 7\,u_3x_1 
                      \,-\, \frac{5}{8}\,u_3x_3 \,-\, 
                    \frac{1}{3}\,N_F\,u_3z_3 \Big) 
                    \,\Big( \,\zeta_{12} \,+\,\zeta_{14} \,\Big) 
              \,-\, \Big( \frac{1}{2}\,\beta_0\,u_3^2 \,-\, 
                \frac{15}{16}\,u_3^2 \Big) \,\zeta_{13} \\*[1mm]
      & & \hspace*{-9mm} -\;
         \Big( \frac{11}{2}\,u_3x_1 \,-\, \frac{5}{8}\,u_3x_3 \,-\, 
                      \frac{1}{3}\,N_F\,u_3z_3 \Big) \,\zeta_{17}       
\end{eqnarray*}   
\begin{eqnarray*}  \quad\qquad & = & -\; \frac{3}{2}\,u_3^2x_3 \,+\, 
                       6\,u_3x_1^2x_3 \,+\, 5\,u_3x_1x_4
                       \,-\, 3\,x_1^2x_5  \,-\, 6\,x_1x_2x_4 \\*[1mm]
             & & +\; N_F\,\Big( \frac{u_3}{w_3} \Big)^2 \,
                  \Big( -\,\frac{2}{9}\,w_3z_1z_4
                     \,+\, \frac{2}{9}\,w_3z_1^2z_3
                    \,-\, \frac{2}{9}\,z_1z_2z_4
                       \,+\, \frac{2}{9}\,z_1^2z_5 \Big)
\end{eqnarray*}
\end{small} }}
\begin{equation} \label{skgv21}
\end{equation}\\
\fbox{\parbox{14.8cm}{
\begin{small}
\begin{eqnarray*} & & \hspace*{-12mm} \Big( \frac{3}{16}\,u_3^2 \,-\, 
                \frac{7}{4}\,u_3x_4 \Big) \,\zeta_1 \,-\, 
             \frac{1}{8}\,u_3x_1\,\Big( \,16\,\zeta_2 \,-\,
                      11\,\zeta_6 \,\Big) \\*[1mm]
      & & \hspace*{-12mm} -\; \Big( 2\,
                    u_3x_1 \,-\, \frac{5}{16}\,u_3x_3 \,-\, 
               \frac{1}{6}\,N_F\,u_3z_3 \,\Big) \,\zeta_3 
                    \,+\, \Big( \frac{1}{4}\,\beta_0\,u_3^2 
                 \,-\,\frac{3}{8}\,u_3^2 \,+\, \frac{7}{4}\,u_3x_4 
                               \Big) \,\zeta_7 \\*[1mm]
      & & \hspace*{-12mm} +\; \Big( \frac{29}{8}\,u_3x_1 \,-\, 
                 \frac{5}{16}\,u_3x_3 \,-\, 
                      \frac{1}{6}\,N_F\,u_3z_3 \Big) \,\zeta_8
             \,+\, \Big( \frac{13}{16}\,u_3x_1 \,+\, 
                 \frac{5}{32}\,u_3x_3 \,+\, 
                   \frac{1}{12}\,N_F\,u_3z_3 \Big) \,\zeta_9 \\*[1mm]
      & & \hspace*{-12mm} +\; \Big( \frac{1}{4}\,u_3x_1 \,-\, 
                 \frac{5}{16}\,u_3x_3 \,-\, 
                      \frac{1}{6}\,N_F\,u_3z_3 \Big) \,\zeta_{12}
             \,-\, \Big( \frac{1}{2}\,\beta_0\,u_3^2 \,-\, 
                 \frac{15}{32}\,u_3^2 \,+\, 
                \frac{13}{8}\,u_3x_4 \Big) \,\zeta_{13} \\*[1mm]
      & & \hspace*{-12mm} -\; \Big( \frac{43}{8}\,u_3x_1 \,-\, 
                 \frac{5}{8}\,u_3x_3 \,-\, 
                   \frac{1}{3}\,N_F\,u_3z_3 \Big) \,\zeta_{14}
             \,+\, \Big( u_3x_1 \,-\, 
                 \frac{5}{16}\,u_3x_3 \,-\, 
                      \frac{1}{6}\,N_F\,u_3z_3 \Big) \,\zeta_{17}       
\end{eqnarray*}
\begin{eqnarray*}  \qquad\qquad & = & -\; \frac{3}{2}\,u_3^2x_3 \,+\, 
                       6\,u_3x_1^2x_3 \,+\, 8\,u_3x_1x_4
                       \,-\, 3\,x_1^2x_5  \,-\, 6\,x_1x_2x_4 \\*[1mm]
             & & +\; N_F\,\Big( \frac{u_3}{w_3} \Big)^2 \,
                  \Big( \frac{1}{9}\,w_3z_1z_4
                     \,-\, \frac{1}{9}\,w_3z_1^2z_3
                    \,+\, \frac{1}{9}\,z_1z_2z_4
                       \,-\, \frac{1}{9}\,z_1^2z_5 \Big)
\end{eqnarray*}
\end{small} }}
\begin{equation} \label{skgv22}
\end{equation}\\
\subsection{$C_B$--$L_{-}$ equations:}
\vspace{6mm}
\fbox{\parbox{14.8cm}{
\begin{small}
\begin{eqnarray*} & & \hspace*{-9mm} \frac{3}{8}\,u_3^2x_3 \,\Big( 
              \,4\,\zeta_1 \,-\, 4\,\zeta_7 \,+\, 5\,\zeta_{13}\,\Big)
            \,-\, \Big( \frac{9}{16}\,u_3^2 \,+\, \frac{3}{2}\,u_3x_4 
                 \Big) \,\Big( \,\zeta_2 \,-\, \zeta_8 \,\Big) 
              \,+\, \frac{3}{4}\,\beta_0\,u_3^2 \,\zeta_3 \\*[1mm]
      & & \hspace*{-9mm} +\; \frac{3}{8}\,u_3x_1 \,\Big( 
                    \,4\,\zeta_4 \,-\, 4\,\zeta_{10} \,+\,
                    \,5\,\zeta_{15} \,\Big) \,-\,
                \Big( \frac{9}{16} \,u_3^2 \,-\, \frac{15}{8} 
           \,u_3x_4 \Big) \,\Big( \,\zeta_6 \,-\,\zeta_{12}\,\Big) 
                  \,+\, \Big( \frac{9}{8}\,\beta_0\,u_3^2 
                  \,-\,\frac{9}{8} \,u_3^2 \Big. \\*[1mm]
      & & \hspace*{-9mm} \Big.\! +\;  \frac{3}{2} 
              \,u_3x_4 \Big) \,\zeta_9 \,-\,\Big( 
               \frac{9}{32}\,u_3^2 \,+\, \frac{15}{8}\,u_3x_4 
              \Big) \,\zeta_{14} \,-\, \Big( \frac{3}{4}\,\beta_0\,u_3^2 
                  \,-\,\frac{9}{16} \,u_3^2 \,+\, \frac{3}{4} 
              \,u_3x_4 \Big) \,\zeta_{17} 
\end{eqnarray*}   
\begin{eqnarray*}  \qquad\qquad & = & -\; 3\,u_3^2x_3^2 \,+\, 3\,u_3x_4^2
                 \,+\, N_F\,\Big( \frac{u_3}{w_3} \Big)^2 \,
                  \Big( -\,\frac{1}{3}\,w_3z_1z_3z_4
                     \,+\, \frac{1}{3}\,w_3z_4^2 \Big. \\*[1mm]
           & &  \Big. \!-\; \frac{1}{3}\,z_1z_4z_5
                       \,+\, \frac{1}{3}\,z_2z_4^2 \Big)
\end{eqnarray*}
\end{small} }}
\begin{equation} \label{skgv23}
\end{equation}\\
\fbox{\parbox{14.8cm}{
\begin{small}
\begin{eqnarray*} & & \hspace*{-11mm} -\; \Big( \frac{9}{16} \,u_3^2 \,+\, 
           \frac{3}{2}\,u_3x_4 \Big) \,\Big( \,\zeta_1 \,-\,
                \zeta_7\,\Big) \,+\, \frac{3}{8}\,u_3x_1\,\Big( \,
              4\,\zeta_2 \,-\, 5\,\zeta_6 \,-\, 4\,\zeta_8 
               \,+\, 5\,\zeta_{12} \,+\, 5\,\zeta_{14} \,\Big) \\*[1mm]
      & & \hspace*{-11mm} -\;\Big( 
               \frac{45}{8}\,u_3x_1 \,-\,\frac{15}{16}\,u_3x_3 
              \,-\, \frac{1}{2} \,N_F\,u_3z_3 \Big) \,\zeta_3 
          \,-\, \Big( \frac{159}{16}\,u_3x_1 \,-\,\frac{45}{32}\,u_3x_3 
              \,-\, \frac{3}{4} \,N_F\,u_3z_3 \Big) \,\zeta_9 \\*[1mm]
      & & \hspace*{-11mm} -\; \Big( \frac{9}{32} \,u_3^2 \,+\, 
           \frac{15}{8}\,u_3x_4 \Big) \,\zeta_{13} \,+\,
               \Big( \frac{51}{8}\,u_3x_1 \,-\, 
                      \frac{15}{16}\,u_3x_3 \,-\, \frac{1}{2}\,N_F 
                      \,u_3z_3 \Big) \,\zeta_{17}  
\end{eqnarray*}   
\begin{eqnarray*}  \qquad\qquad\qquad\qquad & = &  3\,u_3x_1x_4
                 \,+\, N_F\,\Big( \frac{u_3}{w_3} \Big)^2 \,
                  \Big( \frac{1}{3}\,w_3z_1z_4
                     \,-\, \frac{1}{3}\,w_3z_1^2z_3 \Big. \\*[1mm]
           & &  \Big. \!+\; \frac{1}{3}\,z_1z_2z_4
                       \,-\, \frac{1}{3}\,z_1^2z_5 \Big)
\end{eqnarray*}
\end{small} }}
\begin{equation} \label{skgv24} 
\end{equation}\\
\subsection{$C_C$--$L_{0}$ equations:}
\vspace{6mm}
\fbox{\parbox{14.8cm}{
\begin{small}
\begin{eqnarray*} & & \hspace*{-5mm} \frac{9}{8}\,u_3x_4 \,
                      \Big( \,\zeta_3 \,+\, 
               2\,\zeta_8 \,-\,\zeta_{17} \,\Big) \,-\, \frac{9}{8} 
                  \,u_3^2x_3\,\Big( \,2\,\zeta_7 \,-\, \zeta_{13}
                \, \Big) \,-\, \Big( \frac{9}{8}\,\beta_0\,u_3^2 \,-\, 
                      \frac{9}{4}\,u_3x_4 \Big) \,\zeta_9 \\*[1mm]
      & & \hspace*{-5mm} -\; \frac{9}{8}\,u_3x_1 \,\Big( \,
                        2\,\zeta_{10} \,-\,\zeta_{15}\,\Big)  
                  \,-\, \Big( \frac{9}{4}\,\beta_0\,u_3^2 \,-\, 
                     \frac{27}{16}\,u_3^2 \,+\, 
                  \frac{9}{8}\,u_3x_4 \Big) \,\zeta_{12} \,-\,
              \Big( \frac{27}{32}\,\beta_0\,u_3^2 \,+\, 
                      \frac{9}{8}\,u_3x_4 \Big) \,\zeta_{14} 
\end{eqnarray*}\vspace*{1mm}   
\[  \qquad\qquad\qquad\qquad\qquad\qquad\qquad\qquad\qquad
                         \qquad\qquad\qquad  = \quad 0 \]          \\*
\end{small} }}
\begin{equation} \label{skgv25}
\end{equation}\\
\fbox{\parbox{14.8cm}{
\begin{small}
\begin{eqnarray*} & & \hspace*{-8mm} -\; \frac{9}{8} \,u_3x_1 \, \Big( 
           \,\zeta_3 \,+\, 2\,\zeta_7 \,-\, \zeta_{14} \,-\, 
            \zeta_{17} \,\Big) \,+\, \frac{9}{4}\,u_3x_4
              \,\zeta_7 \,+\, \Big( \frac{99}{16}\,u_3x_1 \,-\,
             \frac{45}{32}\,u_3x_3 \,-\,
             \frac{3}{4} \,N_F\,u_3z_3 \Big) \,\zeta_9 \\*[1mm]
      & & \hspace*{-8mm} +\; \Big( 18\,u_3x_1 \,-\, 
                      \frac{45}{16}\,u_3x_3 \,-\, \frac{3}{2}\,N_F 
                      \,u_3z_3 \Big) \,\zeta_{12} \,-\,
                 \Big( \frac{27}{32} \,u_3^2 \,+\, 
          \frac{9}{8}\,u_3x_4 \Big) \,\zeta_{13} \quad\;\; = \;\;\quad 0
\end{eqnarray*}
\end{small} }}
\begin{equation} \label{skgv26} 
\end{equation}\\
\subsection{$C_C$--$L_{+}$ equations:}
\vspace{6mm}
\fbox{\parbox{14.8cm}{
\begin{small}
\begin{eqnarray*} & & \hspace*{-6mm} -\;\Big( \,\frac{3}{8} \,u_3^2 \,-\,
            \frac{7}{8}\,u_3x_4 \Big) \,\Big( \,\zeta_3 \,-\,\zeta_{17} 
                \,\Big) \,-\, \frac{1}{8} \,u_3^2x_3\,\Big( 
                 \,14\,\zeta_7 \,-\, 13\,\zeta_{13}
                \, \Big) \,+\, \Big( \frac{3}{8}\,\beta_0\,u_3^2 \,+\, 
                      \frac{7}{4}\,u_3x_4 \Big) \,\zeta_8 \\*[1mm]
      & & \hspace*{-6mm} +\; \Big( \frac{3}{8}\,\beta_0\,u_3^2 \,-\, 
                     \frac{3}{4}\,u_3^2 \,+\, 
                  \frac{7}{4}\,u_3x_4 \Big) \,\zeta_9 \,-\,
                 \frac{1}{8}\,u_3x_1\,\Big( \,14\,\zeta_{10} \,-\,
                   13\,\zeta_{15} \,\Big) \,-\, \Big( \frac{3}{4}
              \,\beta_0\,u_3^2 \,-\, \frac{15}{16}\,u_3^2 \Big. \\*[1mm]
      & & \hspace*{-6mm} \Big. \! +\; 
                  \frac{13}{8}\,u_3x_4 \Big) \,\zeta_{12} 
              \,-\, \Big( \frac{15}{32} \,u_3^2 \,+\, 
         \frac{13}{8}\,u_3x_4 \Big) \,\zeta_{14} \quad\;\; = \;\;\quad 0
\end{eqnarray*}
\end{small} }}
\begin{equation} \label{skgv27}
\end{equation}\\
\fbox{\parbox{14.8cm}{
\begin{small}
\begin{eqnarray*} & & \hspace*{-8mm} -\; \frac{1}{8} \,u_3x_1 \,\Big( 7\,\zeta_1 
                \,+\, 14\,\zeta_8 \,-\, 13\,\zeta_{14} \,-\,
                 7\,\zeta_{17} \Big) \,+\, \Big( \frac{3}{8}
                   \,u_3^2 \,+\, \frac{7}{4} \,\,u_3x_4 \Big) \,
                 \zeta_7 \,-\, \Big( \frac{73}{16}\,u_3x_1 \,-\,
                            \frac{15}{32}\,u_3x_3 \\*[1mm]
      & & \hspace*{-8mm} -\; \frac{1}{4} \,N_F\,u_3z_3 \Big) \,\zeta_9 
          \,+\, \Big( \frac{29}{4}\,u_3x_1 \,-\,\frac{15}{16}\,u_3x_3 
              \,-\, \frac{1}{2} \,N_F\,u_3z_3 \Big) \,\zeta_{12}
                   \,-\, \Big( \frac{15}{32} \,u_3^2 \,+\, 
           \frac{13}{8}\,u_3x_4 \Big) \,\zeta_{13}
\end{eqnarray*}\vspace*{1mm}
\[  \qquad\qquad\qquad\qquad\qquad\qquad\qquad\qquad\qquad
                            \qquad\qquad\qquad  = \quad 0 \]         \\*
\end{small} }}
\begin{equation} \label{skgv28} 
\end{equation}\\
\subsection{$C_C$--$L_{-}$ equations:}
\vspace{6mm}
\fbox{\parbox{14.8cm}{
\begin{small}
\begin{eqnarray*} & & \hspace*{-6mm} \frac{3}{4} 
                  \,u_3^2x_3\,\Big( \,4\,\zeta_3 \,-\, 4\,\zeta_8
                   \,+\, 8\,\zeta_9 \,-\, 10\,\zeta_{12} \,+\,
                        5\,\zeta_{14} \,-\, 4\,\zeta_{17}
                \, \Big) \,+\, \Big( \frac{3}{4}\,\beta_0\,u_3^2 \,+\, 
                                3\,u_3x_4 \Big) \,\zeta_{10} \\*[1mm]
      & & \hspace*{-6mm} +\; \Big( \, \frac{45}{8} \,u_3x_1 \,-\, 
                     \frac{15}{16}\,u_3x_3 \,-\, 
                  \frac{1}{2}\,N_F\,u_3z_3 \Big) \,\zeta_{11} \,-\,
             \frac{15}{4} \,u_3x_4 \,\zeta_{15} \quad\;\; = \;\;\quad 0
\end{eqnarray*}
\end{small} }}
\begin{equation} \label{skgv29} 
\end{equation}
\fbox{\parbox{14.8cm}{
\begin{small}
\begin{eqnarray*} & & \hspace*{-8mm} \frac{3}{4}\,u_3x_4\,\Big( \,2\,\zeta_3 
               \,-\, 5\,\zeta_{12} \,-\, 5\,\zeta_{14} \,-\, 
                  2\,\zeta_{17}\,\Big) \,-\, \frac{3}{4}\,u_3^2x_3 \, 
                    \Big( \,4\,\zeta_7 \,-\,5\,\zeta_{13} \,\Big)
                      \,+\,\Big( \frac{3}{4}\,\beta_0\,u_3^2 \\*[1mm]
      & & \hspace*{-8mm} \Big. \! +\; 
                     3\,u_3x_4 \Big) \,\Big( \,\zeta_8 \,+\,
                 \zeta_9 \,\Big) \,+\, \Big( \frac{45}{8}\,u_3x_1 \,-\, 
                  \frac{15}{16}\,u_3x_3 \,-\, \frac{1}{2} 
           \,N_F\,u_3z_3 \Big) \, \zeta_{10} \quad\;\; = \;\;\quad 0 
\end{eqnarray*}
\end{small} }}
\begin{equation} \label{skgv30}
\end{equation}\\
\fbox{\parbox{14.8cm}{
\begin{small}
\begin{eqnarray*} & & \hspace*{-8mm}  \Big( \frac{9}{16}\,u_3^2 \,+\,
               \frac{3}{4}\,u_3x_4 \Big) \,\Big( \,\zeta_3 \,-\, 
             \zeta_{17} \,\Big) \,-\, \frac{3}{8} \,u_3^2x_3 \,
                    \Big( \,4\,\zeta_7 \,-\, 5\,\zeta_{13}
                 \Big) \,+\, \Big( \frac{3}{4}\,
                    \beta_0\,u_3^2 \,-\, \frac{9}{16}\,u_3^2 
                     \,+\, \frac{3}{2}\,u_3x_4 \Big) \,\zeta_8  \\*[1mm]
      & & \hspace*{-8mm} -\; \Big( \frac{3}{8}\,
                    \beta_0\,u_3^2 \,-\, \frac{9}{8}\,u_3^2 
                 \,-\, \frac{3}{2}\,u_3x_4 \Big) \,\zeta_9 \,+\,    
            \Big( \frac{57}{8}\,u_3x_1 \,-\, \frac{15}{16}\,u_3x_3 
                    \,-\, \frac{1}{2} \,N_F\,u_3z_3 \Big)
                         \,\zeta_{10} \\*[1mm]
      & & \hspace*{-8mm} -\; \Big( \frac{9}{16}\,u_3^2 
                 \,+\, \frac{15}{8}\,u_3x_4 \Big) \,\zeta_{12} \,+\, 
                 \Big( \frac{9}{32}\,u_3^2 \,-\,\frac{15}{8}
                      \,u_3x_4 \Big) \, \zeta_{14} \,-\,
                   \frac{15}{8}\,u_3x_1 \,\zeta_{15} \quad\;\; = \;\;\quad 0
\end{eqnarray*}
\end{small} }}
\begin{equation} \label{skgv31}
\end{equation}\\
\fbox{\parbox{14.8cm}{
\begin{small}
\begin{eqnarray*} & & \hspace*{-8mm} \frac{3}{4}\,u_3x_1 \,\Big( \,
               2\,\zeta_3 \,-\,5\,\zeta_{12} \,-\, 5\,\zeta_{14}
               \,-\,2\,\zeta_{17} \,\Big) \,+\, \Big( 
               \frac{3}{4}\,\beta_0\,u_3^2 \,-\,\frac{9}{8}\,u_3^2 
                  \Big) \,\zeta_7 \,+\, \Big( \frac{69}{8}\,u_3x_1 \,-\, 
                 \frac{15}{16}\,u_3x_3 \Big.  \\*[1mm]
      & & \hspace*{-8mm} \Big. \! -\; 
                      \frac{1}{2}\,N_F\,u_3z_3 \Big) \,\Big( 
             \,\zeta_8 \,+\,\zeta_9 \,\Big) \,+\, 
             \frac{9}{16}\,u_3^2 \,\zeta_{13} \quad\;\; = \;\;\quad 0       
\end{eqnarray*}
\end{small} }}
\begin{equation} \label{skgv32}
\end{equation}\\
\fbox{\parbox{14.8cm}{
\begin{small}
\begin{eqnarray*} & & \hspace*{-8mm} \frac{3}{8}\,u_3x_1 \,\Big( \,
               2\,\zeta_3 \,-\,5\,\zeta_{12} \,-\, 5\,\zeta_{14}
               \,-\,2\,\zeta_{17} \,\Big) \,+\, \Big( 
               \frac{3}{4}\,\beta_0\,u_3^2 \,-\,\frac{9}{16}\,u_3^2 
                 \,+\,\frac{3}{2} \,u_3x_4  \Big) \,\zeta_7 \\*[1mm]
      & & \hspace*{-8mm} +\; \Big( \frac{57}{8}\,u_3x_1 \,-\, 
                 \frac{15}{16}\,u_3x_3 \,-\, 
                      \frac{1}{2}\,N_F\,u_3z_3 \Big) \,\zeta_8 
              \,-\,  \Big( \frac{21}{16}\,u_3x_1 \,-\, 
                 \frac{15}{32}\,u_3x_3 \,-\, 
                      \frac{1}{4}\,N_F\,u_3z_3 \Big) \,\zeta_9 \\*[1mm]
      & & \hspace*{-8mm} +\; \Big( \frac{9}{32}\,u_3^2 
         \,-\,\frac{15}{8} \,u_3x_4  \Big) \,\zeta_{13} 
                                   \quad\;\; = \;\;\quad 0     
\end{eqnarray*}
\end{small} }}
\begin{equation} \label{skgv33}
\end{equation}\\
\subsection{$C_D$--$L_{0}$ equations:}
\vspace{6mm}
\fbox{\parbox{14.8cm}{
\begin{small}
\begin{eqnarray*} & & \hspace*{-8mm} \frac{9}{8}\,u_3^2x_3\,\Big( \,\zeta_1 
               \,+\,\zeta_7 \,+\, \zeta_{13}\,\Big) \,+\, 
             \Big( \frac{81}{32}\,u_3^2 \,-\, \frac{9}{8}\,u_3x_4
                 \Big)\,\zeta_2 \,+\, \Big( \frac{9}{4}\,
                    \beta_0\,u_3^2 \,+\, \frac{81}{32}\,u_3^2
                         \,+\, \frac{99}{8}\,u_3x_4
                            \Big) \,\zeta_3 \\*[1mm]
      & & \hspace*{-8mm} +\;\frac{9}{8}\,u_3x_1\,\Big( \,\zeta_4 
                      \,+\,\zeta_{10} \,+\,\zeta_{15} \,\Big) \,+\, \Big( 
                    \frac{9}{2}\,\beta_0\,u_3^2 \,-\,
              \frac{27}{32}\,u_3^2 \,+\, \frac{45}{2}\,u_3x_4 
                            \Big) \,\zeta_6  \\*[1mm]
      & & \hspace*{-8mm} - \; \Big( \frac{27}{32}\,u_3^2 
                     \,+\, \frac{9}{8}\,u_3x_4 \Big) 
                    \,\Big( \,\zeta_8 \,+\,\zeta_{14}
                    \,\Big) \,-\, \Big( \frac{27}{32}\,u_3^2 
                 \,+\, \frac{27}{8}\,u_3x_4 \Big) \,\zeta_9 \\*[1mm]
      & & \hspace*{-8mm} -\; \Big( \frac{27}{32}
             \,u_3^2 \,+\, \frac{27}{4}\,u_3x_4 \Big) \,
                 \zeta_{12} \,-\, \frac{9}{8}\, u_3x_4 \, \zeta_{17}
\end{eqnarray*}   
\begin{eqnarray*}  \qquad\qquad & = & -\; \frac{9}{4}\,u_3^2x_3^2 \,+\, 
                       9\,u_3x_1x_3x_4 \,-\, \frac{27}{4}\,u_3x_4^2
                       \,-\, 9\,x_2x_4^2  \,+\, 9\,x_1x_4x_5 \\*[1mm]
             & & +\; N_F\,\Big( \frac{u_3}{w_3} \Big)^2 \,
                  \Big( -\, w_3z_1z_3z_4
                     \,+\, w_3z_4^2
                    \,-\, z_1z_4z_5
                       \,+\, z_2z_4^2 \Big)
\end{eqnarray*}
\end{small} }}
\begin{equation} \label{skgv34}
\end{equation}\\
\fbox{\parbox{14.8cm}{
\begin{small}
\begin{eqnarray*} & & \hspace*{-9mm} \Big( \frac{81}{32}\,u_3^2 \,-\, 
                \frac{7}{8}\,u_3x_4 \Big) \,\zeta_1 \,-\, 
             \frac{9}{8}\,u_3x_1\,\Big( \,\zeta_2 \,-\,
                      \zeta_8 \,+\, 4\,\zeta_9 \,+\,
                   7\,\zeta_{12} \,-\, \zeta_{14} 
                        \,-\, \zeta_{17} \,\Big)  \\*[1mm]
      & & \hspace*{-9mm} -\; \Big( \frac{27}{4}\,u_3x_1 \,-\, 
                 \frac{45}{16}\,u_3x_3 \,-\, 
                      \frac{3}{2}\,N_F\,u_3z_3 \Big) \,\zeta_3 
                   \,-\, \Big( \frac{45}{4}\,u_3x_1 \,-\, 
                 \frac{45}{8}\,u_3x_3 \,-\, 
                      3\,N_F\,u_3z_3 \Big) \,\zeta_6 \\*[1mm]
      & & \hspace*{-9mm} -\; \Big( \frac{27}{32}\,u_3^2 \,+\, 
                \frac{9}{8}\,u_3x_4 \Big) \,\zeta_7 
              \,-\, \Big( \frac{27}{32}\,u_3^2 \,+\, 
                \frac{9}{8}\,u_3x_4 \Big) \,\zeta_{13}       
\end{eqnarray*}   
\begin{eqnarray*}  \quad\qquad & = &  9\,u_3x_1^2x_3 \,-\, 
                            \frac{27}{4}\,u_3x_1x_4
                       \,+\, 9\,x_1^2x_5  \,-\, 9\,x_1x_2x_4 \\*[1mm]
             & & +\; N_F\,\Big( \frac{u_3}{w_3} \Big)^2 \,
                  \Big( w_3z_1z_4
                     \,-\, w_3z_1^2z_3
                    \,+\, z_1z_2z_4
                       \,-\, z_1^2z_5 \Big)
\end{eqnarray*}
\end{small} }}
\begin{equation} \label{skgv35}
\end{equation}\\
\subsection{$C_D$--$L_{+}$ equations:}
\vspace{6mm}
\fbox{\parbox{14.8cm}{
\begin{small}
\begin{eqnarray*} & & \hspace*{-8mm} \frac{1}{8}\,u_3^2x_3\,\Big( \,7\,\zeta_1 
               \,+\,7\,\zeta_7 \,+\,13\,\zeta_{13}\,\Big) \,-\, 
             \Big( \frac{33}{32}\,u_3^2 \,+\, \frac{7}{8}\,u_3x_4
                 \Big)\,\zeta_2 \,-\, \Big( \frac{3}{4}\,
                    \beta_0\,u_3^2 \,+\, \frac{33}{32}\,u_3^2  \\*[1mm]
      & & \hspace*{-8mm} -\; \frac{17}{8}\,u_3x_4
                  \Big) \,\zeta_3 \,+\, \frac{1}{8}\,u_3x_1\,\Big( 
                \,7\,\zeta_4 \,+\, 7\,\zeta_{10} \,+\,
            13\,\zeta_{15} \,\Big) \,+\, \Big( \frac{3}{2}
                    \,\beta_0\,u_3^2 \,-\, \frac{15}{32}\,u_3^2 
                     \,+\, \frac{5}{2}\,u_3x_4  \Big) \,\zeta_6  \\*[1mm]
      & & \hspace*{-8mm} +\;  \Big( \frac{3}{32}\,u_3^2 
                     \,-\, \frac{7}{8}\,u_3x_4 \Big) \,\zeta_8 
                       \,+\, \Big( \frac{3}{32}\,u_3^2 
                 \,-\, \frac{1}{8}\,u_3x_4 \Big) \,\zeta_9 \,-\,    
            \Big( \frac{15}{32}\,u_3^2 
                 \,-\, \frac{1}{4}\,u_3x_4 \Big) \,\zeta_{12} \\*[1mm]
      & & \hspace*{-8mm} -\; \Big( \frac{15}{32}\,u_3^2 
                 \,+\, \frac{13}{8}\,u_3x_4 \Big) \,\zeta_{14} \,+\, 
                 \Big( \frac{3}{8}\,u_3^2 \,-\,
                   \frac{7}{8}\,u_3x_4 \Big) \,\zeta_{17}
\end{eqnarray*}   
\begin{eqnarray*}  \qquad\qquad & = & -\; \frac{7}{4}\,u_3^2x_3^2 \,+\, 
                       3\,u_3x_1x_3x_4 \,-\, \frac{5}{4}\,u_3x_4^2
                       \,-\, 3\,x_2x_4^2  \,+\, 9\,x_1x_4x_5 \\*[1mm]
             & & +\; N_F\,\Big( \frac{u_3}{w_3} \Big)^2 \,
                  \Big( \frac{1}{3}\,w_3z_1z_3z_4
                     \,-\, \frac{1}{3}\,w_3z_4^2
                    \,+\, \frac{1}{3}\,z_1z_4z_5
                       \,-\, \frac{1}{3}\,z_2z_4^2 \Big)
\end{eqnarray*}
\end{small} }}
\begin{equation} \label{skgv36}
\end{equation}\\
\fbox{\parbox{14.8cm}{
\begin{small}
\begin{eqnarray*} & & \hspace*{-9mm} -\; \Big( \frac{33}{32}\,u_3^2 \,+\, 
                \frac{7}{8}\,u_3x_4 \Big) \,\zeta_1 \,+\, 
             \frac{1}{8}\,u_3x_1\,\Big( \,7\,\zeta_2 \,+\,
                      7\,\zeta_8 \,-\,8\,\zeta_9 \,-\,
                    11\,\zeta_{12} \,+\, 13\,\zeta_{14}
              \,+\, 7\,\zeta_{17} \,\Big)  \\*[1mm]
      & & \hspace*{-9mm} +\; \Big( \frac{55}{8}\,u_3x_1
                        \,-\, \frac{15}{16}\,u_3x_3 \,-\, 
                      \frac{1}{2}\,N_F\,u_3z_3 \Big) \,\zeta_3 
                   \,-\, \Big( \frac{83}{8}\,u_3x_1 \,-\, 
                 \frac{15}{8}\,u_3x_3 \,-\, N_F\,u_3z_3 \Big) 
                              \,\zeta_6  \\*[1mm]
      & & \hspace*{-9mm} +\; \Big( \frac{3}{32}\,u_3^2 \,-\, 
                \frac{7}{8}\,u_3x_4 \Big) \,\zeta_7 
              \,-\, \Big( \frac{15}{32}\,u_3^2 \,+\, 
                \frac{13}{8}\,u_3x_4 \Big) \,\zeta_{13}       
\end{eqnarray*}   
\begin{eqnarray*}  \quad\qquad & = & -\; 3\,u_3x_1^2x_3 \,-\,
                       \frac{5}{2}\,u_3x_1x_4
                       \,+\, 3\,x_1^2x_5  \,-\, 3\,x_1x_2x_4 \\*[1mm]
             & & +\; N_F\,\Big( \frac{u_3}{w_3} \Big)^2 \,
                  \Big( -\,\frac{1}{3}\,w_3z_1z_4
                     \,+\, \frac{1}{3}\,w_3z_1^2z_3
                    \,-\, \frac{1}{3}\,z_1z_2z_4
                       \,+\, \frac{1}{3}\,z_1^2z_5 \Big)
\end{eqnarray*}
\end{small} }}
\begin{equation} \label{skgv37}
\end{equation}\\
\subsection{$C_D$--$L_{-}$ equations:}
\vspace{6mm}
\fbox{\parbox{14.8cm}{
\begin{small}
\begin{eqnarray*} & & \hspace*{-12mm} \frac{3}{4}\,u_3^2x_3 \,
                      \Big( \,2\,\zeta_2 \,+\, 
              2\,\zeta_3 \,+\, 5\,\zeta_6 \,+\, 2\,\zeta_8 \,+\, 
            2\,\zeta_9 \,+\, 5\,\zeta_{12} \,+\, 5\,\zeta_{14} \,-\, 
              4\,\zeta_{17} \,\Big) \,-\, \Big( \frac{3}{2} 
                     \,\beta_0 \,u_3^2 \\*[1mm]
      & & \hspace*{-12mm} +\; \frac{3}{2}\,u_3x_4 \Big) 
                    \,\zeta_4 \,-\,  \Big( \frac{45}{4}\,u_3x_1 \,-\, 
             \frac{15}{8}\,u_3x_3 \,-\, N_F\,u_3z_3 \Big)
               \,\zeta_5 \,-\, \frac{3}{4}\,u_3x_4 \,\Big(
                  \,2\,\zeta_{10} \,+\, 5\,\zeta_{15} \,\Big)
\end{eqnarray*}   
\[  \qquad\qquad = \quad N_F\,\Big( \frac{u_3}{w_3} \Big)^2 \,
                  \Big( 2\,w_3^2z_3^3
                     \,-\, 4\,w_3z_3z_4^2
                       \,+\, 2\,z_4^2z_5 \Big)  \]          \\* 
\end{small} }}
\begin{equation} \label{skgv38} 
\end{equation}
\fbox{\parbox{14.8cm}{
\begin{small}
\begin{eqnarray*} & & \hspace*{-14mm} \frac{3}{4}\,u_3^2x_3\,
                      \Big( \,2\,\zeta_1 
               \,+\,2\,\zeta_7 \,+\,5\,\zeta_{13}\,\Big) \,-\, 
             \Big( \frac{3}{2}\,\beta_0\,u_3^2 \,+\,
                \frac{27}{8}\,u_3^2 \,+\, \frac{3}{2}\,u_3x_4
                 \Big) \,\Big( \,\zeta_2 \,+\, \zeta_3 \,\Big)  \\*[1mm]
      & & \hspace*{-14mm} -\; \Big( \frac{45}{4}\,u_3x_1 \,-\, 
                 \frac{15}{8}\,u_3x_3  \,-\, N_F\,u_3z_3 
                       \Big) \,\zeta_4 \,-\, \frac{3}{4} \,
                       u_3x_4 \,\Big( \,5\,\zeta_6 \,-\,
                5\,\zeta_{12} \,+\, 5\,\zeta_{14}
                                \,+\, 2\,\zeta_{17} \,\Big) \\*[1mm]
      & & \hspace*{-14mm} +\; \Big( \frac{9}{8}\,u_3^2 
                 \,-\, \frac{3}{2}\,u_3x_4 \Big) \,\zeta_8
                \,+\, \Big( \frac{9}{8}\,u_3^2 \,+\, 
                  \frac{3}{2}\,u_3x_4 \Big) \, \zeta_9
\end{eqnarray*}   
\begin{eqnarray*}  \qquad\qquad & = &  N_F\,\Big( \frac{u_3}{w_3} \Big)^2 \,
                \Big( 2\,w_3^2z_3^2 \,-\, 2\,w_3z_1z_3z_4
                     \,-\, 2\,w_3z_4^2 \,+\, 
                              2\,z_1z_4z_5 \Big)
\end{eqnarray*}
\end{small} }}
\begin{equation} \label{skgv39}
\end{equation}\\
\fbox{\parbox{14.8cm}{
\begin{small}
\begin{eqnarray*} & & \hspace*{-18mm} \frac{3}{8}\,u_3^2x_3\,
                      \Big( \,2\,\zeta_1 
               \,+\,2\,\zeta_7 \,+\,5\,\zeta_{13}\,\Big) \,-\, 
            \Big( \frac{3}{2}\, \beta_0\,u_3^2 \,+\, 
                  \frac{9}{16}\, u_3^2 \,+\,
                    \frac{3}{4}\,u_3x_4 \Big) \,\zeta_2 \\*[1mm]
      & & \hspace*{-18mm} +\; \Big( \frac{3}{4}\,\beta_0 \,u_3^2 \,-\, 
                    \frac{9}{16}\,u_3^2 \,+\, \frac{3}{2}\,u_3x_4 
                \Big) \,\zeta_3 \,-\, \Big( 12\,
                    u_3x_1 \,-\, \frac{15}{8}\,u_3x_3 \,-\, 
                        N_F\,u_3z_3 \Big) \,\zeta_4  \\*[1mm]
      & & \hspace*{-18mm} +\; \Big( \frac{9}{32}\,u_3^2 
                 \,+\, \frac{15}{4}\,u_3x_4 \Big) \,\zeta_6 \,+\,    
                      \Big( \frac{9}{16}\,u_3^2 
                 \,-\, \frac{3}{4}\,u_3x_4 \Big) \,\zeta_8 \,+\,
                  \frac{9}{32}\,u_3^2 \,\Big( \,2\,\zeta_9
                           \,+\, \zeta_{12} \,\Big)  \\*[1mm]
      & & \hspace*{-18mm} -\; \frac{3}{8}\,u_3x_1\,\Big( \,
                        2\,\zeta_{10} \,+\, 5\,\zeta_{15} \,\Big) \,+\,
                 \Big( \frac{9}{32}\,u_3^2 
                 \,-\, \frac{15}{8}\,u_3x_4 \Big) \,\zeta_{14} \,-\, 
                 \Big( \frac{9}{16}\,u_3^2 \,+\,
                     \frac{3}{4}\,u_3x_4 \Big) \, \zeta_{17}
\end{eqnarray*}   
\begin{eqnarray*}  \qquad\qquad & = & \frac{3}{2}\,u_3^2x_3^2
                        \,-\, \frac{3}{2}\,u_3x_4^2 \,+\, 
                 N_F\,\Big( \frac{u_3}{w_3} \Big)^2 \,
                                   \Big( 2\,w_3^2z_3^2 \\*[1mm]
             & & -\; 3\,w_3z_1z_3z_4
                     \,-\, w_3z_4^2 \,+\, z_1z_4z_5
                       \,+\, z_2z_4^2 \Big)
\end{eqnarray*}
\end{small} }}
\begin{equation} \label{skgv40}
\end{equation}\\
\fbox{\parbox{14.8cm}{
\begin{small}
\begin{eqnarray*} & & \hspace*{-9mm} -\; \Big( \frac{3}{2}\,
                      \beta_0\,u_3^2 \,+\, 
                \frac{9}{8}\,u_3^2 \Big) \,\zeta_1 \,-\, 
              \Big( \frac{51}{4}\, u_3x_1 \,-\, \frac{15}{8}\,u_3x_3 \,-\, 
                        N_F\,u_3z_3 \,\Big) \,\zeta_2
                         \,-\, \Big( \frac{33}{4}\,u_3x_1 \,-\, 
                                   \frac{15}{8}\,u_3x_3 \\*[1mm]
      & & \hspace*{-9mm} -\;  N_F\,u_3z_3 \Big) 
                \,\zeta_3  \,+\, \frac{3}{4}\,u_3x_1
                    \,\Big( \,10\,\zeta_6 \,-\, 2\,\zeta_8 
                    \,-\, 5\,\zeta_{14} \,-\, 2\,\zeta_{17} \,\Big)
                         \,+\, \frac{9}{16}\,u_3^2 
                  \,\Big( \,2\,\zeta_7 \,+\, \zeta_{13} \,\Big)      
\end{eqnarray*}   
\begin{eqnarray*}  \qquad\qquad & = & 3\,u_3^2x_3
                        \,-\, 3\,u_3x_1x_4 \,+\, 
                 N_F\,\Big( \frac{u_3}{w_3} \Big)^2 \,
                                   \Big( 2\,w_3^2z_3 \\*[1mm]
             & & -\; 2\,w_3z_1z_4
                     \,-\, 2\,w_3z_1^2z_3
                       \,+\, 2\,z_1z_2z_4 \Big)
\end{eqnarray*}
\end{small} }}
\begin{equation} \label{skgv41}
\end{equation}\\
\fbox{\parbox{14.8cm}{
\begin{small}
\begin{eqnarray*} & & \hspace*{-12mm} -\; \Big( \frac{3}{2}\,\beta_0\,u_3^2 
                   \,+\, \frac{63}{16}\,u_3^2 \,+\, 
                \frac{3}{4}\,u_3x_4 \Big) \,\zeta_1 \,-\, \Big( 12\,
                    u_3x_1 \,-\, \frac{15}{8}\,u_3x_3 \,-\, 
                        N_F\,u_3z_3 \,\Big) \,\zeta_2  \\*[1mm]
      & & \hspace*{-12mm} +\; \Big( \frac{39}{8}\,u_3x_1 \,-\, 
                 \frac{15}{16}\,u_3x_3 \,-\, 
                      \frac{1}{2}\,N_F\,u_3z_3 \Big) \,\zeta_3
             \,-\, \frac{3}{8}\,u_3x_1 \,\Big( \,
                    5\,\zeta_6 \,+\, 2\,\zeta_8 \,-\, 2
                      \,\zeta_9 \,-\, 5 \,\zeta_{12} \\*[1mm]
      & & \hspace*{-12mm} +\; 5\,\zeta_{14} \,+\, 2\,\zeta_{17}\, \Big)
                 \,+\, \Big( \frac{27}{16}\,u_3^2 \,-\, 
                \frac{3}{4}\,u_3x_4 \Big) \,\zeta_7
             \,+\, \Big(  \frac{9}{32}\,u_3^2 \,-\, 
                \frac{15}{8}\,u_3x_4 \Big) \,\zeta_{13}       
\end{eqnarray*}
\begin{eqnarray*}  \qquad\qquad & = & 3\,u_3^2x_3
                        \,-\, \frac{3}{2}\,u_3x_1x_4 \,+\, 
                 N_F\,\Big( \frac{u_3}{w_3} \Big)^2 \,
                                   \Big( 2\,w_3^2z_3 \\*[1mm]
             & & -\; 3\,w_3z_1z_4
                 \,-\, w_3z_1^2z_3 \,+\, z_1z_2z_4 
                           \,+\, z_1^2z_5  \Big)
\end{eqnarray*}
\end{small} }}
\begin{equation} \label{skgv42}
\end{equation}
\subsection{$C_E$--$L_{0}$ equations:}
\vspace{6mm}
\fbox{\parbox{14.8cm}{
\begin{small}
\begin{eqnarray*} & & \hspace*{-10mm} -\; \frac{3}{4}\,u_3^2x_3 \,\Big( 
                \,2\,\zeta_2 \,+\, 2\,\zeta_3 \,+\, 5\,\zeta_6 
               \,+\, 2\,\zeta_8 \,+\, 2\,\zeta_9 \,+\, 
               5\,\zeta_{12} \,+\, 5\,\zeta_{14} \,-\, 
              4\,\zeta_{17} \,\Big) \,+\, \Big( \frac{3}{2}
                              \,\beta_0\,u_3^2 \\*[1mm]
      & & \hspace*{-10mm} +\; \frac{3}{2}\,u_3x_4 \Big)
                    \,\zeta_4 \,+\, \Big( \frac{45}{4}\,u_3x_1 \,-\, 
             \frac{15}{8}\,u_3x_3 \,-\, N_F\,u_3z_3 \Big)
                  \,\zeta_5 \,+\, \frac{3}{4}\,u_3x_4 \,
                   \Big( \,2\,\zeta_{10} \,+\, 5\,\zeta_{15} \,\Big)
\end{eqnarray*}   
\[  \qquad\qquad = \quad N_F\,\Big( \frac{u_3}{w_3} \Big)^2 \,
                  \Big( -\, 2\,w_3^2z_3^3
                     \,+\, 4\,w_3z_3z_4^2
                       \,-\, 2\,z_4^2z_5 \Big)  \]           \\*
\end{small} }}
\begin{equation} \label{skgv43} 
\end{equation}\\
\fbox{\parbox{14.8cm}{
\begin{small}
\begin{eqnarray*} & & \hspace*{-8mm} \frac{3}{4}\,u_3^2x_3\,\Big( \,\zeta_1 
               \,+\,\zeta_7 \,+\,\zeta_{13}\,\Big) \,+\, 
             \Big( \frac{3}{2}\,\beta_0\,u_3^2 \,-\, 
              \frac{9}{16}\,u_3^2 \,-\, \frac{3}{4}\,u_3x_4
                 \Big)\,\zeta_2 \,-\, \Big( \frac{3}{2}\,
                    \beta_0\,u_3^2 \,+\, \frac{9}{16}\, u_3^2  \\*[1mm]
      & & \hspace*{-8mm} -\; \frac{15}{4}\,u_3x_4
                \Big) \,\zeta_3 \,+\, \Big( \frac{27}{2}\,u_3x_1 \,-\, 
                 \frac{15}{8}\,u_3x_3  \,-\, N_F\,u_3z_3 
                       \Big) \,\zeta_4 \,+\, \Big( 3\,\beta_0\,u_3^2 \,-\, 
              \frac{9}{8}\,u_3^2 \,+\, \frac{15}{4}\,u_3x_4
                                      \Big)\,\zeta_6 \\*[1mm]
      & & \hspace*{-8mm} -\; \Big( \frac{9}{16}\,u_3^2  
                  \,+\, \frac{3}{4}\,u_3x_4  
                       \Big) \,\Big( \,\zeta_8 \,+\,\zeta_9 \,\Big)
                      \,+\, \frac{9}{4}\,u_3x_1 \,\Big( \,
                            \zeta_{10} \,+\, 2\,\zeta_{15} \,\Big) \\*[1mm]
      & & \hspace*{-8mm} -\; \Big( \frac{9}{8}\,u_3^2 
                 \,+\, \frac{3}{4}\,u_3x_4 \Big) \, \Big( 
                       \,\zeta_{12} \,+\,\zeta_{14} \,\Big)
                     \,+\, \Big( \frac{9}{8}\,u_3^2 \,-\, 
                  \frac{3}{4}\,u_3x_4 \Big) \, \zeta_{17}
\end{eqnarray*}   
\begin{eqnarray*}  \qquad\qquad & = & -\; \frac{9}{2}\,u_3^2x_3^2 \,+\, 
                       6\,u_3x_1x_3x_4 \,-\, \frac{3}{2}\,u_3x_4^2
                       \,-\, 6\,x_2x_4^2  \,+\, 6\,x_1x_4x_5 
                 \,+\, N_F\,\Big( \frac{u_3}{w_3} \Big)^2 \\*[1mm]
             & & \cdot \;
                  \Big( -\,2\,w_3^2z_3^2 \,+\,
                       \frac{10}{3}\,w_3z_1z_3z_4
                     \,+\, \frac{2}{3}\,w_3z_4^2
                    \,-\, \frac{2}{3}\,z_1z_4z_5
                       \,-\, \frac{4}{3}\,z_2z_4^2 \Big)
\end{eqnarray*}
\end{small} }}
\begin{equation} \label{skgv44}
\end{equation}\\
\fbox{\parbox{14.8cm}{
\begin{small}
\begin{eqnarray*} & & \hspace*{-7mm} -\; \frac{3}{8}\,u_3^2x_3\,
                      \Big( \,5\,\zeta_1 
               \,+\,5\,\zeta_7 \,+\,11\,\zeta_{13}\,\Big) \,+\, 
             \Big( \frac{3}{2}\, \beta_0\,u_3^2 \,+\,
                   \frac{81}{32}\,u_3^2 \,+\, \frac{15}{8}\,u_3x_4
                 \Big)\,\zeta_2 \,+\, \Big( \frac{3}{4}\,
                    \beta_0\,u_3^2 \,+\, \frac{81}{32}\,u_3^2 \\*[1mm]
      & & \hspace*{-7mm} -\; \frac{21}{8}\,u_3x_4 \Big) \,\zeta_3
                  \,+\, \Big( \frac{87}{8}\,u_3x_1 \,-\, 
                       \frac{15}{8}\,u_3x_3 \,-\,N_F\,u_3z_3
                \Big) \,\zeta_4 \,-\, \Big( \frac{3}{2}\,
                    \beta_0\,u_3^2 \,-\, \frac{9}{32}\,u_3^2 
                     \,+\, \frac{15}{4}\,u_3x_4 \Big) \,\zeta_6  \\*[1mm]
      & & \hspace*{-7mm} -\; \Big( \frac{27}{32}\,u_3^2 
                 \,-\, \frac{15}{8}\,u_3x_4 \Big) \,\zeta_8 \,-\, 
                 \Big( \frac{27}{32}\,u_3^2 \,+\,
                \frac{3}{8}\,u_3x_4 \Big) \,\zeta_9 \,-\,
                 \frac{3}{8}\, u_3x_1\,\Big( \,\zeta_{10}
                                   \,+\, \zeta_{15} \,\Big) \\*[1mm]
      & & \hspace*{-7mm} +\; \Big( \frac{9}{32}\,u_3^2 \,-\,
              \frac{3}{2}\,u_3x_4 \Big) \,\zeta_{12} \,+\, 
                 \Big( \frac{9}{32}\,u_3^2 \,+\, \frac{33}{8}
                \,u_3x_4 \Big) \,\zeta_{14} \,+\, \frac{15}{8}
                \,u_3x_4 \,\zeta_{17} 
\end{eqnarray*}   
\begin{eqnarray*}  \qquad\qquad & = & \frac{3}{4}\,u_3^2x_3^2 \,-\, 
                       3\,u_3x_1x_3x_4 \,+\, \frac{9}{4}\,u_3x_4^2
                       \,+\, 3\,x_2x_4^2  \,-\, 3\,x_1x_4x_5 
                   \,+\, N_F\,\Big( \frac{u_3}{w_3} \Big)^2 \\*[1mm]
             & & \cdot \; \Big( -\, 2\,w_3^2z_3^2 \,+\,
                      \frac{7}{3}\,w_3z_1z_3z_4
                     \,+\, \frac{5}{3}\,w_3z_4^2
                    \,-\, \frac{5}{3}\,z_1z_4z_5
                       \,-\, \frac{1}{3}\,z_2z_4^2 \Big)
\end{eqnarray*}
\end{small} }}
\begin{equation} \label{skgv45}
\end{equation}\\
\fbox{\parbox{14.8cm}{
\begin{small}
\begin{eqnarray*} & & \hspace*{-9mm} \Big( \frac{3}{2}\,\beta_0\,u_3^2 
                 \,+\, \frac{81}{16}\,u_3^2 \,+\, 
                \frac{9}{4}\,u_3x_4 \Big) \,\zeta_1 \,+\, 
                       \Big( \frac{21}{2}\,
                    u_3x_1 \,-\, \frac{15}{4}\,u_3x_3 \,-\, 
                         N_F\,u_3z_3 \,\Big) \,\zeta_2 \\*[1mm]
      & & \hspace*{-9mm} -\; \Big( \frac{57}{4}\,u_3x_1 \,-\, 
                 \frac{15}{16}\,u_3x_3 \,-\, 
                         N_F\,u_3z_3 \Big) \,\zeta_3 
                   \,+\, \Big( \frac{39}{2}\,u_3x_1 \,-\, 
                 \frac{15}{8}\,u_3x_3 \,-\, 
                      2\,N_F\,u_3z_3 \Big) \,\zeta_6 \\*[1mm]
      & & \hspace*{-9mm} -\; \Big( \frac{27}{16}\,u_3^2 \,-\, 
                \frac{9}{4}\,u_3x_4 \Big) \,\zeta_7  
              \,-\, \frac{3}{4}\,u_3x_1 \,\Big( \,\zeta_8 \,-\,
                      2\,\zeta_9 \,-\,2\, \zeta_{12} \,+\,
               \zeta_{14} \,+\,\zeta_{17} \,\Big) \\*[1mm]
      & & \hspace*{-9mm} +\;
         \Big( \frac{9}{16}\,u_3^2 \,+\, 
                      \frac{9}{2}\,u_3x_4 \Big) \,\zeta_{13}       
\end{eqnarray*}   
\begin{eqnarray*}  \quad\qquad & = & -\; 3\,u_3^2x_3 \,-\, 
                       6\,u_3x_1^2x_3 \,+\, \frac{9}{2}\,u_3x_1x_4
                       \,-\, 6\,x_1^2x_5  \,+\, 6\,x_1x_2x_4
                      \,+\, N_F\,\Big( \frac{u_3}{w_3} \Big)^2  \\*[1mm]
             & & \cdot \; \Big( -\, 2\,w_3^2z_3 \,+\,
                        \frac{10}{3}\,w_3z_1z_4
                     \,+\, \frac{2}{3}\,w_3z_1^2z_3
                    \,-\, \frac{2}{3}\,z_1z_2z_4
                       \,-\, \frac{4}{3}\,z_1^2z_5 \Big)
\end{eqnarray*}
\end{small} }}
\begin{equation} \label{skgv46}
\end{equation}\\
\fbox{\parbox{14.8cm}{
\begin{small}
\begin{eqnarray*} & & \hspace*{-9mm} \Big( \frac{3}{2}\,\beta_0\,u_3^2 
                 \,+\, \frac{63}{32}\,u_3^2 \,-\, 
                \frac{3}{8}\,u_3x_4 \Big) \,\zeta_1 \,+\, 
                       \Big( \frac{105}{8}\,
                    u_3x_1 \,-\, \frac{15}{8}\,u_3x_3 \,-\, 
                         N_F\,u_3z_3 \,\Big) \,\zeta_2 \\*[1mm]
      & & \hspace*{-9mm} +\; \Big( \frac{51}{8}\,u_3x_1 \,-\, 
                 \frac{15}{16}\,u_3x_3 \,-\, \frac{1}{2}\,
                         N_F\,u_3z_3 \Big) \,\zeta_3 
                   \,-\, \Big( \frac{93}{8}\,u_3x_1 \,-\, 
                 \frac{15}{8}\,u_3x_3 \,-\, 
                      N_F\,u_3z_3 \Big) \,\zeta_6 \\*[1mm]
      & & \hspace*{-9mm} -\; \Big( \frac{45}{32}\,u_3^2 \,+\, 
                \frac{3}{8}\,u_3x_4 \Big) \,\zeta_7  
              \,+\, \frac{3}{8}\,u_3x_1 \,\Big( \,5\,\zeta_8 \,-\,
                      4\,\zeta_9 \,-\,7\, \zeta_{12} \,+\,11\,
               \zeta_{14} \,+\,5\,\zeta_{17} \,\Big) \\*[1mm]
      & & \hspace*{-9mm} -\;
         \Big( \frac{27}{32}\,u_3^2 \,+\, 
                      \frac{3}{8}\,u_3x_4 \Big) \,\zeta_{13}       
\end{eqnarray*}
\begin{eqnarray*}  \quad\qquad & = & -\; 3\,u_3^2x_3 \,+\, 
                       3\,u_3x_1^2x_3 \,+\, \frac{3}{4}\,u_3x_1x_4
                       \,+\, 3\,x_1^2x_5  \,-\, 3\,x_1x_2x_4 
             \,+\, N_F\,\Big( \frac{u_3}{w_3} \Big)^2 \\*[1mm]
             & & \cdot \; \Big( -\, 2\,w_3^2z_3 \,+\,
                        \frac{7}{3}\,w_3z_1z_4
                     \,+\, \frac{5}{3}\,w_3z_1^2z_3
                    \,-\, \frac{5}{3}\,z_1z_2z_4
                       \,-\, \frac{1}{3}\,z_1^2z_5 \Big)
\end{eqnarray*}
\end{small} }}
\begin{equation} \label{skgv47}
\end{equation}\\
\subsection{$C_E$--$L_{+}$ equations:}
\vspace{6mm}
\fbox{\parbox{14.8cm}{
\begin{small}
\begin{eqnarray*} & & \hspace*{-10mm} \frac{1}{4}\,u_3^2x_3 \,
                      \Big( \,2\,\zeta_2 \,+\, 
              2\, \zeta_3 \,+\, 5\,\zeta_6 \,+\, 2\,\zeta_8 \,+\, 
            2\,\zeta_9 \,+\, 5\,\zeta_{12} \,+\, 5\,\zeta_{14} \,-\, 
              4\,\zeta_{17} \,\Big) \,-\, \Big( 
                         \frac{1}{2} \,\beta_0 \,u_3^2  \\*[1mm]
      & & \hspace*{-10mm} +\;  \frac{1}{2}\,u_3x_4 \Big)
             \,\zeta_{4} \,-\, \Big( \frac{15}{4}\,u_3x_1 \,-\, 
             \frac{5}{8}\,u_3x_3 \,-\, \frac{1}{3} \,N_F\,u_3z_3 \Big)
             \,\zeta_5 \,-\, \frac{1}{4}\,u_3x_4 \,\Big( 
                        \,2\,\zeta_{10} \,+\, 5\,\zeta_{15} \,\Big) 
\end{eqnarray*}   
\[  \qquad\qquad = \quad N_F\,\Big( \frac{u_3}{w_3} \Big)^2 \,
                  \Big( \frac{2}{3}\,w_3^2z_3^3
                     \,-\, \frac{4}{3}\,w_3z_3z_4^2
                       \,+\, \frac{2}{3}\,z_4^2z_5 \Big)  \]   \\*
\end{small} }}
\begin{equation} \label{skgv48} 
\end{equation}\\
\fbox{\parbox{14.8cm}{
\begin{small}
\begin{eqnarray*} & & \hspace*{-8mm} \frac{1}{12}\,u_3^2x_3\,
                      \Big( \,7\,\zeta_1 
               \,+\,7\,\zeta_7 \,+\,13\,\zeta_{13}\,\Big)
                 \,-\, \Big( \frac{1}{2}\,\beta_0\,u_3^2 \,-\, 
                   \frac{1}{16}\,u_3^2 \,+\,\frac{7}{12}\,u_3x_4
                            \Big) \,\zeta_2 \\*[1mm]
      & & \hspace*{-8mm} +\; \Big( \frac{1}{2}\,\beta_0\,u_3^2 \,+\, 
                   \frac{1}{16}\,u_3^2 \,+\,\frac{35}{12}\,u_3x_4
                            \Big) \,\zeta_3 \,-\, \Big(
                      \frac{11}{3}\,u_3x_1 \,-\, 
                 \frac{5}{8}\,u_3x_3  \,-\, \frac{1}{3} 
                           \,N_F\,u_3z_3\Big) \,\zeta_4  \\*[1mm]
      & & \hspace*{-8mm} +\; \Big( \beta_0\,u_3^2 \,-\, 
                   \frac{1}{8}\,u_3^2 \,+\,\frac{65}{12}\,u_3x_4 
                       \Big) \,\zeta_6 \,+\, \Big( \frac{1}{16}
                     \,u_3^2 \,-\, \frac{7}{12}\,u_3x_4 \Big) 
                  \,\Big( \,\zeta_8 \,+\,\zeta_9 \,\Big) \\*[1mm]
      & & \hspace*{-8mm} +\; \frac{1}{12}\,u_3x_1 
                     \,\Big( \,\zeta_{10} \,-\,2\,\zeta_{15} \,\Big) 
                   \,-\,\Big( \frac{1}{8}\,u_3^2 
                          \,+\, \frac{13}{12}\,u_3x_4 \Big) 
                  \,\Big( \,\zeta_{12} \,+\,\zeta_{14} \,\Big) 
                     \,-\, \Big( \frac{1}{8}\,u_3^2 \,+\, 
                  \frac{7}{12}\,u_3x_4 \Big) \, \zeta_{17}
\end{eqnarray*}   
\begin{eqnarray*}  \qquad\qquad & = & -\; \frac{1}{6}\,u_3^2x_3^2 \,+\, 
                       2\,u_3x_1x_3x_4 \,-\, \frac{11}{6}\,u_3x_4^2
                       \,-\, 2\,x_2x_4^2  \,+\, 2\,x_1x_4x_5 
                   \,+\, N_F\,\Big( \frac{u_3}{w_3} \Big)^2 \\*[1mm]
             & & \cdot \; \Big( \frac{2}{3}\,w_3^2z_3^2
                     \,-\, \frac{10}{9}\,w_3z_1z_3z_4
                     \,-\, \frac{2}{9}\,w_3z_4^2
                    \,+\, \frac{2}{9}\,z_1z_4z_5
                       \,+\, \frac{4}{9}\,z_2z_4^2 \Big)
\end{eqnarray*}
\end{small} }}
\begin{equation} \label{skgv49}
\end{equation}\\
\fbox{\parbox{14.8cm}{
\begin{small}
\begin{eqnarray*} & & \hspace*{-8mm} \frac{1}{24}\,u_3^2x_3\,
                      \Big( \,5\,\zeta_1 
               \,+\,5\,\zeta_7 \,+\,17\,\zeta_{13}\,\Big) \,-\, 
             \Big( \frac{1}{2}\, \beta_0\,u_3^2 \,+\, 
                \frac{25}{32}\,u_3^2 \,+\, \frac{5}{24}\,u_3x_4
                       \Big)\,\zeta_2 \\*[1mm]
      & & \hspace*{-8mm} -\; \Big( \frac{1}{4}\,
                    \beta_0\,u_3^2 \,+\, \frac{25}{32}\,u_3^2
                    \,+\, \frac{29}{24}\,u_3x_4
                \Big) \,\zeta_3 \,-\, \Big( \frac{97}{24}\,u_3x_1 
                        \,-\, \frac{5}{8}\,u_3x_3 
                    \,-\, \frac{1}{3} \,N_F\,u_3z_3 
                      \Big) \,\zeta_4 \\*[1mm]
      & & \hspace*{-8mm} -\; \Big( \frac{1}{2}\,
                    \beta_0\,u_3^2 \,-\, \frac{5}{32}\,u_3^2 
                     \,+\, \frac{25}{12}\,u_3x_4 \Big) \,\zeta_6
                     \,+\, \Big( \frac{11}{32}\,u_3^2 
                 \,-\, \frac{5}{24}\,u_3x_4 \Big) \,\zeta_8 \,+\,    
            \Big( \frac{11}{32}\,u_3^2  \,+\, \frac{13}{24}
                        \,u_3x_4 \Big) \,\zeta_9 \\*[1mm]
      & & \hspace*{-8mm} -\; \frac{1}{24}\,u_3x_1 
                 \,\Big( \,7\,\zeta_{10} \,+\, 13 \,\zeta_{15} 
                   \,\Big) \,+\, \Big( \frac{5}{32}\,u_3^2 \,+\,
                   \frac{7}{6}\,u_3x_4 \Big) \, \zeta_{12} \,+\, 
               \Big( \frac{5}{32}\,u_3^2 \,-\,
                   \frac{17}{24}\,u_3x_4 \Big) \, \zeta_{14} \\*[1mm]
      & & \hspace*{-8mm} -\;  \Big( \frac{1}{8}\,u_3^2 \,+\,
                   \frac{5}{24}\,u_3x_4 \Big) \, \zeta_{17}
\end{eqnarray*}   
\begin{eqnarray*}  \qquad\qquad & = & \frac{7}{12}\,u_3^2x_3^2 \,-\, 
                       u_3x_1x_3x_4 \,+\, \frac{5}{12}\,u_3x_4^2
                       \,+\, x_2x_4^2  \,-\, x_1x_4x_5 
                 \,+\, N_F\,\Big( \frac{u_3}{w_3} \Big)^2 \\*[1mm]
             & & \cdot \; \Big( \frac{2}{3}\,w_3^2z_3^2
                     \,-\, \frac{7}{9}\,w_3z_1z_3z_4
                     \,-\, \frac{5}{9}\,w_3z_4^2
                    \,+\, \frac{5}{9}\,z_1z_4z_5
                       \,+\, \frac{1}{9}\,z_2z_4^2 \Big)
\end{eqnarray*}
\end{small} }}
\begin{equation} \label{skgv50}
\end{equation}\\
\fbox{\parbox{14.8cm}{
\begin{small}
\begin{eqnarray*} & & \hspace*{-10mm} -\; \Big( \frac{1}{2}\,\beta_0\,u_3^2 
                 \,+\, \frac{25}{16}\,u_3^2 \,-\, 
                \frac{1}{12}\,u_3x_4 \Big) \,\zeta_1  \,-\, 
                       \Big( \frac{13}{3}\,
                    u_3x_1 \,-\, \frac{5}{8}\,u_3x_3 \,-\, 
               \frac{1}{3}\,N_F\,u_3z_3 \,\Big) \,\zeta_2 \\*[1mm]
      & & \hspace*{-10mm} +\; \Big( \frac{17}{12}\,
                    u_3x_1 \,-\, \frac{5}{8}\,u_3x_3 \,-\, 
               \frac{1}{3}\,N_F\,u_3z_3 \Big) \,\zeta_3 
                   \,+\, \Big( \frac{19}{6}\,u_3x_1 \,-\, 
                 \frac{5}{4}\,u_3x_3 \,-\, 
                      \frac{2}{3}\,N_F\,u_3z_3 \Big) \,\zeta_6  \\*[1mm]
      & & \hspace*{-10mm} +\; \Big( \frac{11}{16}\,u_3^2 \,+\, 
                \frac{1}{12}\,u_3x_4 \Big) \,\zeta_7  \,-\, 
                 \frac{1}{12}\,u_3x_1 \, \Big( \,7\,\zeta_8 \,-\,
                   14\,\zeta_9  \,-\, 26\,\zeta_{12} \,+\, 
                    13\,\zeta_{14} \,+\, 7\,\zeta_{17} \,\Big) \\*[1mm]
      & & \hspace*{-10mm} +\; \Big( \frac{5}{16}\,u_3^2 \,-\, 
                     \frac{1}{6}\,u_3x_4 \Big) \,\zeta_{13}
\end{eqnarray*}   
\begin{eqnarray*}  \quad\qquad & = & u_3^2x_3 \,-\, 
                       2\,u_3x_1^2x_3 \,+\, \frac{5}{6}\,u_3x_1x_4
                       \,-\, 2\,x_1^2x_5  \,-\, x_1x_2x_4
               \,+\, N_F\,\Big( \frac{u_3}{w_3} \Big)^2 \\*[1mm]
             & & \cdot \; \Big( \frac{2}{3}\,w_3^2z_3 \,-\,
                        \frac{10}{9}\,w_3z_1z_4
                     \,-\, \frac{2}{9}\,w_3z_1^2z_3
                    \,+\, \frac{2}{9}\,z_1z_2z_4
                       \,+\, \frac{4}{9}\,z_1^2z_5 \Big)
\end{eqnarray*}
\end{small} }}
\begin{equation} \label{skgv51}
\end{equation}\\
\fbox{\parbox{14.8cm}{
\begin{small}
\begin{eqnarray*} & & \hspace*{-12mm} -\; \Big( \frac{1}{2}\,\beta_0\,u_3^2 
                 \,+\,\frac{23}{32}\,u_3^2 \,+\, \frac{7}{24}\,u_3x_4 
                       \Big) \,\zeta_1 \,-\, \Big( \frac{95}{24}\,
                    u_3x_1 \,-\, \frac{5}{8}\,u_3x_3 \,-\, 
               \frac{1}{3}\,N_F\,u_3z_3 \,\Big) \,\zeta_2 \\*[1mm]
      & & \hspace*{-12mm} -\; \Big( \frac{11}{24}\,u_3x_1 \,-\, 
                 \frac{5}{16}\,u_3x_3 \,-\, 
                      \frac{1}{6}\,N_F\,u_3z_3 \Big) \,\zeta_3
             \,-\, \Big( \frac{23}{24}\,u_3x_1 \,-\, 
                 \frac{5}{8}\,u_3x_3 \,-\, 
                   \frac{1}{3}\,N_F\,u_3z_3 \Big) \,\zeta_6 \\*[1mm]
      & & \hspace*{-10mm} +\; \Big( \frac{13}{32}\,u_3^2 \,-\, 
                \frac{7}{24}\,u_3x_4 \Big) \,\zeta_7  \,-\, 
                 \frac{1}{24}\,u_3x_1 \, \Big( \,5\,\zeta_8 \,+\,
                   8\,\zeta_9  \,+\, 11\,\zeta_{12} \,+\, 
                    17\,\zeta_{14} \,+\, 5\,\zeta_{17} \,\Big) \\*[1mm]
      & & \hspace*{-10mm} +\; \Big( \frac{1}{32}\,u_3^2 \,-\, 
                     \frac{13}{24}\,u_3x_4 \Big) \,\zeta_{13}
\end{eqnarray*}
\begin{eqnarray*}  \quad\qquad & = & 3\,u_3^2x_3 \,+\, 
                       u_3x_1^2x_3 \,-\, \frac{17}{12}\,u_3x_1x_4
                       \,+\, x_1^2x_5  \,-\, x_1x_2x_4 
              \,+\, N_F\,\Big( \frac{u_3}{w_3} \Big)^2 \\*[1mm]
             & & \cdot \; \Big( \frac{2}{3}\,w_3^2z_3 \,-\,
                        \frac{7}{9}\,w_3z_1z_4
                     \,-\, \frac{5}{9}\,w_3z_1^2z_3
                    \,+\, \frac{5}{9}\,z_1z_2z_4
                       \,+\, \frac{1}{9}\,z_1^2z_5 \Big)
\end{eqnarray*}
\end{small} }}
\begin{equation} \label{skgv52}
\end{equation}\\
\subsection{$C_E$--$L_{-}$ equations:}
\vspace{6mm}
\fbox{\parbox{14.8cm}{
\begin{small}
\begin{eqnarray*} & & \hspace*{-6mm} \frac{1}{8}\,u_3^2x_3\,\Big( \,2\,\zeta_1 
               \,+\,2\,\zeta_7 \,+\,5\,\zeta_{13}\,\Big) \,-\, 
             \Big( \frac{15}{16}\,u_3^2 \,+\, \frac{1}{4}\,u_3x_4
                 \Big)\,\zeta_2 \,-\, \Big( \frac{3}{4}\,
                    \beta_0\,u_3^2 \,+\, \frac{15}{16}\,u_3^2 
                     \,+\, u_3x_4 \Big) \,\zeta_3 \\*[1mm]
      & & \hspace*{-6mm} +\; \frac{1}{8}\,u_3x_1\,
                        \Big( \,2\,\zeta_4 \,+\,2\,\zeta_{10} 
                     \,+\,5\,\zeta_{15}\,\Big) \,-\, \Big( 
                    \frac{3}{32}\,u_3^2 \,+\, \frac{5}{2}\,u_3x_4 
                \Big) \,\zeta_6 \,+\, \Big( \frac{3}{16}\,u_3^2 
                     \,-\, \frac{1}{4}\,u_3x_4 \Big) \,\Big(
                        \,\zeta_8 \,+\,\zeta_{17}\,\Big)  \\*[1mm]
      & & \hspace*{-6mm} +\; \Big( \frac{3}{16}\,u_3^2 
                 \,+\, \frac{1}{2}\,u_3x_4 \Big) \,\zeta_9
                 \,-\, \Big( \frac{3}{32}\,u_3^2 
                 \,-\, \frac{5}{4}\,u_3x_4 \Big) \,\zeta_{12}
                 \,-\, \Big( \frac{3}{32}\,u_3^2 
                 \,+\, \frac{5}{8}\,u_3x_4 \Big) \,\zeta_{14}
\end{eqnarray*}   
\begin{eqnarray*}  \qquad\qquad & = & -\; \frac{1}{2}\,u_3^2x_3^2 \,+\, 
                      \frac{1}{2}\,u_3x_4^2
                       \,-\, 3\,x_2x_4^2 \,+\, N_F\,\Big( 
                            \frac{u_3}{w_3} \Big)^2 \,
                  \Big( \frac{1}{3}\,w_3z_1z_3z_4  \\*[1mm]
             & & -\; \frac{1}{3}\,w_3z_4^2 \,+\, 
                    \frac{1}{3}\,z_1z_4z_5
                       \,-\, \frac{1}{3}\,z_2z_4^2 \Big)
\end{eqnarray*}
\end{small} }}
\begin{equation} \label{skgv53}
\end{equation}\\
\fbox{\parbox{14.8cm}{
\begin{small}
\begin{eqnarray*} & & \hspace*{-9mm} -\; \Big( \frac{15}{16}\,u_3^2 \,+\, 
                \frac{1}{4}\,u_3x_4 \Big) \,\zeta_1 \,+\, 
             \frac{1}{8}\,u_3x_1\,\Big( \,2\,\zeta_2 \,-\,
                    25\,\zeta_6 \,+\, 2\,\zeta_8 \,+\, 
            2\,\zeta_9 \,+\, 5\,\zeta_{12} \,+\, 5\,\zeta_{14}
                           \,+\, 2\,\zeta_{17} \,\Big)  \\*[1mm]
      & & \hspace*{-9mm} +\; \Big( \frac{35}{8}\,u_3x_1
                        \,-\, \frac{15}{16}\,u_3x_3 \,-\, 
                      \frac{1}{2}\,N_F\,u_3z_3 \Big) \,\zeta_3 
                   \,+\, \Big( \frac{3}{16}\,u_3^2 \,-\,
                  \frac{1}{4}\,u_3x_4 \Big) \,\zeta_7 
                 \,-\, \Big( \frac{3}{32}\,u_3^2 \,+\, 
                \frac{5}{8}\,u_3x_4 \Big) \,\zeta_{13}       
\end{eqnarray*}  
\begin{eqnarray*}  \quad\qquad & = & \frac{1}{2}\,u_3x_1x_4
                       \,+\, N_F\,\Big( \frac{u_3}{w_3} \Big)^2 \,
                  \Big( -\,\frac{1}{3}\,w_3z_1z_4
                     \,+\, \frac{1}{3}\,w_3z_1^2z_3 \\*[1mm]
             & & -\; \frac{1}{3}\,z_1z_2z_4
                       \,+\, \frac{1}{3}\,z_1^2z_5 \Big)
\end{eqnarray*}
\end{small} }}
\begin{equation} \label{skgv54}
\end{equation}

\end{appendix}
%
%                                                      List of References
%                                                      ------------------  
\newpage                                              
\begin{center}

\end{center}
%                               Inclusion of Postscript Figures 1 -- 11
%                               ---------------------------------------  
%
% Insertion of Postscript figures 1 - 11 for 4-gluon-vertex paper
%----------------------------------------------------------------
%
%\pagestyle{plain}
%
%\vspace*{-3cm}
%
%\noindent
%
\begin{figure}[t]                                            % Figure 1
%hspace{mm}
\begin{center}
\epsfig{file=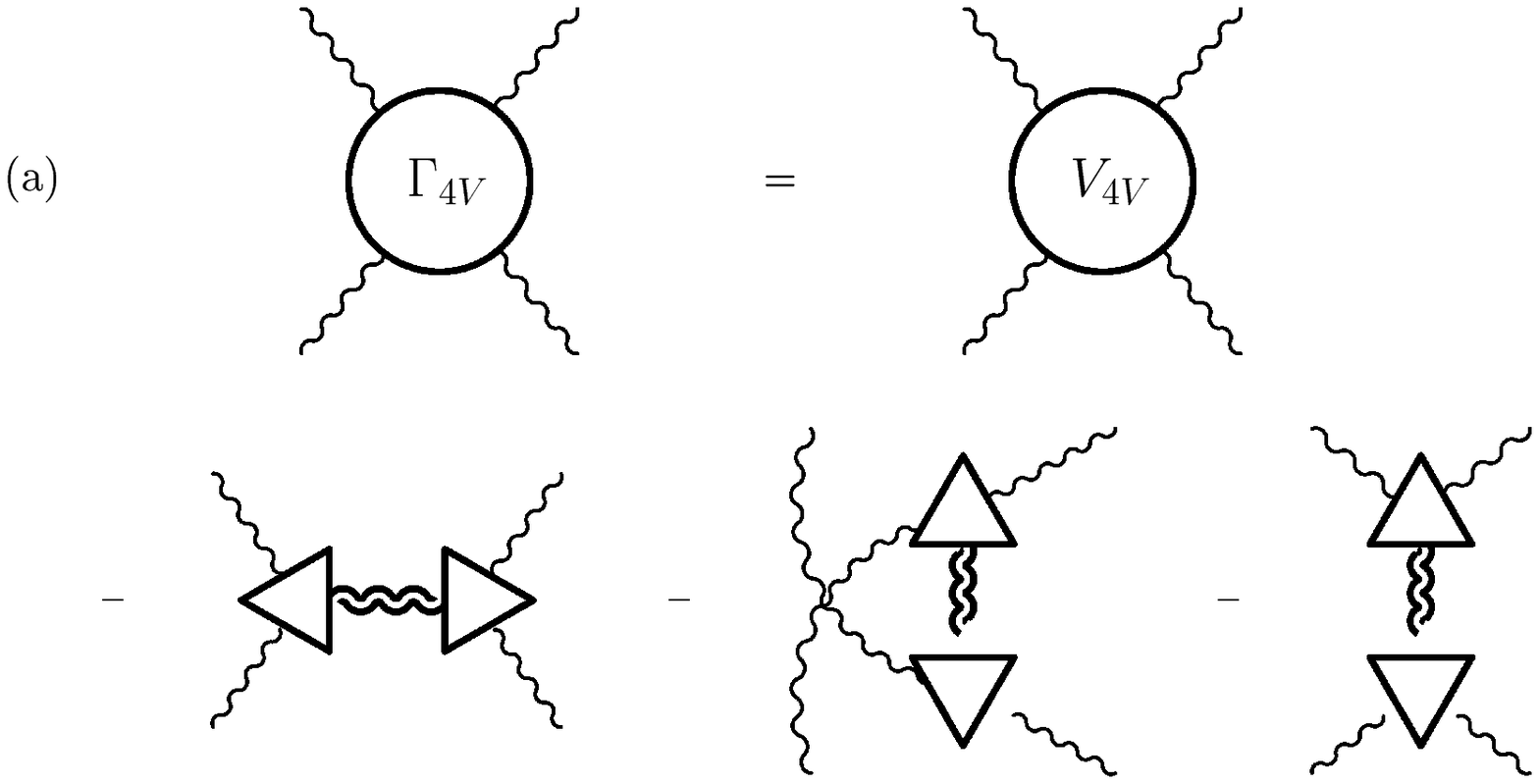,width=15cm}
\vspace{-13cm}

\noindent
\epsfig{file=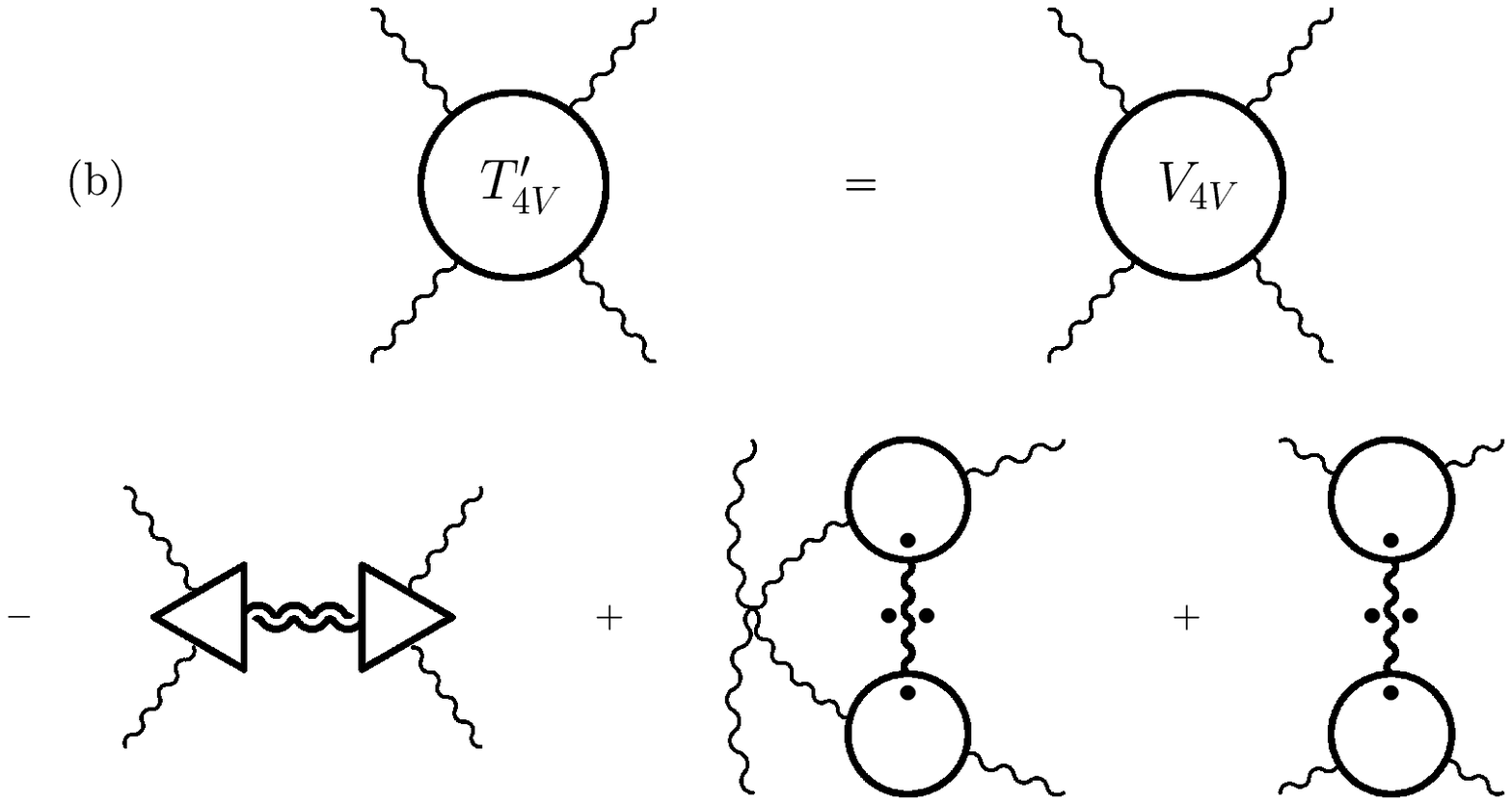,width=15cm}
\vspace{-11cm}

\caption{ \textbf{(a)} Decomposition of 4-gluon vertex into reduced
vertex and negative-shadow (``compensating'') poles. \textbf{(b)} 
Decomposition of partially irreducible 4-gluon amplitude $T'$ into
reduced vertex, compensating pole, and softened 1-gluon exchanges. }
\end{center}
\end{figure}
\pagestyle{plain}
%
%---------------------------------------------------------------------
%
\begin{figure}[t]                                           % Figure 2
%hspace{mm}
\begin{center}
\epsfig{file=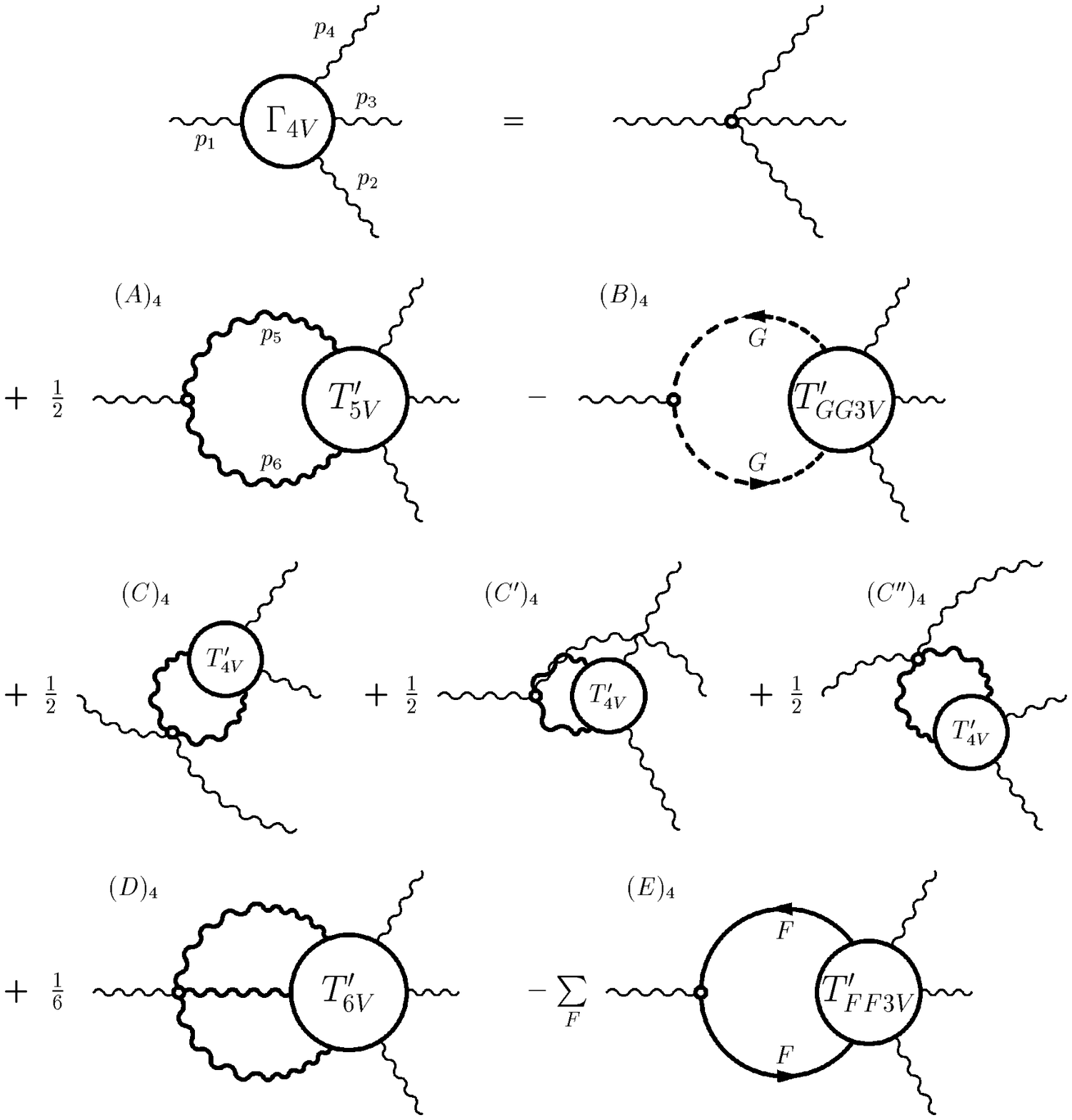,width=17cm}
\vspace{-7cm}

\caption{ DS equation for proper 4-gluon vertex in compact form, featuring
partially irreducible $T$ matrices. }
\end{center}
\end{figure}
%
%---------------------------------------------------------------------
%
\begin{figure}[t]                                           % Figure 3
%hspace{mm}
\begin{center}
\epsfig{file=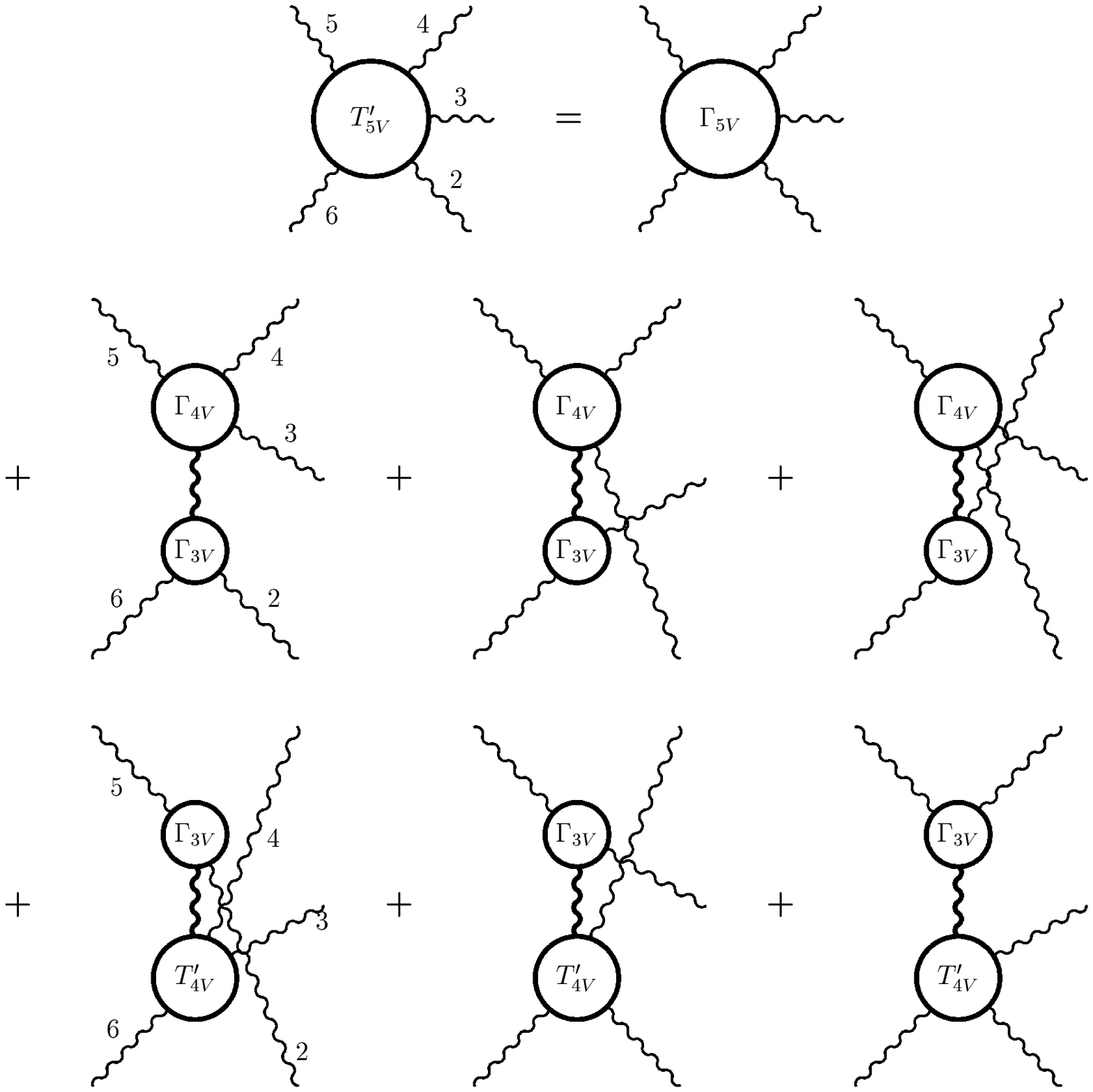,width=17cm}
\vspace{-7cm}

\caption{ Representation of partially 1-gluon-irreducible 5-gluon $T$ matrix
$T'_{5V}$. $\Gamma_{5V}$ denotes the proper 5-gluon vertex. An equivalent
representation has the roles of $\Gamma_{4V}$ and $T'_{4V}$ interchanged. }
\end{center}
\end{figure}
%
%---------------------------------------------------------------------
%
\begin{figure}[t]                                           % Figure 4
%hspace{mm}
\begin{center}
\epsfig{file=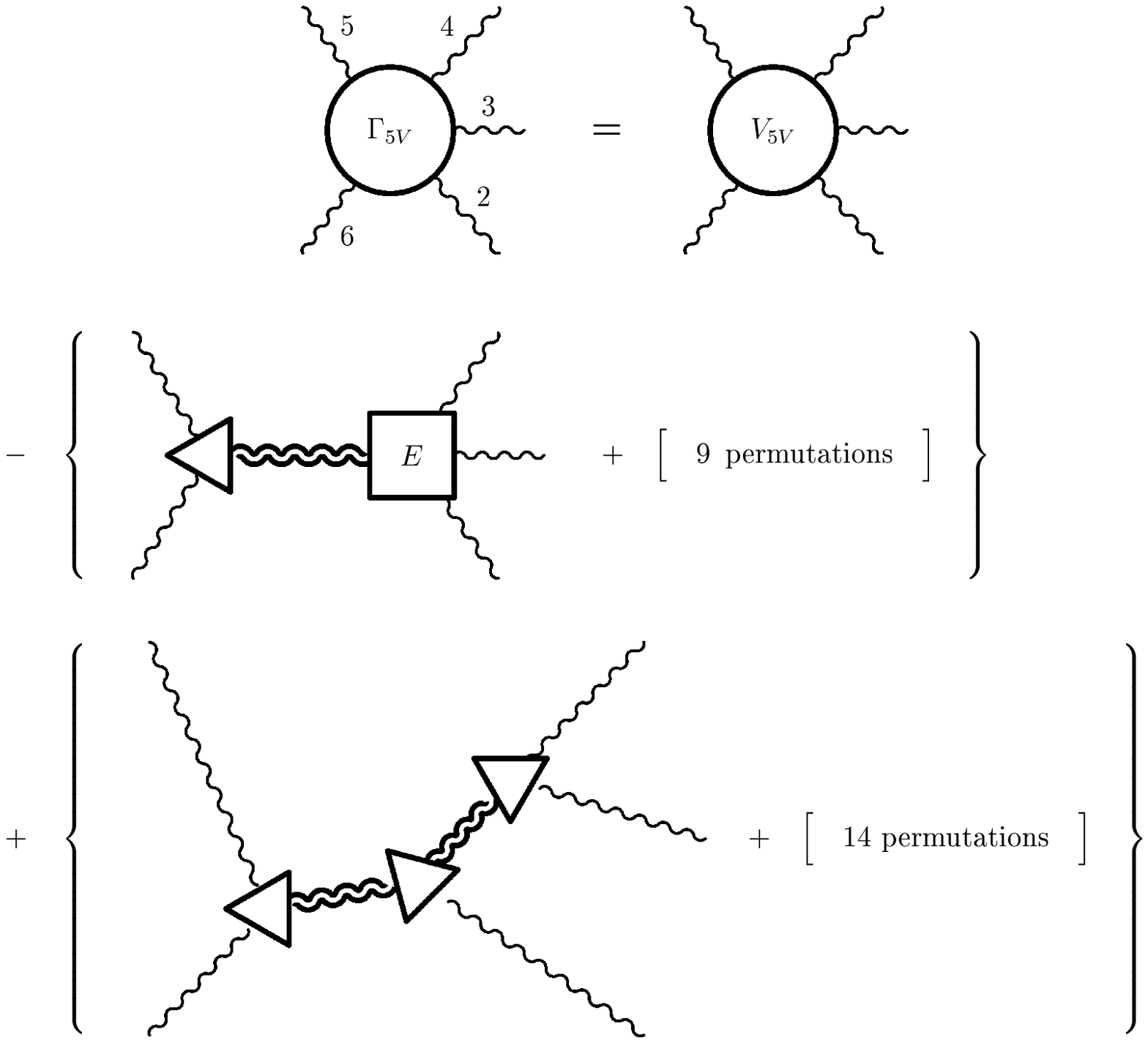,width=16cm}
\vspace{-8cm}

\caption{ Decomposition of proper 5-gluon vertex into reduced ( 1-gluon-
and 1-shadow-irreducible ) vertex $V_{5V}$, negative one-shadow terms in
the 10 channels, and positive two-shadows terms for the 15 (2+1+2)
partitions of the external legs. }
\end{center}
\end{figure}
%
%---------------------------------------------------------------------
%
\begin{figure}[t]                                           % Figure 5
%hspace{mm}
\begin{center}
\epsfig{file=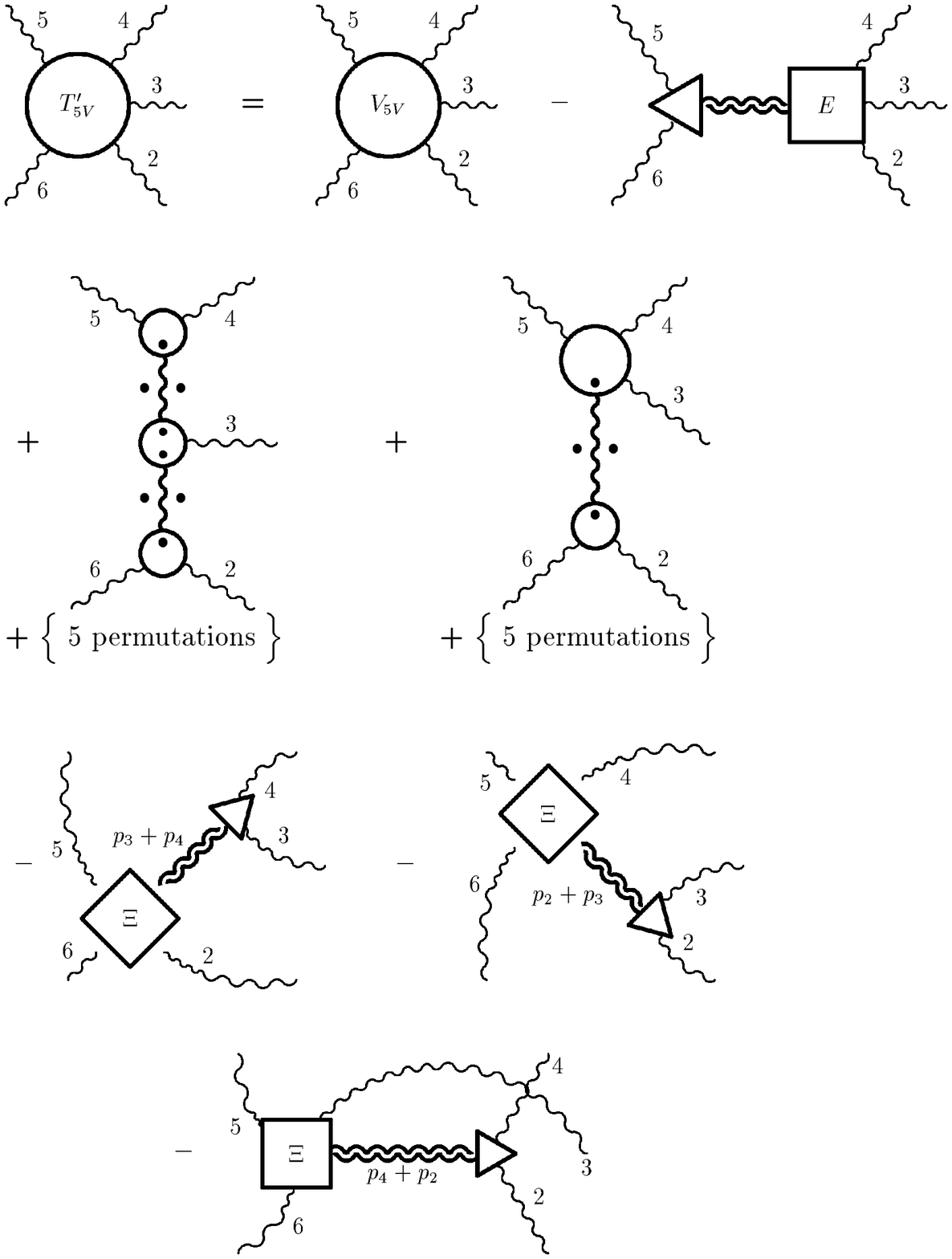,width=16cm}
\vspace{-3cm}

\caption{ Another representation of the $T'_{5V}$ amplitude. The $\Xi$
amplitudes in the last two lines are defined by Fig.7. }
\end{center}
\end{figure}
%
%---------------------------------------------------------------------
%
\begin{figure}[t]                                           % Figure 6
%hspace{mm}
\begin{center}
\epsfig{file=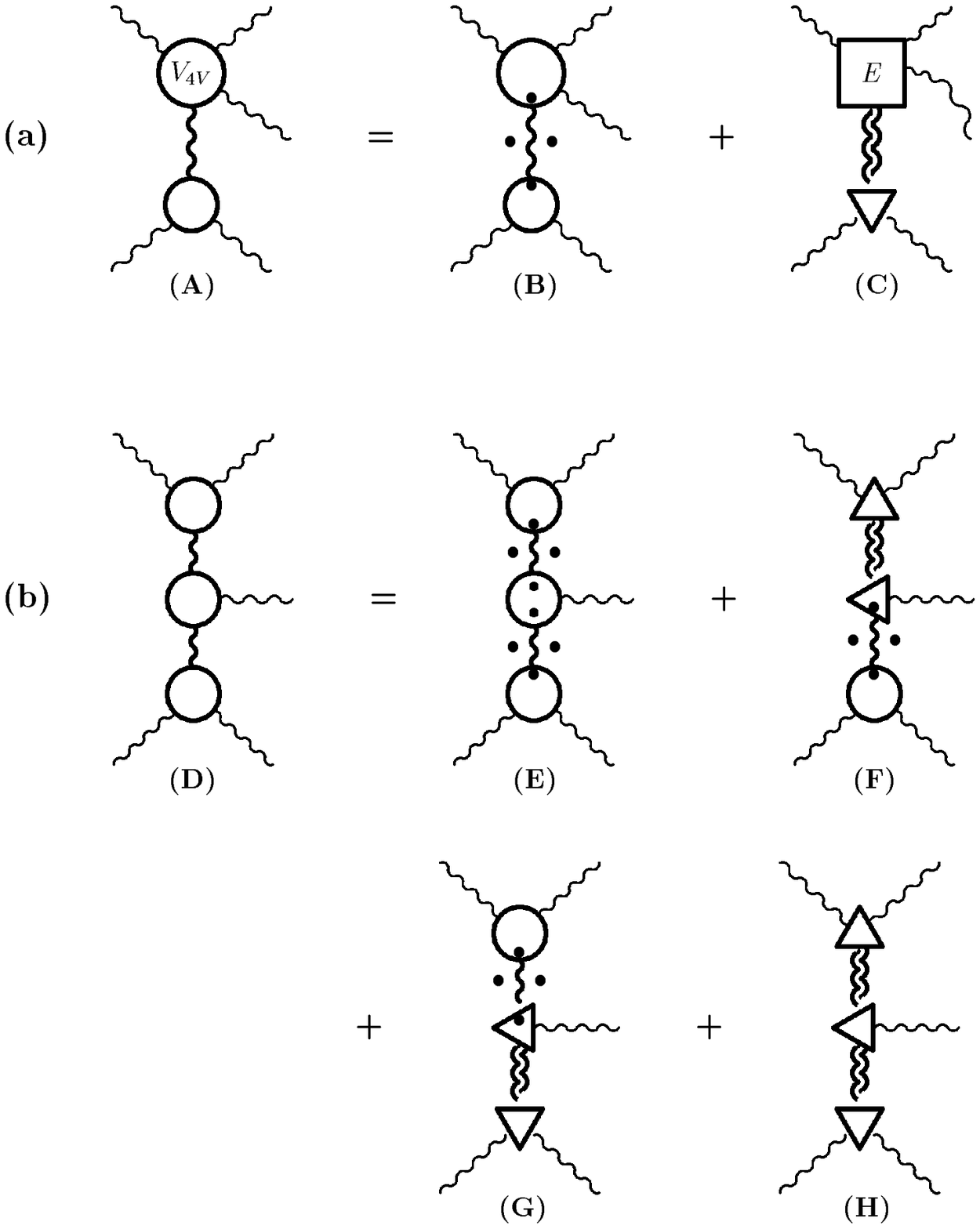,width=18cm}
\vspace{-7cm}

\caption{ Definition of (a) single and (b) double 1-shadow-irreducible
(``softened'') gluon exchanges, marked by dots, through extraction of
shadow terms from ordinary single and double gluon exchanges. }
\end{center}
\end{figure}
%
%---------------------------------------------------------------------
%
\begin{figure}[t]                                           % Figure 7
%hspace{mm}
\begin{center}
\epsfig{file=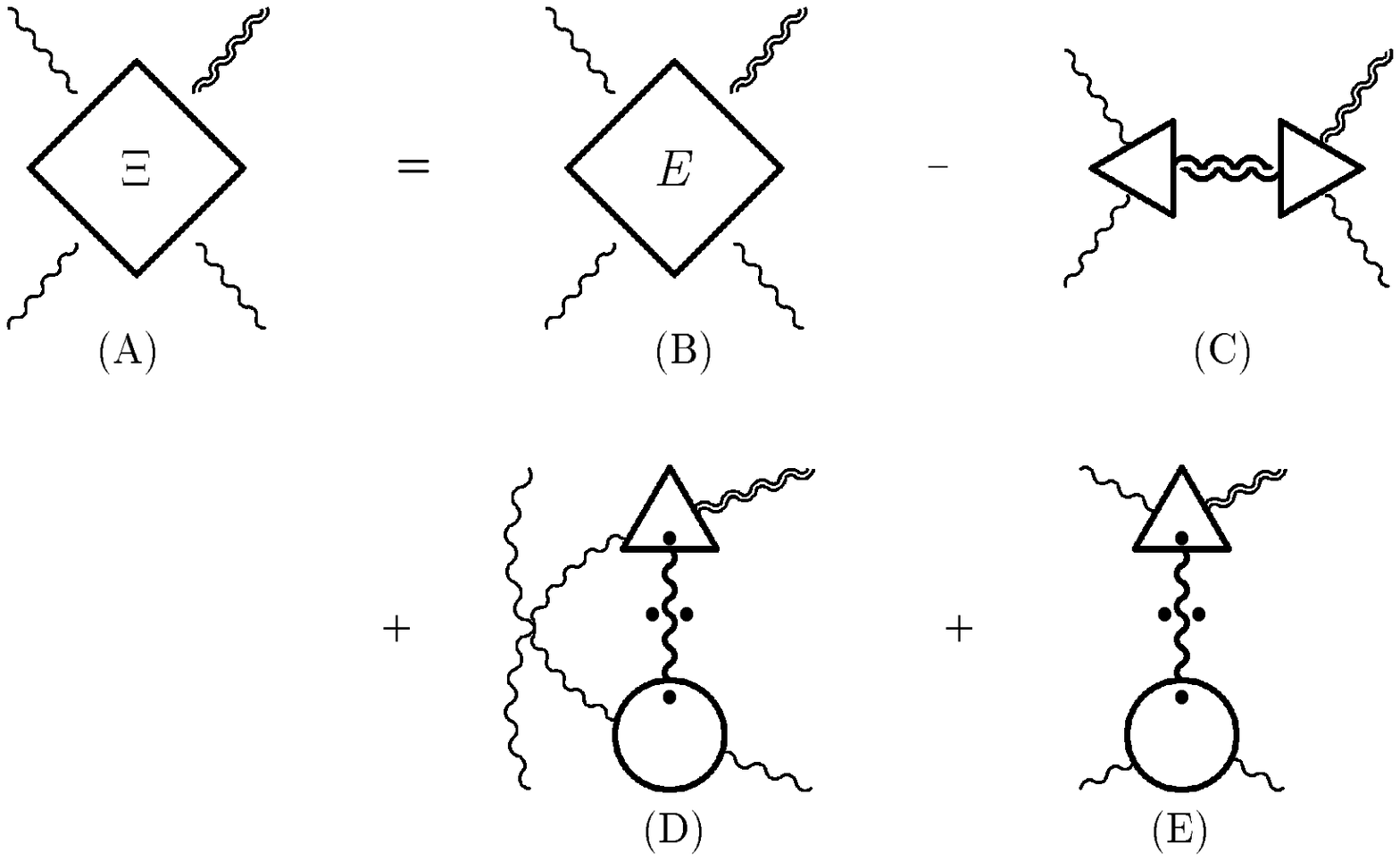,width=15cm}
\vspace{-12cm}

\caption{ Definition of 3-gluon-1-shadow auxiliary amplitudes $\Xi$. }
\end{center}
\end{figure}
%\vspace{cm}
%
%\noindent
%
\begin{figure}[hb]                                           % Figure 8
%hspace{mm}
\begin{center}
\epsfig{file=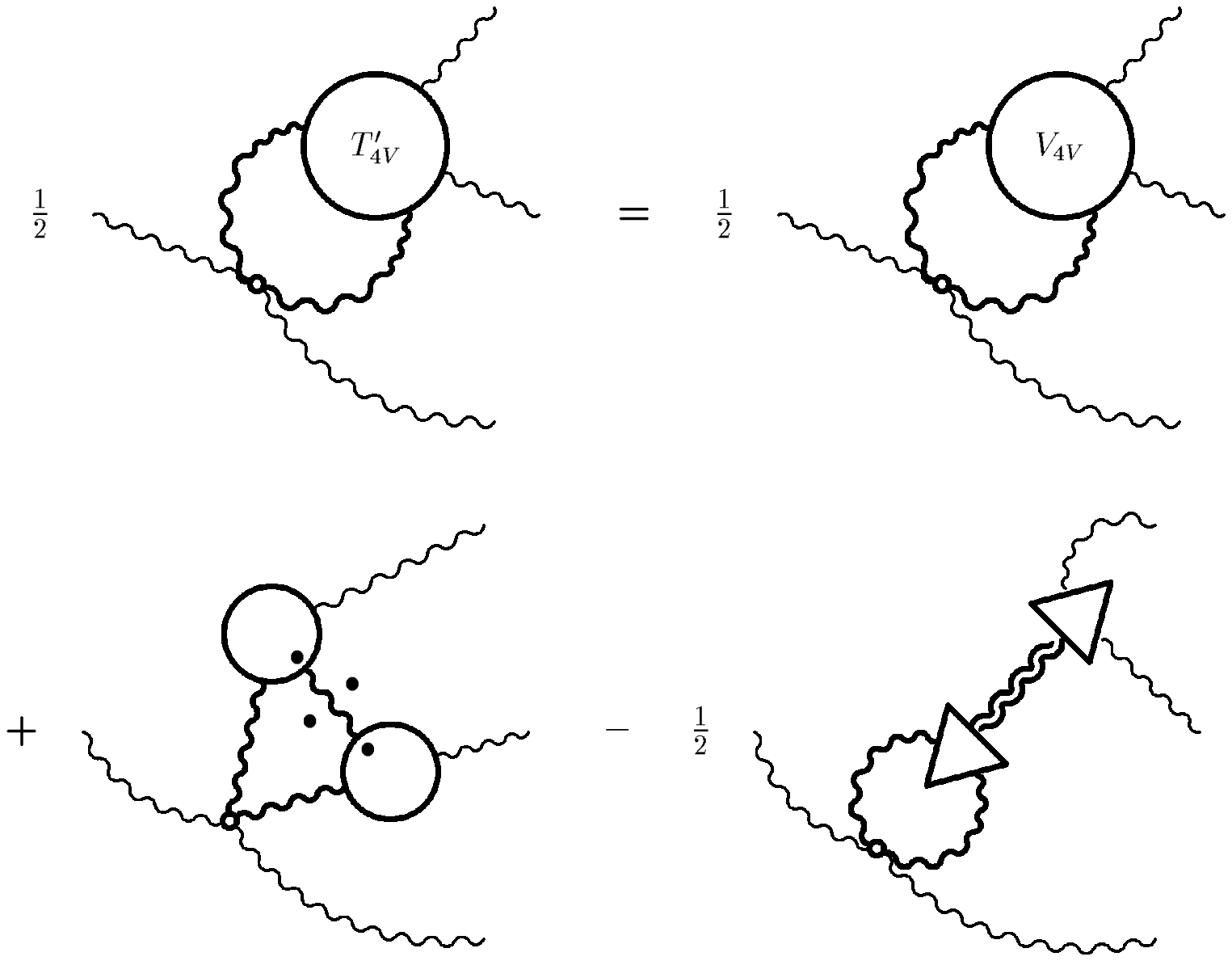,width=14cm}
\vspace{-10cm}

\caption{ Decomposition of the $(C)_4$ term of Fig.2, obtained by use of
Fig.1(b). }
\end{center}
\end{figure}
%
%---------------------------------------------------------------------
%

\vspace*{-12cm}

\begin{figure}[t]                                           % Figure 9
%hspace{mm}
\begin{center}
\epsfig{file=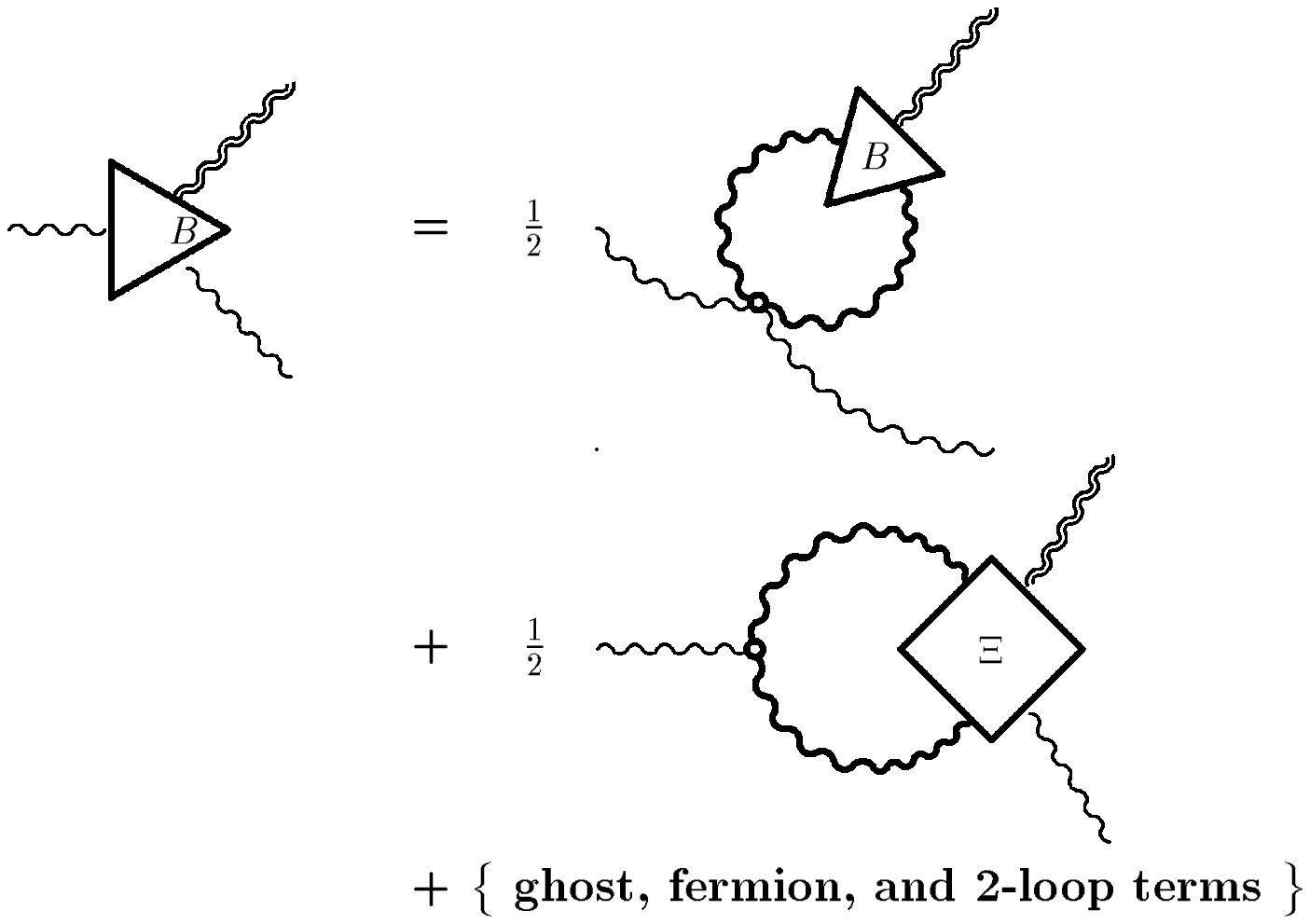,width=17cm}

\vspace{-13cm}

\caption{ DS-like integral equation for partial amplitudes $B_n$ $(n\geq 1)$
of 3-gluon vertex $\Gamma_{3V}$. The auxiliary amplitude $\Xi$ in the second
line is defined in Fig.7. }
\end{center}
\end{figure}
%
%
%\setcounter{figure}{10}
%\begin{figure}[hb]
%\begin{center}
%\epsfig{file=figu11.ps,width=19cm}
%
%\vspace*{-14cm}  
%
%\noindent
%\caption{ Diagrammatic representation of Bethe-Salpeter normalization
%condition. The partial differentiations with respect to $P_{\mu}$ act
%on the portions in dashed brackets. }
%\end{center}
%\end{figure}
%
%
%----------------------------------------------------------------------
%
%\setcounter{figure}{9}
%
\begin{figure}[t]                                           % Figure 10
%hspace{mm}
\begin{center}
\epsfig{file=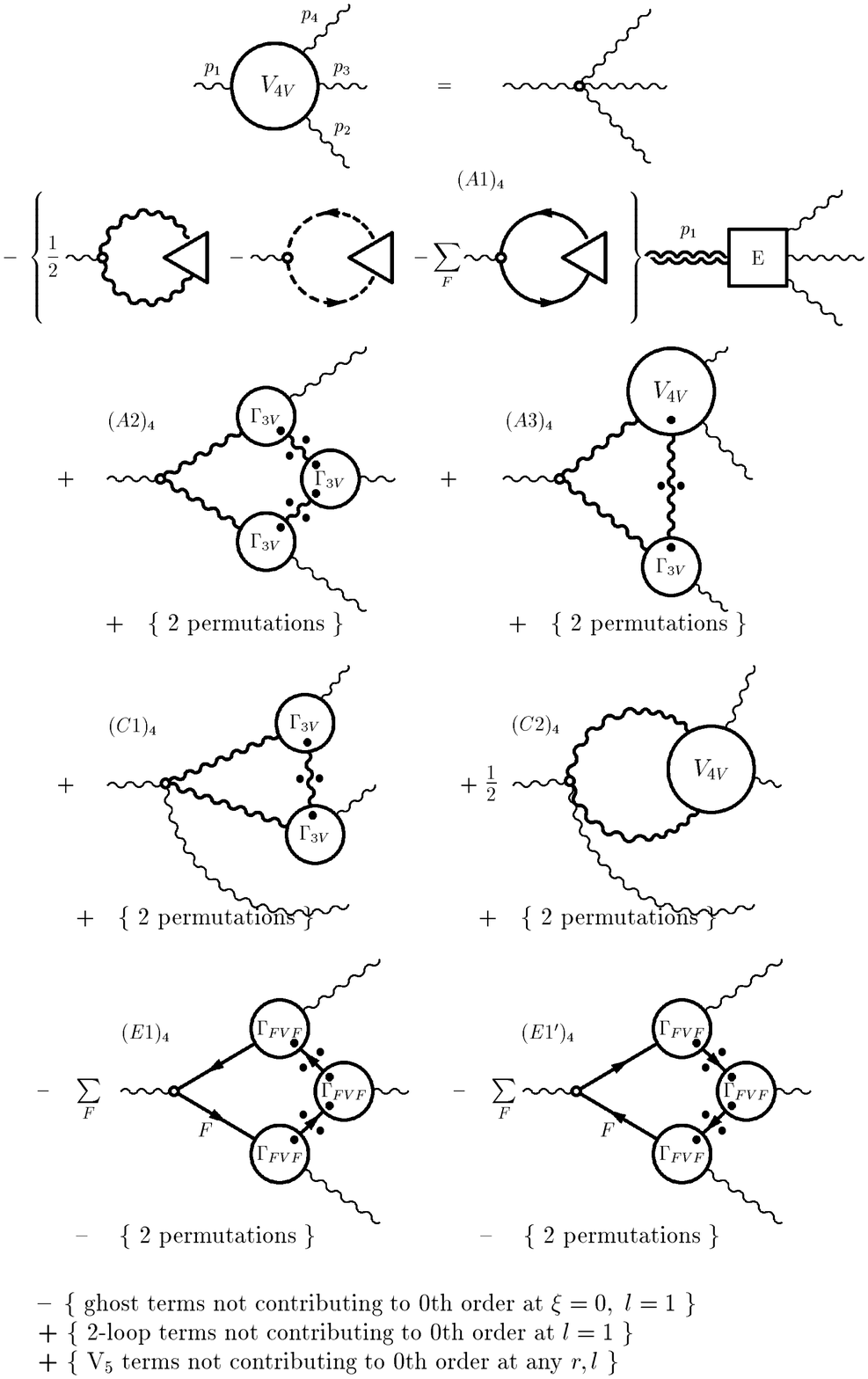,height=21cm}
\vspace{3mm}

\caption{ DS equation for reduced 4-gluon vertex $V_{4V}$, obtained
from Fig. 2 upon using Figs. 5, 8, and 9. }
\end{center}
\end{figure}
%
%----------------------------------------------------------------------
%
\begin{figure}[t]                                          % Figure 11
\begin{center}
\epsfig{file=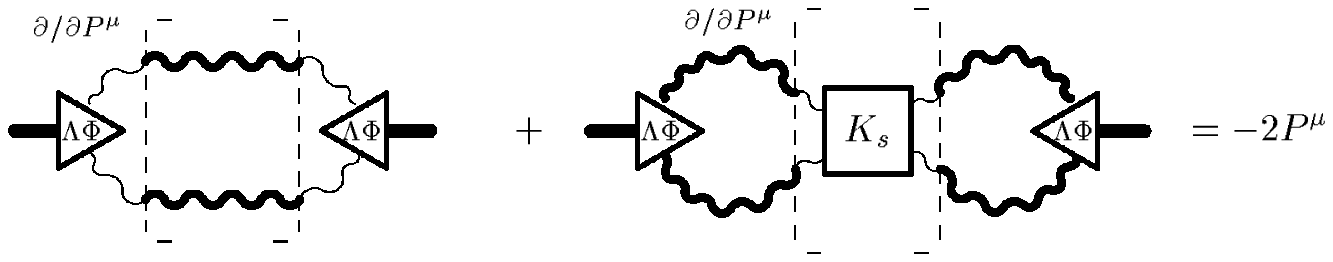,width=17cm}
\vspace{-16cm}

\caption{ Diagrammatic representation of Bethe-Salpeter normalization
condition. The partial differentiations with respect to $P_{\mu}$ act
on the portions in dashed brackets. }
\end{center}
\end{figure}
%
%----------------------------------------------------------------------
%
% End of PS-figures insertion

%
\end{document}